\pdfoutput=1

\documentclass[preprint]{aastex62}

\usepackage{soul}
\usepackage{colortbl}
\graphicspath{{./}{figures/}}

\received{April 4th, 2018}
\revised{May 14th, 2018}
\accepted{June 27th, 2018}
\submitjournal{ApJ}

\shorttitle{A comprehensive model of the meteoroid environment around Mercury}
\shortauthors{Pokorn\'{y} et al.}

\begin{document}

\title{A comprehensive model of the meteoroid environment around Mercury}

\correspondingauthor{Petr Pokorn\'{y}}
\email{petr.pokorny@nasa.gov}

\author{Petr Pokorn\'{y}}
\affiliation{Department of Physics, The Catholic University of America, Washington, DC 20064, USA}
\affiliation{Heliophysics Science Division, NASA Goddard Space Flight Center, Greenbelt, MD 20771}

\author{Menelaos Sarantos}
\affiliation{Heliophysics Science Division, NASA Goddard Space Flight Center, Greenbelt, MD 20771}

\author{Diego Janches}
\affiliation{Heliophysics Science Division, NASA Goddard Space Flight Center, Greenbelt, MD 20771}

\begin{abstract}



To characterize the meteoroid environment around Mercury and its contribution to the planet's exosphere, we combined four distinctive sources of meteoroids in the solar system: main-belt asteroids, Jupiter family comets, Halley-type comets, and Oort Cloud comets. All meteoroid populations are described by currently available dynamical models. We used a recent calibration of the meteoroid influx onto Earth as a constraint for the combined population model on Mercury. We predict vastly different distributions of orbital elements, impact velocities and directions of arrival for all four meteoroid populations at Mercury.  We demonstrate that the most likely model of Mercury's meteoroid environment- in the sense of agreement with Earth -provides good agreement with previously reported observations of Mercury's exosphere by the MESSENGER spacecraft and is not highly sensitive to variations of uncertain parameters such as the ratio of these populations at Earth, the size frequency distribution, and the collisional lifetime of meteoroids. Finally, we provide a fully calibrated model consisting of high-resolution maps of mass influx and surface vaporization rates for different values of Mercury's true anomaly angle. 

\end{abstract}



\section{Introduction}

In this manuscript we aim to address three questions regarding the Hermean meteoroid environment: 1) how the distribution of meteoroid impacts on Mercury's surface changes with local time, latitude and true anomaly angle, 2) what are the dynamical characteristics of the Hermean meteoroid environment, and 3) what is the current calibration of the Hermean meteoroid environment and what is the sensitivity of such a calibration to uncertainties and free parameters of dynamical meteoroid models.
Knowing the calibrated distribution of meteoroid impacts on surfaces of airless bodies is crucial for exosphere models. Here, we aim to provide the first calibrated dynamical model for all major sources of meteoroids in the inner solar system. 

This manuscript generalizes the work reported by \citet{Pokorny_etal_2017_APJL}, who showed for the first time that the meteoroid environment at Mercury is strongly directional and differs significantly from that at Earth due to the non-zero eccentricity of Mercury's orbit.   \replaced{In this work}{In the current work}, we combined four different populations of meteoroids, assumed to be the major contributors to the meteoroid complex in the inner solar system, and  quantified their respective influx rates at Mercury in terms of tons/day, whereas only relative contributions of just two populations were given in \citet{Pokorny_etal_2017_APJL}. \added{These four meteoroid source populations are: main-belt asteroids, Jupiter Family comets (generally comets with orbital periods below 20 years and inclinations below $30^\circ$, Halley type comets (orbital periods between 20 and 200 years), and Oort cloud comets (orbital periods larger than 200 years).} Furthermore, we examined the influence of collisions and the size-frequency distribution, which are uncertain, on the predicted environment around Mercury. To test predictions from this Hermean environment model we used measurements of Mercury's exosphere obtained by the MESSENGER spacecraft, which show evidence that impact vaporization is an important source mechanism for this exosphere. Our calculations provide not only improved total vaporization rates from impacts by estimating velocity distribution functions of meteoroids beyond those reported by \citet{Cintala_1992}, but also high-resolution maps of impact vaporization rates and how they change with the true anomaly angle. Additionally, our calculations provide clues to the transport time of different meteoroids from their source to Mercury, which could improve estimates of volatile influx to Mercury.

Meteoroids originating from main-belt asteroids (AST) have been studied in the past in various size ranges and configurations. \citet{Marchi_etal_2005} extensively studied AST meteoroids with diameter $D>2$ cm, a size range that is not significantly influenced by radiative forces from the sun, namely the radiation pressure and the Poynting-Robertson (PR) drag force. Hence, the AST meteoroids in \citet{Marchi_etal_2005} are assumed to be subject only to gravitational forces, reflecting the influence of the mean-motion and secular resonances. Compared to \citet{Cintala_1992} this paper reported a broad meteoroid impact velocity range on Mercury ranging from 20 to 80 km s$^{-1}$. Since the effects of PR drag are negligible in this size range, such an outcome is expected because in order for the AST meteoroids to impact Mercury, they would need to be ejected on highly eccentric orbits similar to near-Earth asteroids with low perihelia. Later, \citet{Borin_etal_2009,Borin_etal_2016b} focused on AST meteoroids with $D<2$ cm. These authors showed that PR drag effectively circularizes the AST meteoroid orbits, resulting in a narrow impact velocity distribution peaking around 12 km s$^{-1}$, which is less than those reported by \citet{Cintala_1992}. Furthermore, \citet{Borin_etal_2016b} predicted that the majority of AST meteoroid impacts are in latitudes between $-40^\circ \mathrm{
~and~} 40^\circ$ for meteoroids with $D=5 - 100 ~\mu$m. Because  our previous work (\citet{Pokorny_etal_2017_APJL}) omitted AST meteoroids, and they were considered important by Borin et al., we utilized in this work one of the most elaborate models for the AST meteoroids in the solar system based upon \citet{Nesvorny_etal_2010}.


A second population of meteoroids, Jupiter Family comet (JFC) meteoroids, were extensively studied in the past decade and recently \citet{Nesvorny_etal_2010,Nesvorny_etal_2011JFC} concluded that they dominate the inner solar system in mass flux, number flux, and total cross-section in the micrometer to millimeter range. Consequently, there have been many studies focusing mainly on Mercury and its proximity \citep[e.g.][]{Borin_etal_2016a,Pokorny_etal_2017_APJL, Borin_etal_2017}. \citet{Pokorny_etal_2017_APJL} used a combination of JFC meteoroids and Halley-type comet (HTC) meteoroids with the mixing ratio reported by \citet{Carrillo-Sanchez_etal_2016}, while \citet{Borin_etal_2016a,Borin_etal_2017} treated the JFC population separately. In particular, \citet{Pokorny_etal_2017_APJL} showed that the impact directions of JFC meteoroids experience significant motion in the local time reference frame during Mercury's orbit. This is due to the non-zero eccentricity of Mercury and low impact velocities compared to Mercury's orbital velocity \citep{Borin_etal_2016a,Pokorny_etal_2017_APJL}. However,\citep {Pokorny_etal_2017_APJL} used only one meteoroid size, $100 ~\mu$m, and this is something we amend here.

\replaced{A third population, long-period comets,}{Third and fourth source populations,} namely Halley-type comets (HTC) and Oort Cloud comets (OCC), were until recently omitted as a considerable source of meteoroids for the Hermean meteoroid environment. However, \citet{Pokorny_etal_2017_APJL} demonstrated that HTC meteoroids must be considered in any model of the Hermean meteoroid environment since they can easily reach orbits intersecting Mercury at any point of its eccentric orbit. While long period comets are not considered to be a dominant part of the inner solar system meteoroid budget in terms of mass flux, number flux or total meteoroid cross-section \citep{Nesvorny_etal_2011JFC, Pokorny_etal_2014, Carrillo-Sanchez_etal_2016}, their intrinsically high impact velocities may make them the dominant source of a number of physical phenomena. This is due to their scattered disk/Oort cloud origin \citep{Levison_etal_2006,Nesvorny_etal_2017} and possibility of acquiring retrograde orbits. Though the mass flux of long period comet meteoroids at Mercury compared to JFC meteoroids could be small ($\sim 5\%$), their impact velocities reported over 100 km s$^{-1}$ can be dominant in terms of the impact vaporization or the impact yield \citep{Pokorny_etal_2017_APJL, Janches_etal_2018}.

And, finally, the relative contribution of different meteoroid populations at Mercury has seldom been studied, something that motivated the present work. \citet{Borin_etal_2017} came closest to fulfilling this need when they combined their previous studies and presented a comparative study for the whole inner solar system. Using their dynamical models they were able to estimate the mass flux ratio between the inner solar system bodies with Mercury/Earth ratio equal to 35.14. Unfortunately, none of these studies included collisions between modeled meteoroids and the zodiacal cloud, entirely ignoring the fact that a large portion of meteoroids might be lost between the Earth crossing and Mercury crossing orbits. Additionally, long-period comet meteoroids were also omitted in Borin et al. studies, which as pointed out by \citet{Pokorny_etal_2017_APJL} may lead to incorrect conclusions since the long-period comet meteoroids impact Mercury with higher impact velocities as compared to AST and JFC meteoroids. Leaving out the high velocity component of impacting populations severely underestimates the amount of energy deposited into the surface.

\section{Overview of Meteoroid models}
\label{SEC:Models}
Our meteoroid environment model consists of four different meteoroid populations; meteoroids originating from main belt asteroids (AST), Jupiter-Family Comets (JFC), Halley-type comets (HTC), and Oort-Cloud Comets (OCC). All meteoroids are released from their parent objects and tracked on their path through the solar system until they reach one of the following conditions: 1) the meteoroid impacts one of eight planets, 2) the meteoroid is too close to the Sun ($<$0.05 au), or 3) the heliocentric distance of the meteoroid is $>$ 10,000 au. Once one of these conditions is met, the meteoroid is removed from the simulation.  We use the SWIFT\_RMVS3 integrator \citep{Levison_Duncan_2013} to track meteoroids in our simulations. The integrator time steps are the following: ASTs and JFCs are integrated with 9 day time step, while HTCs and OCCs are integrated with 1 day time step. This difference stems from the minimum perihelion distances that meteoroids usually experience during the simulation. Additionally, our use of previously published models in order to decrease the computational burden prevents us from changing the integration time step. The meteoroids are subjected to the effects of gravitation of all eight planets, the effects of radiation pressure, Poynting-Robertson (PR) drag and solar wind \citep[30\% addition to PR drag; see Fig. 11 in][]{Burns_etal_1979}. All particles in our model are considered to be perfectly spherical, with Mie coefficient $Q_{PR}=1$ \replaced{\citep{Burns_etal_1979}}{\citep[see][for an excellent review of radiative forces and complex nature of their effects]{Burns_etal_1979}.}

 As the meteoroids are traced during the simulations, their orbital elements are recorded every 100 years, constructing a history of their pathway through the solar system.
 Assuming that the source populations stay active in time, we create a steady-state model by considering every record of each meteoroid as an individual particle that can collide with another object in the solar system. This approach allows us to treat the collisions between meteoroids and the zodiacal cloud after the integration is completed, considering the resulting steady-state model as a collisionless representation of the meteoroid environment in the solar system.

Our model for AST meteoroids follows that reported by \citet{Nesvorny_etal_2010}. We used the 5,000 largest asteroids in the Main Belt as sources of meteoroids in our model. We chose meteoroid diameters in the range $D=10, 20, 50, 100, 200, 400, 800, 1000, 1200, 1500, 2000~\mu$m, where $D$ is the meteoroid diameter. For each size we generated 20,000 meteoroids with the initial semimajor axis, eccentricity, inclination, and the argument of pericenter taken from the seed population of 5,000 asteroids, where each asteroid has an equal probability to be selected as the parent body. The initial longitude of the ascending node and true anomaly of each meteoroid were set to be a random number uniformly distributed between 0 and $2\pi$. 

For JFC meteoroids we applied the model reported by \citet{Nesvorny_etal_2011JFC}, with the following meteoroid diameters: $D=10, 30, 50, 100, 200, 500, 800, 1000, 1200, 1500, 2000 ~\mu$m. We opted to use the simulations performed by \citet{Nesvorny_etal_2011JFC} instead of creating new ones in order to save the computational time and because the original simulations have suitable resolution for the purpose of this paper. We selected these 11 size bins because they match the best simulations for other parent populations we currently have available.

The HTC meteoroid model uses the simulations performed by \citet{Pokorny_etal_2014}, who followed meteoroids originating from the steady-state population of HTCs \citep{Levison_etal_2006}. For this work we simulated 20,000 individual meteoroids for each diameter, $D= 10, 20, 50, 100, 200, $ $400, 800, 1000, 1200, 1500, 2000 ~\mu$m. 

\citet{Nesvorny_etal_2011OCC} created a model for OCC meteoroids, but we substantially enhanced their original simulations for this work. Here we simulated meteoroids having three different values of the initial semimajor axis: 300 au, 1000 au, 3000 au in order to identify how the initial semimajor axis shapes the final appearance of the meteoroid cloud. We followed meteoroids of eight different diameters with values, $D=10, 20, 50, 100, 200, 400, 800, 1200 ~\mu$m, where we simulated 20,000 meteoroids for each size and initial semimajor axis. This produces almost half a million different trajectories (streams) of meteoroids that are sampled every 100 years and thus allowing us to study the properties of the OCC meteoroid cloud in much greater detail than the original work of \citet{Nesvorny_etal_2011OCC} who used only 1,000 meteoroids for four different sizes. 

Additional information about the initialization of particles from each population can be found in \citet{Janches_etal_2018}.

\section{Methods}
We amend this collisionless, steady state model of meteoroid populations with additional effects that shape the characteristics of the meteoroid environment in the solar system. In this steady state model, sources (asteroids and comets) and sinks (evaporation near the Sun, escape on unbound orbits, and collisions with planets) of meteoroids are assumed to be in balance over at least the last few million years of the solar system's history. With this assumption each recorded orbital state vector of a simulated meteoroid is treated as an individual particle, thus a simulation with 20,000 initial meteoroids results in a cloud of millions of meteoroids that each have their own probability to impact any planet in the solar system. Keeping track of the dynamical age of each individual record allows us to add collisions into the model. From the resulting collisionally evolved model, we then calculate the impact probabilities of all meteoroid records with an object of interest, where the collisional probabilities are weighted based on the collisional fading of meteoroid populations.

Additionally, constructing a meteoroid complex from individual meteoroid populations comprised of meteoroids of various sizes requires setting of the size-frequency distribution (SFD) for all populations. 

\subsection{Collisions}
Mutual collisions between dust grains effectively shape the characteristics of the entire meteoroid cloud. Collisional fading is more significant for meteoroids with longer dynamical lifetimes thus affecting mostly larger meteoroids. This is because their dynamical evolution is much slower under the influence of PR drag which is inversely proportional to the particle's size. In this manuscript, we use a method developed by \citet{Steel_Elford_1986} [henceforth SE86] that uses several analytic expressions to estimate the collisional lifetime between a meteoroid and the meteoroid cloud in the solar system. To calibrate the density of the meteoroid cloud in the solar system used to estimate collisions, we use {\it LDEF} (Long-Duration Exposure Facility) measurements that were recently re-investigated by \citet{Cremonese_etal_2012}. SE86 also discusses the mass ratio between two colliding meteoroids at characteristic impact velocities of 10 km s$^{-1}$ that can cause catastrophic disruptions. The authors concluded that a $5 \times 10^4$ less massive body (30--40 times smaller in radius) caused catastrophic fragmentation of both projectile and target. In this manuscript we assume that meteoroids in our simulations are destroyed by projectiles 30 times smaller in radius or larger. Using results from \citet{Cremonese_etal_2012} and previous assumption of catastrophic collisions we parameterize the decadic logarithm of the spatial density $\psi$ of projectiles at 1 AU for a given diameter $D$ in micrometers by the seventh degree polynomial:
\begin{equation}
\mathrm{log_{10}}(\psi(D)) = a_0+a_1 D+a_2 D^2+a_3 D^3+a_4 D^4+a_5 D^5+a_6 D^6+a_7 D^7,
\end{equation}
with coefficients: 
	$a_0 = 24.5740501811233$
	$a_1 = -0.00273729871839826$
	$a_2 = 2.42366145666713\times 10^{-06}$
	$a_3 = -1.55353572997867\times 10^{-09}$
	$a_4 = 6.15333399708131\times 10^{-13}$
	$a_5 = -1.42903699974705\times 10^{-16}$
	$a_6 = 1.76861418259828\times 10^{-20}$
	$a_7 = -8.9675339812721\times 10^{-25}$, \added{where $\psi$ is in units number of particles per m$^2$ s, i.e. the particle flux in SI units.}
	
As in SE86, the spatial density of meteoroids is assumed to decrease with heliocentric distance $r$ as $\psi \propto r^{-1.3}$, and varies with the ecliptic latitude $\beta$ as $(1-\sin \beta)$. Recalling the conclusions of \citet{Nesvorny_etal_2011JFC} and \citet{Pokorny_etal_2014}, who reported that the nominal collisional lifetime is too short for meteoroids from short-period comets to dynamically evolve enough to obtain the observed distribution of orbital elements, we adopt a fudge factor $F_\mathrm{coll}$ that extends the nominal collisional lifetime. In the following calculations we used six different values of $F_\mathrm{coll} = 1, 5 ,10, 20, 50, \mathrm{~and~} \infty$, where $F_\mathrm{coll}=20$ was previously used in \citet{Pokorny_etal_2014} and \citet{Pokorny_etal_2017_APJL}, and values in the range $F_\mathrm{coll}=20-50$ are comparable to those in \citet{Nesvorny_etal_2011JFC} who used a different collisional approach.

Since meteoroids in our model have a wide variety of orbits, we must determine their collisional lifetimes at different positions of their orbits in order to take into account the effects of the heliocentric distance $r$ and inclination $I$. Such an approach prevents us from under/overestimating the collisional lifetime for eccentric and/or high-inclination orbits where the instantaneous collisional lifetime at a meteoroid's pericenter might be an order of magnitude different from the value at aphelion.
To estimate the collisional probability for each meteoroid we use 50 points uniformly distributed in true anomaly $f$ along the meteoroid's orbit. Then we calculate a weighted average of the collisional lifetime considering non-uniform variation of the true anomaly $f$ with time $t$, i.e. the average is weighted by $df/dt$. We select 50 points along the orbit for two reasons: the precision does not significantly increase for more points along the orbit and the calculation would be more computationally demanding.

Collisions in our model are applied in a rather simple manner. At the beginning of the simulation each meteoroid is assigned a collisional weight $W_c=1$. As time elapses the weight of each meteoroid is reduced to $W_c^\mathrm{new}=W_c*\exp(-T_\mathrm{step}/T_\mathrm{coll})$, i.e. a simple exponential decay, sometimes called collisional fading. We compared this method to that of \citet{Nesvorny_etal_2011JFC} and \citet{Pokorny_etal_2014}, and found that the collisional fading used in this manuscript works better for smaller samples of simulated particles.

\subsection{Impact probability calculation}
\label{SEC_IMPACT_PROBABILITY}
Despite the significant number of simulated meteoroids of various sizes, we record only a limited number of  direct impacts on Earth and other planets in our simulations. This is not surprising, because the actual number of meteoroids presently orbiting in the solar system is many orders of magnitude higher than the number of particles we can simulate. In order to overcome this limitation we implemented a method developed by \citet{Kessler_1981} that allowed us to calculate the intrinsic collisional probability of every meteoroid in our model at each time step with the target of interest, in this case Mercury. Kessler's method was selected because it grants us the freedom to choose the target's position and velocity vector contrary to other methods \citep{Opik_1951, Wetherill_1967, Greenberg_1982, Vokrouhlicky_etal_2012, Pokorny_Vokrouhlicky_2013}. As pointed out by \citet{Rickman_etal_2014}, Kessler's method contains singularities for certain orbital configurations leading to overestimating the collisional probability between the target and projectile. However, the high computational speed of Kessler's purely analytic method, combined with the high statistics in our model, outweighs these limitations.

An additional improvement from the version of the model presented in \citet{Pokorny_etal_2017_APJL} and \citet{Janches_etal_2018} is that we include the effect of gravitational focusing on the collisional cross-sectional area $\sigma$ between the target and the projectile \citep[Eq 3. in][]{Kessler_1981}:
\begin{equation}
    \sigma = \pi (r_1 + r_2)^2 \left[1+\frac{V_\mathrm{esc}^2}{V_\mathrm{rel}^2+\epsilon^2} \right],
    \label{EQ_GRAVFOC}
\end{equation}
where $r_1$ and $r_2$ are the average radii of target and projectile respectively, $V_\mathrm{esc}$ is the escape velocity of the target, $V_\mathrm{rel}$ is the relative velocity of the target and the projectile, and $\epsilon$ is the softening parameter. We chose $\epsilon=0.1$ km s$^{-1}$ to avoid extreme values of $\sigma$ for individual meteoroids with $V_\mathrm{rel}$ close to 0, e.g. AST meteoroids crossing Earth's orbit at very low eccentricities. Gravitational focusing is very important for the correct determination of the meteoroid mass flux on Earth, since meteoroids originating from main-belt asteroids and JFCs have in general $V_\mathrm{rel}$ smaller than $V_\mathrm{esc}$ of Earth. Due to Mercury's lower escape velocity $V_\mathrm{esc}=4.25$ km s$^{-1}$ and larger orbital velocities of all meteoroid populations in Mercury's vicinity this effect has a smaller impact for the Hermean meteoroid environment. Therefore, this effect was found to significantly affect (by a factor of two) the relative numbers of AST and JFC meteoroids intercepted by Earth and Mercury and influenced the calculation of total influx and vaporization rates for Mercury.

In this manuscript we investigated meteoroid impacts on Mercury for one Hermean year ($\sim$ 88 days) from \replaced{February 17th, 2013}{2013 February 17} at 00:00 UTC ($\mathrm{TAA}=359.4^\circ$) until \replaced{May 16th, 2013}{2013 May 16} at 00:00 UTC ($\mathrm{TAA}=359.5^\circ$), where TAA is Mercury's True Anomaly Angle. Impacts of meteoroids were calculated every 12 hours along Mercury's orbit, thus sampling the environment at 177 uniformly separated positions of Mercury in time. Mercury's orbital elements were taken from JPL HORIZONS Web-Interface for Target body: Mercury (199), Center body: Sun (10), DE431mx. We evaluated the collisional probability of each meteoroid in our model using Kessler's method for each of the 177 positions of Mercury separately to precisely capture changes of the Hermean meteoroid environment.

\subsection{Size-frequency distribution (SFD)}
The size or mass frequency distribution (SFD/MFD) of meteoroids is an essential characteristic of the meteoroid environment, yet for many source populations only a handful of independent measurements are available (as explained in this section). In this work, we treat SFD/MFD as a free parameter in order to investigate how sensitive or robust our resulting model is to variations of the assumed SFD/MFD. One of the simplest descriptions of SFD/MFD is a single power law distribution, where the number of meteoroids, $\mathrm{d}N$, with diameter between $D$ and $D+\mathrm{d}D$ is expressed as:
\begin{equation}
    \mathrm{d}N \propto D^{-\alpha} \mathrm{d}D,
    \label{EQ_SFD}
\end{equation}
where the exponent $\alpha$ is the differential size index. We can rewrite Eq. \ref{EQ_SFD} for the mass instead of size as
\begin{equation}
    \mathrm{d}N \propto M^{-\beta} \mathrm{d}M \propto D^{-3\beta} D^2\mathrm{d}D,
    \label{EQ_MFD}
\end{equation}
where the exponent $\beta$ is the differential mass index with the conversion formula $\alpha=3\beta-2$.
Integrating Eqs. \ref{EQ_SFD} and \ref{EQ_MFD} leads to the total number of meteoroids with diameters/masses in selected diameter/mass ranges $(D_1,D_2)$ or $(M_1,M_2)$
\begin{equation}
    N(D,\alpha) \propto \int_{D_1}^{D_2} D^{-\alpha} \mathrm{d}D \propto \frac{1}{\alpha+1} \left[ D_2^{-(\alpha-1)} - D_1^{-(\alpha-1)} \right],
    \label{EQ_SFD_CUM}
\end{equation}
\begin{equation}
    N(M,\beta) \propto \int_{M_1}^{M_2} M^{-\beta} \mathrm{d}M \propto \frac{1}{\beta+1} \left[M_2^{-(\beta-1)} - M_1^{-(\beta-1)} \right],
    \label{EQ_MFD_CUM}
\end{equation}

In order to simplify our calculation and to decrease the size of the large free parameter space, we treated all SFD/MFDs in this manuscript as single-power law distributions. Although more complex distributions, such as broken-power law distributions, have been used to describe dynamical models in the past \citep[e.g][]{Nesvorny_etal_2011JFC, Pokorny_etal_2014,Poppe_2016}, recent observations of the Earth meteoroid complex by radar and optical systems \citep{Blaauw_etal_2011,Pokorny_Brown_2016} suggest that a single power law may provide a good approximation of the meteoroid complex SFD/MFD for the diameter range considered in this manuscript ($D=10-2000~\mu$m).

It is important to distinguish between SFD/MFD at the source (i.e. where meteoroids are released) and SFD/MFD measured at different locations in the solar system. Due to complex dynamical evolution, scaling of the effects of radiation pressure and PR drag with meteoroid diameter, the SFD/MFD of different populations might vary throughout the solar system.
Discussions about the values of $\alpha$ and $\beta$ have appeared before. \citet{Dohnanyi_1969} reported that systems in collisional equilibrium have $\alpha=3.5, \beta=11/6$, while a population with evenly distributed mass per logarithmically spaced mass bins has $\alpha=4.0, \beta=2.0$.  \citet{Grun_etal_1985} compiled all measurements available at the time to produce a comprehensive empirical model for the SFD/MFD observed at 1 au, and reported $\beta=2.35$ for the size range considered in this manuscript. \citet{Cremonese_etal_2012} revisited the original LDEF results from \citet{Love_Brownlee_1993} and inferred a varying SFD/MFD that can be approximated by a broken power-law with $\beta=1.6$ for $D<200~\mu$m and $\beta=3.6$ for $D>200~\mu$m. The sporadic meteoroid background, as observed at Earth from the latest meteor radar and optical measurements reported by \citet{Blaauw_etal_2011,Pokorny_Brown_2016}, suggest $\beta=2.17\pm0.07$ and $\beta=2.10\pm 0.08$ respectively, which translates to $\alpha=4.51 \pm 0.21$ and $\alpha=4.30 \pm 0.28$. Nevertheless, all these values are referring to the SFD/MFD of a dynamically and collisionally processed meteoroid population that is a mixture of meteoroids from different sources.

The SFD/MFD at the source of each meteoroid population, i.e. the required model input, is more difficult to assess. The in-situ observations made by the {\it Rosetta} mission in the coma of 67P/Churyumov-Gerasimenko reported the SFD/MFD to vary with time and to be rather complex but approximated by a double broken power-law function \citep{Rotundi_etal_2015,Fulle_etal_2016}. For meteoroids with $D>1$ mm the differential SFD/MFD index was $\alpha=4,\beta=2$, while for $D<1$ mm the differential index varied from $\alpha=2,\beta=1.33$ at comet's aphelion to $\alpha=3.7,\beta=1.9$ at the comet's perihelion. Measurements from {\it Giotto} \citep[1P/Halley - HTC][]{Fulle_etal_1995} and {\it Stardust} \citep[81P/Wild - JFC][]{Green_etal_2007} provided similar values for the differential SFD/MFD indices.

In our models we perform all meteoroid integrations for each meteoroid diameter separately and we apply the SFD/MFD at the source regions afterwards. This gives us the flexibility to evaluate the effect of a wide range of differential SFD/MFD indices where we selected the following ranges, $\alpha=(1,7),\beta=(1,3)$, covering all currently available measurements.



\section{Results}


First, we will study the access of each meteoroid population to Mercury separately, aiming to evaluate how properties such as distributions of orbital elements change with different sizes and collisional lifetimes. 

\subsection{Results - Asteroidal meteoroids}
AST meteoroids impacting Mercury were the subject of several studies in the last decade. \citet{Marchi_etal_2005} investigated the flux of AST meteoroids on Mercury's surface for meteoroids with $D>1$~cm for which we can assume that PR drag is negligible. Due to the absence of PR drag and its induced dynamical evolution, \citet{Marchi_etal_2005} placed AST meteoroids into the $\nu_6$ secular resonance and the 3:1 mean motion resonance (MMR) with Jupiter. They found that the impact velocity of AST meteoroids at Mercury ranges widely between 15--80 km s$^{-1}$, the impacts are much faster at Mercury's perihelion, while the dawn-dusk asymmetry is more pronounced at perihelion and almost disappears at aphelion, and that both the $\nu_6$ secular resonance and the 3:1 MMR with Jupiter, had comparable efficiency of perturbing AST meteoroids into crossing orbits with terrestrial planets. In contrast,
\citet{Borin_etal_2009, Borin_etal_2016a} focused on smaller meteoroids with diameters in the range $D\in (20,200)  ~\mu$m with semimajor axes, $a$, randomly selected between $2.1$ and $3.3$ au, eccentricities $e<0.4$, and inclinations $I<20^\circ$. Their findings differ from those of \citet{Marchi_etal_2005} since their velocity distribution is much narrower with a peak at 17 km s$^{-1}$, which is similar to the results of \citet{Cintala_1992}.

AST meteoroids in our model ($D<2$~mm) originate from the largest main-belt asteroids and thus their initial orbits are predominantly distributed close to the ecliptic and have low eccentricity. Hence, there are no initial intersections between the young AST meteoroid population and Mercury. AST meteoroids are shepherded to the inner solar system by PR drag, where the dynamical timescale of the meteoroid transport from the original location towards Mercury crossing orbits is a linear function of meteoroid diameter $D$ \citep[Eq. 47 in][]{Burns_etal_1979}. For meteoroids on initially circular orbits, the time before meteoroids reach the Sun is $\tau_{\mathrm{PR}} = 400 R^2/\beta_{\mathrm{PR}} $, where $R$ is the heliocentric distance in au and  $\beta_{\mathrm{PR}} = 5.7 \times 10^{-5} / (\rho s)$, where $\rho, s $ are the meteoroids' density and radius both in cgs units \citep[Eq. 50 in][]{Burns_etal_1979}. Using these expressions we recognize that AST meteoroids in our model have a wide range of dynamical timescales ranging from  $\tau_{\mathrm{PR}} \approx 60$ ky for meteoroids with $D=10~\mu$m to $\tau_{\mathrm{PR}} \approx 12$ My for $D=2$~mm meteoroids (here we assumed that meteoroids start at heliocentric distances $R = 3$ au).  Hence, not only must AST meteoroids be dynamically evolved in order to decrease their perihelion distances to be able to reach Mercury, but also the planet generally encounters more evolved AST meteoroids at its perihelion while the less evolved meteoroids are able to reach the planet even at its aphelion. 

\begin{figure}[h]
\centering
\includegraphics[width=0.9\textwidth]{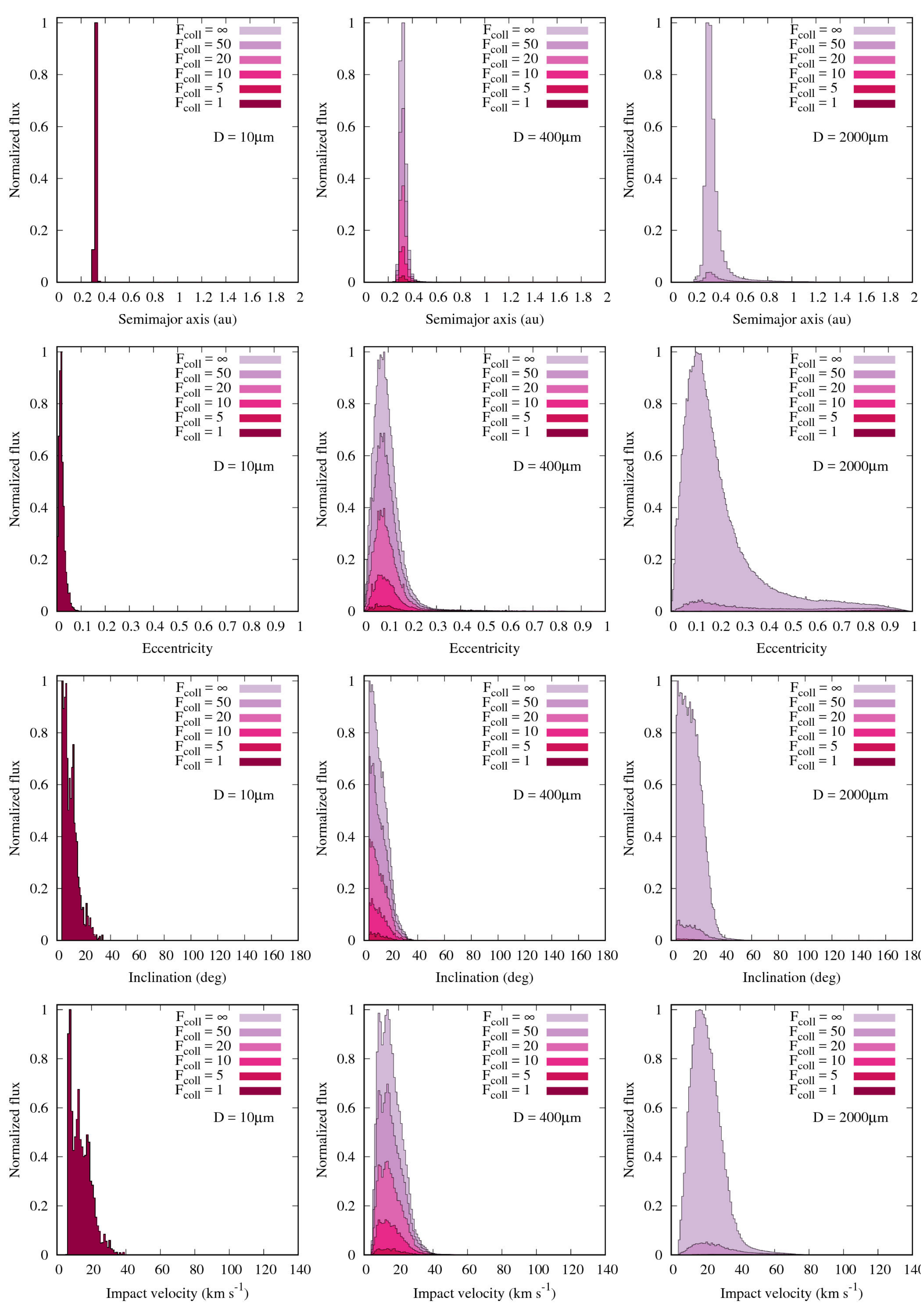}
\caption{Distribution of orbital elements and impact velocities for AST meteoroids with $D=10,400,\mathrm{~and ~2000}~\mu$m for TAA$=0^\circ$. Different collisional lifetime multipliers $F_\mathrm{coll}$ are represented by the range of colors. All distributions are normalized to unity. \added{All colors are superimposed on top of each other with the smallest $F_\mathrm{coll}$ on the top to show all features at once. The flux can only decrease with decreasing $F_\mathrm{coll}$.}
}
\label{FIG_AST_ELEMS_10_400_2000}
 \end{figure}

First, we explore the influence of the collisional lifetime multiplier $F_\mathrm{coll}$ on AST meteoroid population with access to Mercury. Figure \ref{FIG_AST_ELEMS_10_400_2000} shows the semimajor axis, eccentricity, inclination, and the impact velocity of meteoroids with $D=10,~400,\mathrm{~and~2000}~\mu$m impacting Mercury's surface at perihelion. 

We readily recognize from this figure that the collisional lifetime is not a significant agent in the dynamical evolution of the smallest AST meteoroids, $D=10~\mu$m (Fig. \ref{FIG_AST_ELEMS_10_400_2000}, left column). This is a consequence of their shorter PR evolution time, $\tau_{\mathrm{PR}}$, which is approximately $40$ times shorter than for $D=400~\mu$m meteoroids, leaving a much shorter time for mutual collisions to influence the meteoroid population. The smallest AST meteoroids have a very narrow distribution of eccentricities with a peak at $e=0.02$ at Mercury's perihelion, which has a direct influence on the distribution of semimajor axes of impactors of this size, making them concentrated around Mercury's heliocentric distance. The almost circularized nature of orbits of impacting meteoroids with $D=10~\mu$m is also a direct consequence of their short dynamical lifetime since these particles effectively pass through mean-motion resonances, secular resonances and areas of close-encounters with terrestrial planets. 
The inclination distribution of AST meteoroids is fairly similar along Mercury's whole orbit where the only significant difference is the minimum inclination cutoff caused by the non-zero inclination of Mercury's orbit ($I=7.005^\circ$ with respect to the ecliptic) for different TAAs. The impact velocities are fairy low in range $V_\mathrm{imp} = 5 - 30$~km~s$^{-1}$ where the $V_\mathrm{imp}$ is higher when Mercury is closer to the pericenter. 

AST meteoroids with $D=400~\mu$m (Fig. \ref{FIG_AST_ELEMS_10_400_2000}, middle column) arrive with  low eccentricities with a median value close to $e=0.1$, regardless of Mercury's true anomaly and the collisional lifetime multiplier $F_\mathrm{coll}$. The eccentricity distribution is wider than for $D=10~\mu$m meteoroids mostly \replaced{due to interior MMRs with Jupiter}{due to eccentricity pumping when passing through interior MMRs with Jupiter and exterior MMRs with Earth}.
The inclination distribution is similar to that of the smallest AST meteoroids. The impact velocity is correlated with the meteoroid eccentricity and is broader than for smaller AST meteoroids.
When no collisions are assumed ($F_\mathrm{coll} = \infty$) the relative flux of asteroidal meteoroids of this size is 3-4 orders of magnitude larger than for the nominal collisional lifetime $F_\mathrm{coll}=1$.
Despite such a rapid decay of the meteoroid population with shorter characteristic collisional lifetimes, it is still possible to transport material from the main belt to Mercury. However, shorter collisional lifetimes might require unrealistically high production rates of the source population in order to achieve the same volume of accreted mass or flux in the inner solar system. 

The orbital characteristics of the largest AST particles in our sample, $D=2000~\mu$m, are significantly different from their smaller counterparts. Such large meteoroids evolve, on average, over intervals 200 times longer than the smallest grains considered here, $D=10~\mu$m, before they arrive at Mercury. Because of that, collisions and mean-motion and secular resonances shape the orbital distribution of larger AST meteoroids impacting Mercury's surface (Fig. \ref{FIG_AST_ELEMS_10_400_2000}, right column). Even for the longest collisional lifetime considered here, $F_\mathrm{coll}=50$, the comparison with the collision-less population shows a severe decay where more than $95\%$ of meteoroids were destroyed in collisions. MMRs, secular resonances, and close-encounters with terrestrial planets (mainly with the Earth) effectively excite the eccentricity of impactors that is now exceeding the original population allowing meteoroids to access very eccentric orbits resembling trends reported by \citet{Marchi_etal_2005}. The tail in their eccentricity distribution allows them to reach Mercury from larger semimajor axes, however the majority of impacts have $a$ close to the heliocentric distance of Mercury. Inclinations and impact velocities of impactors with $D=2000~\mu$m are fairly immune to the dynamical influence from other planets, skewing the distributions of inclinations and impact velocities toward $I=40^\circ$ and $V_\mathrm{imp}=40$~km~s$^{-1}$.

\begin{figure}[h]
\centering
\includegraphics[width=0.9\textwidth]{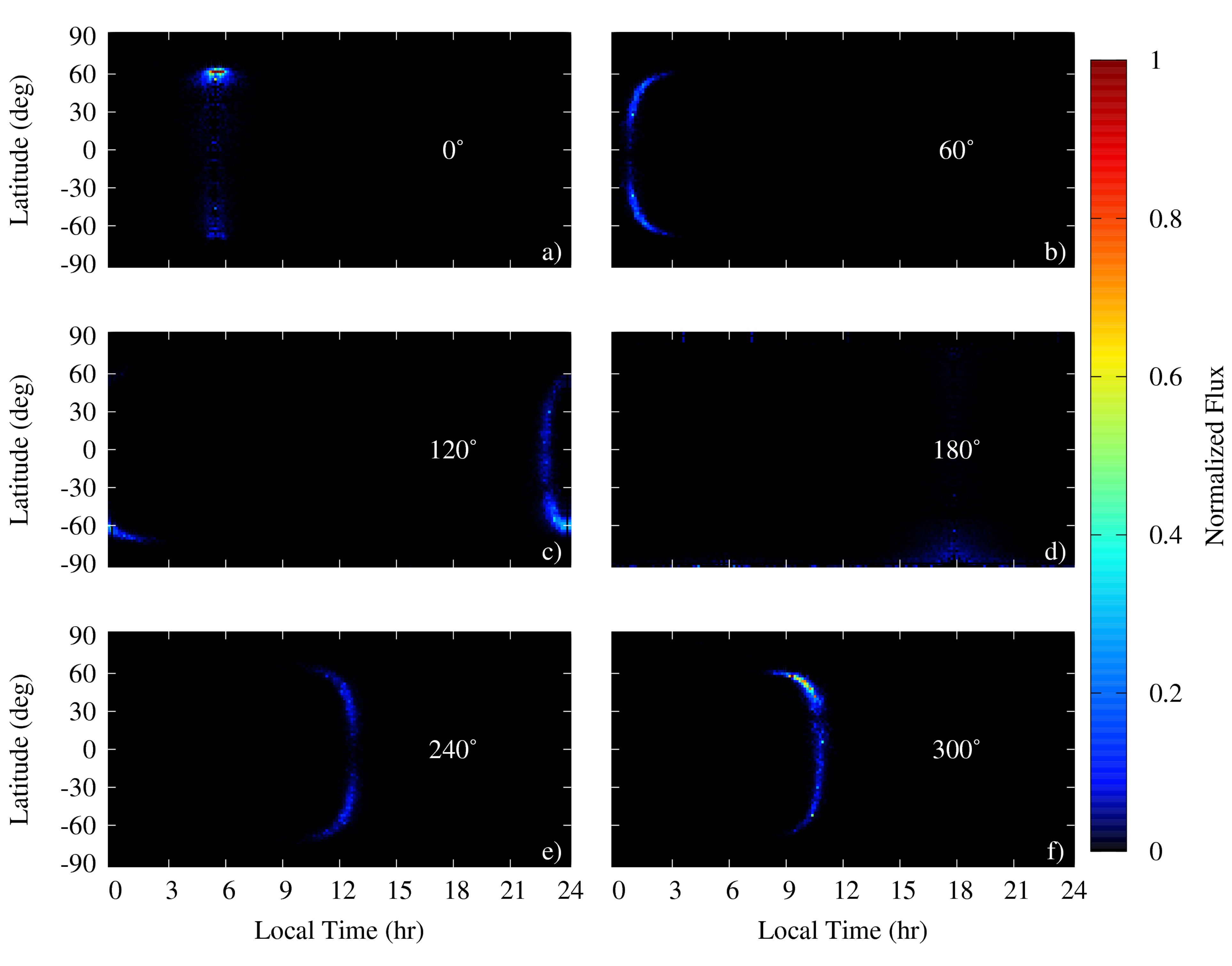}
\caption{Normalized radiant distribution of $D=10~\mu$m AST meteoroids impacting Mercury's surface for six different TAA (white number at 18 hr, $0^\circ$ in all panels). The mutual meteoroid collisions are not considered in this case. The x-axis represents the local time on Mercury, and it is fixed with regards to sub-solar point (12 hr). Due to Mercury's eccentricity, the location of the apex (approximately at 6 hr) changes along Mercury's orbit. The latitude is measured from Mercury's orbital plane (not the ecliptic).} 
\label{FIG_AST_RADIANTS_10}
 \end{figure}

The radiant distributions of impacting AST meteoroids are reflecting the distribution of orbital elements for different sizes and different collisional lifetimes. For the smallest meteoroids, $D=10-50~\mu$m, the collisional lifetime plays a negligible role and the radiant distribution is fundamentally driven by the orbital evolution of AST meteoroids. Figure \ref{FIG_AST_RADIANTS_10} shows radiant distributions for 6 different phases of Mercury's orbit for $D=10~\mu$m meteoroids. At perihelion (TAA$=0^\circ$) AST meteoroids impact predominantly from the apex direction (local time, or LT, of 6 hr) with a significant concentration at high northern latitudes (Panel a in Fig. \ref{FIG_AST_RADIANTS_10}). This north/south asymmetry in the radiant distribution at perihelion is caused by Mercury's non-zero velocity with respect to the ecliptic,
due to Mercury's argument of pericenter value $\omega=29.17^\circ$ and inclination $I=7.005^\circ$. This fact causes the planet to move through the meteoroid cloud upwards and thus enhances the north hemisphere contribution. The radiant map at perihelion would be symmetric around the apex if the planet's inclination was zero or if its argument of pericenter was $\omega=90^\circ~\mathrm{or}~270^\circ$. 
Ten days later  Mercury approaches TAA$=60^\circ$ and the radiant distributions of AST meteoroids change significantly, with the majority of impacts now coming from the night side (at 1.5 hr LT). This effect is caused by the change of direction of Mercury's velocity vector with respect to the radial vector. At perihelion the angle between these two vectors is always $90^\circ$ simply because the change of the heliocentric distance with respect to time $\mathrm{d}r/\mathrm{d}t$ is zero; however, for eccentric bodies once the body leaves perihelion (or aphelion) the $90^\circ$ symmetry is disturbed. At TAA = $60^\circ$ Mercury's velocity vector is pointing $100^\circ$ from the Sun thus the radiant flux is skewed towards the night side. Interestingly, at TAA = $60^\circ$ Mercury reaches the highest point of its orbit above the ecliptic and its vertical velocity is close to 0, resulting in a symmetric radiant distribution with respect to the ecliptic/Mercury's equator (Panel b in Fig. \ref{FIG_AST_RADIANTS_10}). 
Twenty four days after perihelion passage (TAA$=120^\circ$) the smallest AST meteoroids impact mainly around midnight (i.e. from the anti-helion direction) and the overall flux decreases compared to the flux during the perihelion passage (panel c in Fig. \ref{FIG_AST_RADIANTS_10}). The flux of AST meteoroids decreases to the minimum at Mercury's aphelion and is focused in the southern part of the anti-apex region (18 hr LT, $-75^\circ$ latitude). This is caused by the lower orbital velocity of Mercury in its aphelion compared to orbital velocities of surrounding AST meteoroids on nearly circular orbits. At aphelion Mercury is moving slower than the circularized AST meteoroid cloud thus the planet is being impacted from behind with respect to its velocity vector. Due to low relative impact velocities the overall flux of impacting meteoroids is also minimized. Once Mercury leaves its aphelion the AST meteoroids impact mostly from the day side (12 hr LT, panel e in Fig. \ref{FIG_AST_RADIANTS_10}), where the situation for TAA=$240^\circ$ reflects the meteoroid environment for TAA=$60^\circ$ but now mirrored to the day side. Closer to perihelion the overall flux increases and once again it is concentrated in northern latitudes (panel f) in Fig. \ref{FIG_AST_RADIANTS_10}).

\begin{figure}[h]
\centering
\includegraphics[width=0.9\textwidth]{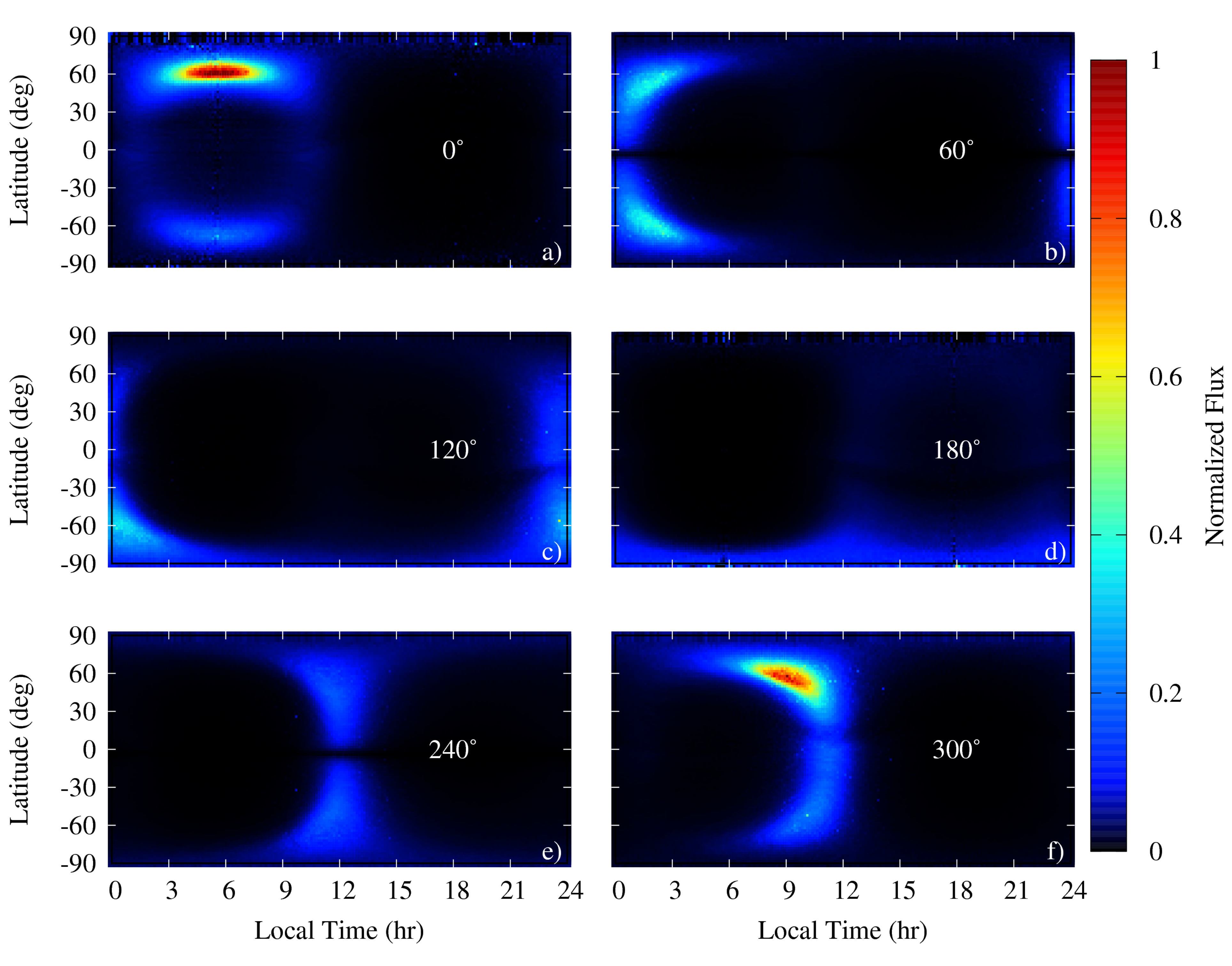}
\caption{The same as Fig. \ref{FIG_AST_RADIANTS_10} but now for AST meteoroids with $D=2000~\mu$m with no collisions assumed, $F_\mathrm{coll}=\infty$.}
\label{FIG_AST_RADIANTS_2000}
 \end{figure}

The radiant distribution of the largest AST meteoroids, $D=2000~\mu$m, is more complex than for their smaller cousins due to their more complex dynamical evolution. The higher eccentricity achieved by eccentricity pumping in MMRs during their path from the main belt allows larger AST meteoroids to impact Mercury from a variety of directions (Fig. \ref{FIG_AST_RADIANTS_2000}). At perihelion (panel a in Fig. \ref{FIG_AST_RADIANTS_2000}) the north/south asymmetry is still present, however, the pattern is similar to that of the toroidal structure observed at the Earth \citep[e.g.][]{Pokorny_etal_2014}. The closer the meteoroid radiants are to the ecliptic the higher is their eccentricity, which introduces an additional effect when the collisions are taken into account. Since meteoroids with higher eccentricities are able to impact Mercury sooner after their release from the source population, collisions are more effective at removing low eccentricity meteoroids and thus for short collisional lifetimes the AST meteoroid environment for populations affected by collisions will be more concentrated around the ecliptic. The AST meteoroid environment for the largest grains follows the same pattern as the smallest meteoroids along Mercury's orbit, where the smallest particles represent the low eccentricity core and the broader structures of larger meteoroids reflect meteoroids with more eccentric orbits.
 
\begin{figure}[h]
\centering
\includegraphics[width=0.9\textwidth]{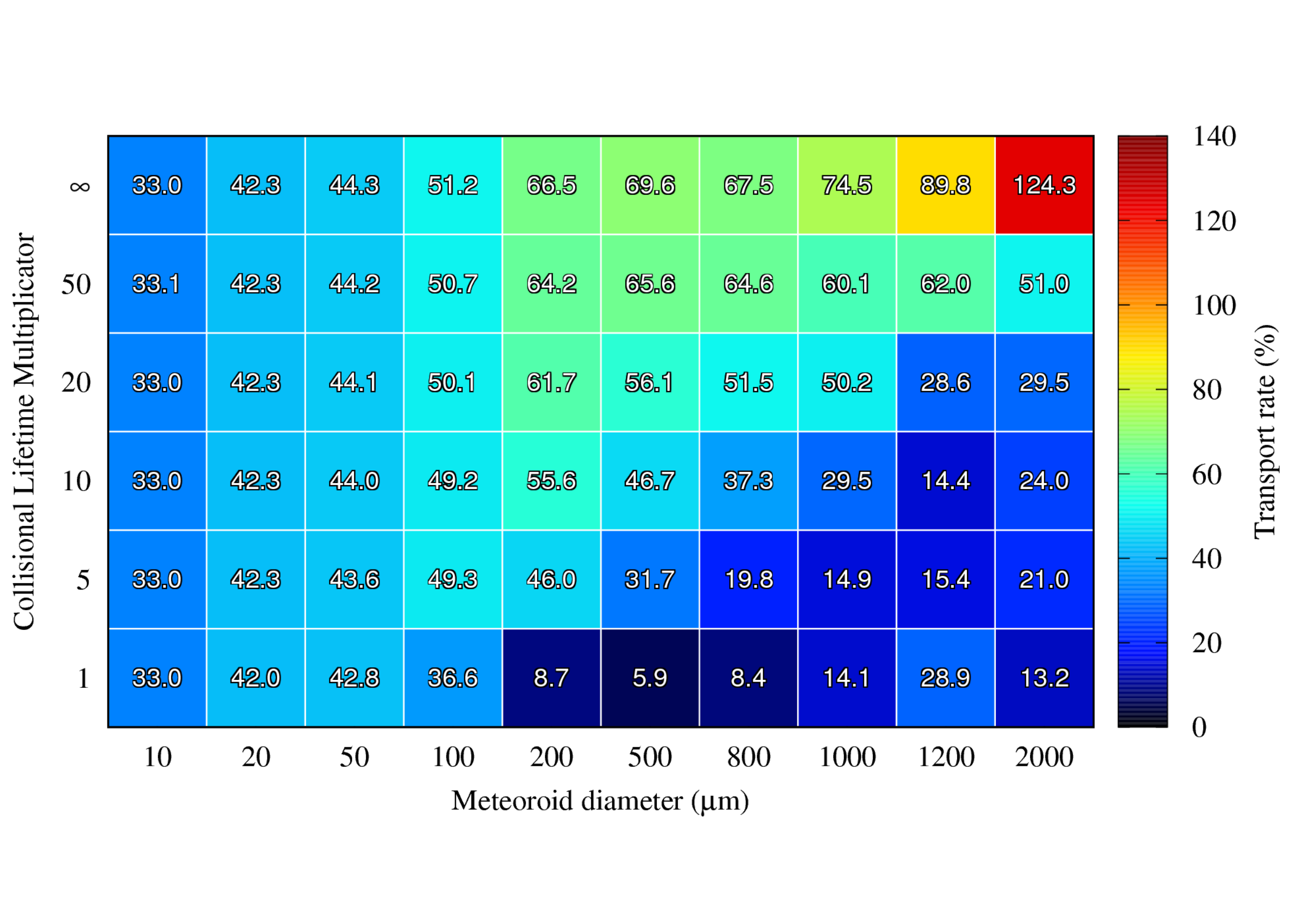}
\caption{The transport rate (i.e., the ratio between the AST meteoroid flux averaged over Mercury's orbit and the AST meteoroid flux at Earth) as a function of the meteoroid diameter $D$ and the collision lifetime multiplier $F_\mathrm{coll}$. A transport rate of 100\% means that the meteoroid flux averaged over Mercury's orbit and the flux at Earth are the same. The transport rate for AST meteoroids reflects the complex dynamical evolution of AST meteoroids and the effect of collisions.}
\label{FIG_AST_EARTH_MERCURY_TRANS}
 \end{figure}

 To combine all AST meteoroids and quantify their flux at Mercury, we must know 1) the production rate and SFD/MFD of AST meteoroids, and 2) the relative efficiency of transport of AST meteoroids from Earth to Mercury. Because no in-situ observations of the Main Belt meteoroid production rates exist currently, for 1) we use constraints on the AST particle flux at Earth's proximity (this is addressed in the next paragraph of this section). For 2), hereby we compare the modeled ratio between the AST meteoroid flux averaged over Mercury's orbit and the AST meteoroid flux at Earth, which we will call the transport rate (Fig. \ref{FIG_AST_EARTH_MERCURY_TRANS}).
 First, we see that without collisions ($F_\mathrm{coll} =\infty$) different sizes have different  transport rates, where at Mercury the meteoroid flux is lower than at the Earth \deleted{by a factor of ten to twenty} with the exception of $D=2000~\mu$m meteoroids. This is due to the effect of gravitational focusing which is stronger at Earth than at Mercury for two reasons: a) Earth has a deeper gravitation well and a higher escape velocity $V_\mathrm{esc}=11.2$ km s$^{-1}$ as compared to Mercury with $V_\mathrm{esc}=4.2$ km s$^{-1}$ (see Eq. \ref{EQ_GRAVFOC}), and b) the relative velocity of AST meteoroids is smaller at Earth than at Mercury. The increasing transport rate for larger meteoroids reflects a higher fraction of high eccentricity meteoroids at Earth and then subsequently higher relative velocities and smaller gravitational focusing effects. 
 We note that the total accreted mass will only be $<1/7$ compared to that at Earth, because Mercury's surface area is smaller by 1/7. Once collisions are taken into account the transfer rate of meteoroids with larger diameters decreases rapidly, emphasizing the significant role that collisions have on shaping the AST meteoroid environment around Mercury. Without addressing collisions accurately, one can easily overestimate the contribution of larger meteoroids by a factor of 2-3. Fig. \ref{FIG_AST_EARTH_MERCURY_TRANS} allows us to calculate the total mass influx on Mercury if we have an estimated value from the Earth and we know the SFD/MFD of the meteoroid population.

\begin{figure}[h]
\centering
\includegraphics[width=0.9\textwidth]{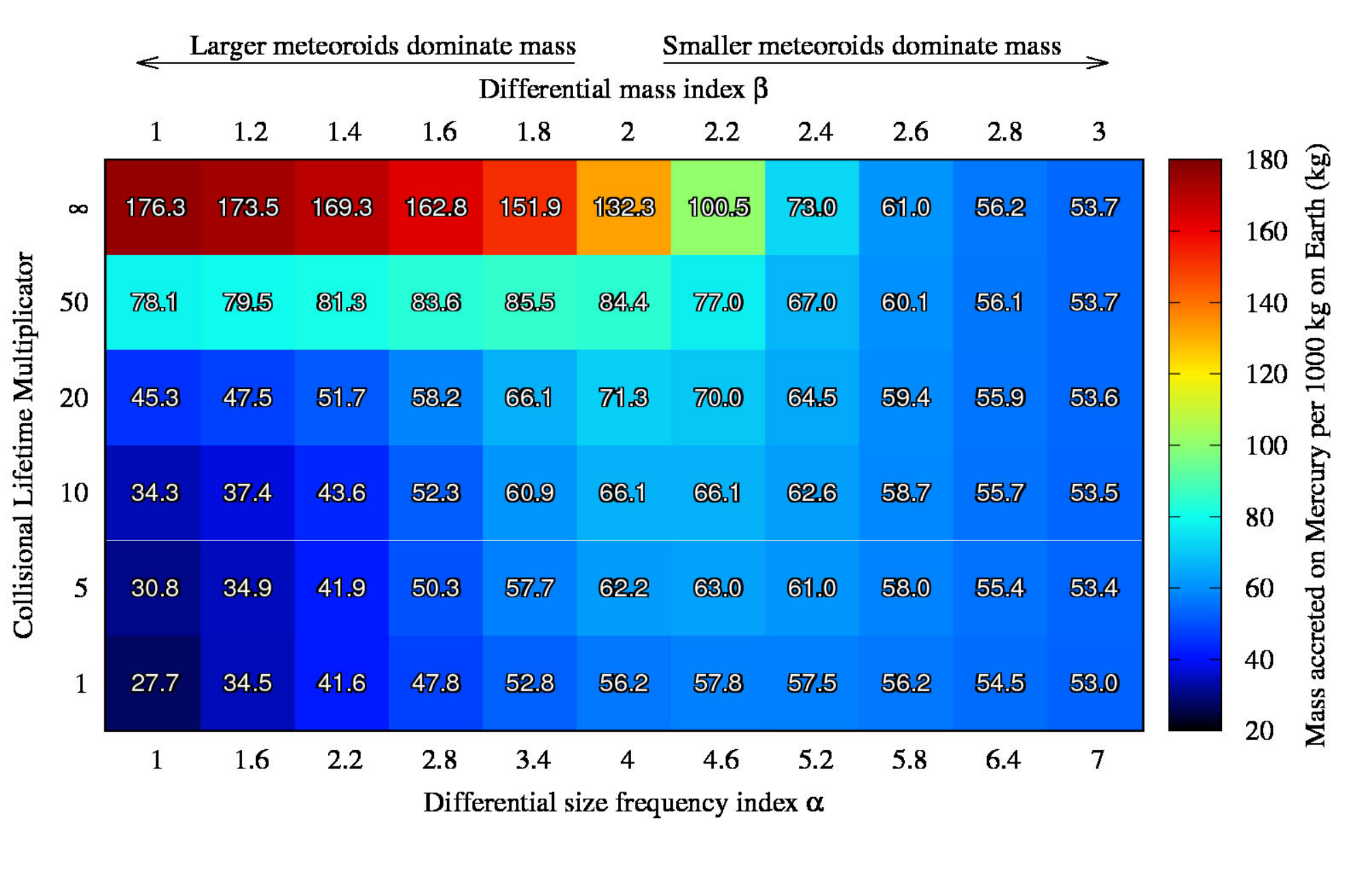}
\caption{The mass in kg of AST meteoroids accreted on Mercury per 1000 kg at Earth as a function of the assumed SFD/MFD indices $\alpha,\beta$ and the collisional lifetime multiplier $F_\mathrm{coll}$.}
\label{FIG_AST_MASSINDEX}
 \end{figure}

 Quantifying the mass influx of AST meteoroids at Earth has proven to be rather challenging due to their small relative velocity with respect to Earth \citep{Carrillo-Sanchez_etal_2016}. Despite the advancements in detection techniques and our current understanding of the chemical processes that meteoroids experience when entering Earth's atmosphere, the mass influx of AST meteoroids is uncertain, where the most recent estimate is $3.7 \pm 2.1$ t~day$^{-1}$. In this Section, we assume that regardless on the collisional lifetime or the SFD/MFD of the AST meteoroid population, the mass influx in the modeled size range is 1000 kg d$^{-1}$ over the whole surface of the Earth. We opt to use this number for scalability across different populations. Finally, by combining the transport rates from Fig. \ref{FIG_AST_EARTH_MERCURY_TRANS}, considering all sizes together, and taking into account the surface area of the Earth and Mercury, we obtain the total mass of AST impactors on Mercury's surface (Fig. \ref{FIG_AST_MASSINDEX}). As expected, applying no collisions results in the largest amount of material accreted at Mercury which decreases when the smaller meteoroids contribution in SFD/MFD is increased (i.e. larger values of $\alpha$, and, $\beta$). The maximum mass influx obtained from AST meteoroids at Mercury is approximately 17\% of that at Earth, but the median from our model is around 7\%. Interestingly, in Fig. \ref{FIG_AST_MASSINDEX} we can observe that even for the differential size frequency index $\alpha \sim 4,~\beta \sim 2$ the accreted mass is fairly constant for different collisional lifetimes, where we can see the gradual decrease in accreted mass with decreasing differential size frequency/mass index. This is a consequence of collisional break up of larger AST meteoroids during their dynamical evolution from Earth crossing orbits towards Mercury crossing orbits, since the dynamical evolution of AST meteoroids before they can impact Earth is not reflected in Fig. \ref{FIG_AST_MASSINDEX}.

 In summary, our AST meteoroid model provides similar orbital element distributions to those reported by \citet{Borin_etal_2009,Borin_etal_2016a} and velocities like \citet{Cintala_1992}. Hence, the vaporization rates from this population will be similar to their estimates per unit mass. However, we additionally report a tail for $V_\mathrm{imp} > 25$ km s$^{-1}$ that becomes more prominent for larger AST meteoroids, $D>200~\mu$m. Owing to their low $V_\mathrm{imp}$ and the non-zero eccentricity of Mercury, AST meteoroids experience significant motion for their radiant locations along Mercury's orbit. Mutual collisions with the zodiacal cloud are important for shaping the meteoroid environment around Mercury for $D>200~\mu$m. The mass accretion of AST meteoroids at Mercury is at most 17\% compared to values at Earth, but could  decrease down to 5\% for shallow SFDs/MFDs ($\alpha \leq 2.8, \beta \leq 1.6$), and shorter collisional lifetimes $F_\mathrm{coll}\leq 20$.
 
 \subsection{Results - Jupiter Family Comet meteoroids}
 \label{sec:JFC}
\begin{figure}[h]
\centering
\includegraphics[width=0.9\textwidth]{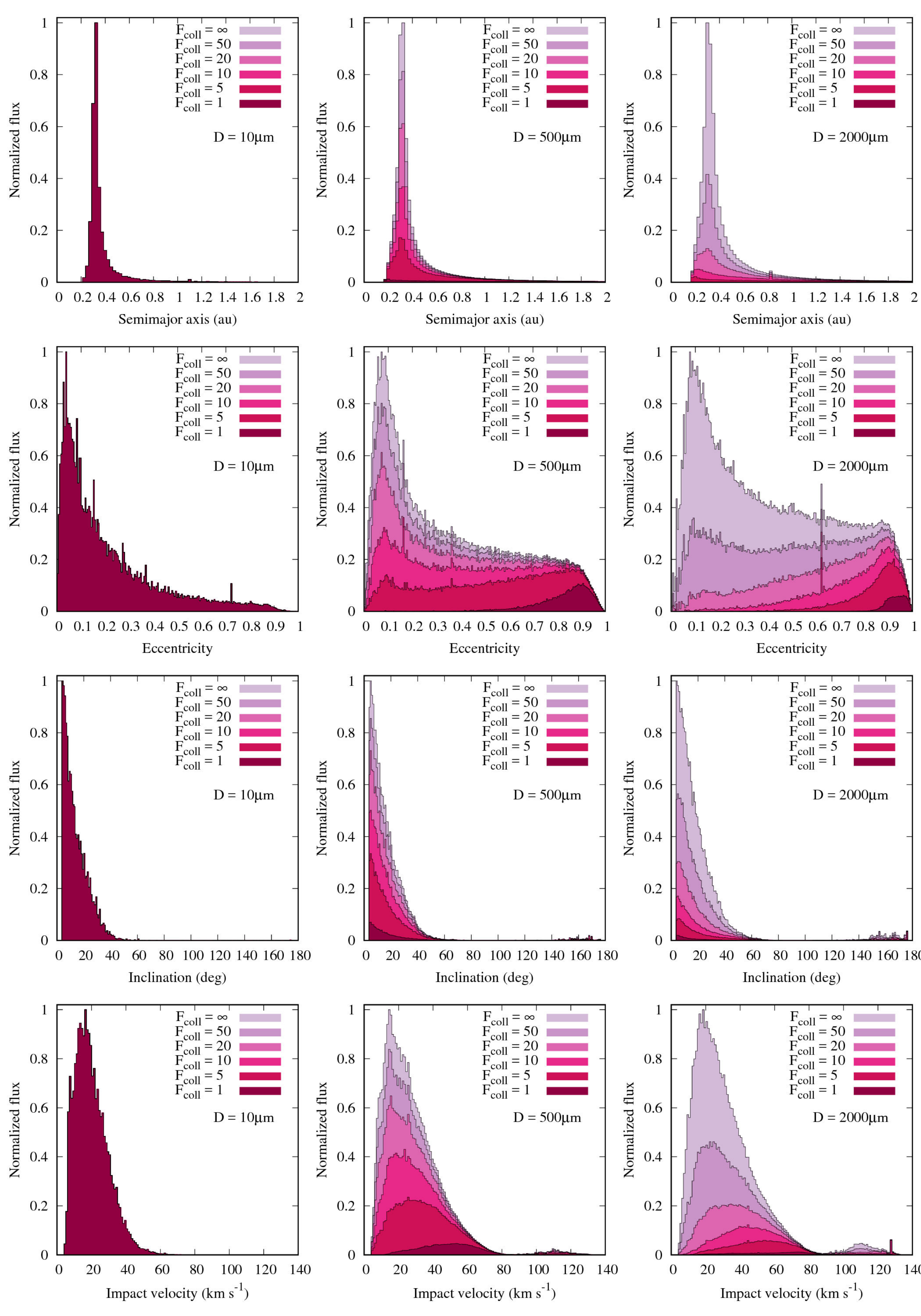}
\caption{Distribution of orbital elements and impact velocities for JFC meteoroids with $D=10,~500, \mathrm{~and~2000}~\mu$m for TAA$=0^\circ$. The range of colors denotes different collisional lifetime multipliers $F_\mathrm{coll}$. All distributions are normalized to unity.
}
\label{FIG_JFC_ELEMS_10_500_2000}
 \end{figure}

 Unlike AST meteoroids, JFC meteoroids can cross Mercury's orbit almost immediately after their ejection from their parent bodies, resulting in particles with a mixture of ages impacting Mercury's surface. Consequently, in this population, dynamically young meteoroids with large eccentricities are unaffected by collisions, while dynamically older meteoroids undergo collisional decay. Figure \ref{FIG_JFC_ELEMS_10_500_2000} shows the distributions of orbital elements of $D=10, 500,$ and $2000~\mu$m JFC meteoroids at Mercury's perihelion.
 
 The smallest JFC meteoroids considered in our model ($D=10~\mu$m) are unaffected by collisions and impact Mercury with preferentially lower eccentricities with a distribution that peaks around $e=0.08$ (Fig.  \ref{FIG_JFC_ELEMS_10_500_2000}, left column). However, due to the wide range of eccentricities of JFC meteoroids, the semimajor axis distribution is wider than for AST meteoroids. JFC meteoroid inclinations are slightly higher compared to those of AST meteoroids. The velocity distribution reflects higher intrinsic eccentricities of JFC meteoroids, peaking at 15-20 km s$^{-1}$.
 
 For larger JFC meteoroids, $D=500~\mu$m (Fig.  \ref{FIG_JFC_ELEMS_10_500_2000}, middle column), the meteoroid eccentricity is significantly influenced by the mutual collisions with the zodiacal cloud. If no collisions are assumed, JFC meteoroids impact Mercury preferentially with lower eccentricities, resulting in a distribution that peaks around $e=0.08$, similar to the smallest JFC meteoroids. By increasing the rate of collisions (decreasing $F_\mathrm{coll}$) the low eccentricity portion of this population wanes while the high eccentricity portion becomes more prominent. 
 Due to a wide range of eccentricities the semimajor axis distribution is wider than for AST meteoroids. The meteoroid population still peaks at the heliocentric distance of Mercury, similarly to AST meteoroids; however, for shorter collisional lifetimes a significant fraction of impacting meteoroids arrives with $a>0.5$ au. The inclination distribution is unaffected by the collisional lifetime in terms of the overall shape, but demonstrates very clearly the decay of the relative flux with shorter collisional lifetimes. A negligible fraction of JFC meteoroids was perturbed into retrograde orbits by close encounters with Jupiter (a weak signal around $I\sim 160^\circ$). Finally, the velocity distribution reflects higher intrinsic eccentricities and while peaking around 15-20 km s$^{-1}$, a significant portion of JFC meteoroids arrive with velocities over 50 km s$^{-1}$ for TAA$=0^\circ$ and over 35 km s$^{-1}$ for TAA$=180^\circ$.
 
 For the largest JFC meteoroids in our sample, $D=2000~\mu$m, collisions heavily shape the orbital element distributions (Fig.  \ref{FIG_JFC_ELEMS_10_500_2000}, right column). It is apparent that even for larger values of the collisional lifetime ($F_\mathrm{coll}=20)$, the collisional rates are too high for meteoroids to reach lower eccentricities before they are destroyed in our model. Overall, the largest JFC meteoroids, $D=2000~\mu$m, resemble $D=500~\mu$m, but they require $F_\mathrm{coll}$ four times higher in order to experience similar dynamical evolution since their PR drag timescale is four times longer.
 
\begin{figure}[h]
\centering
\includegraphics[width=0.9\textwidth]{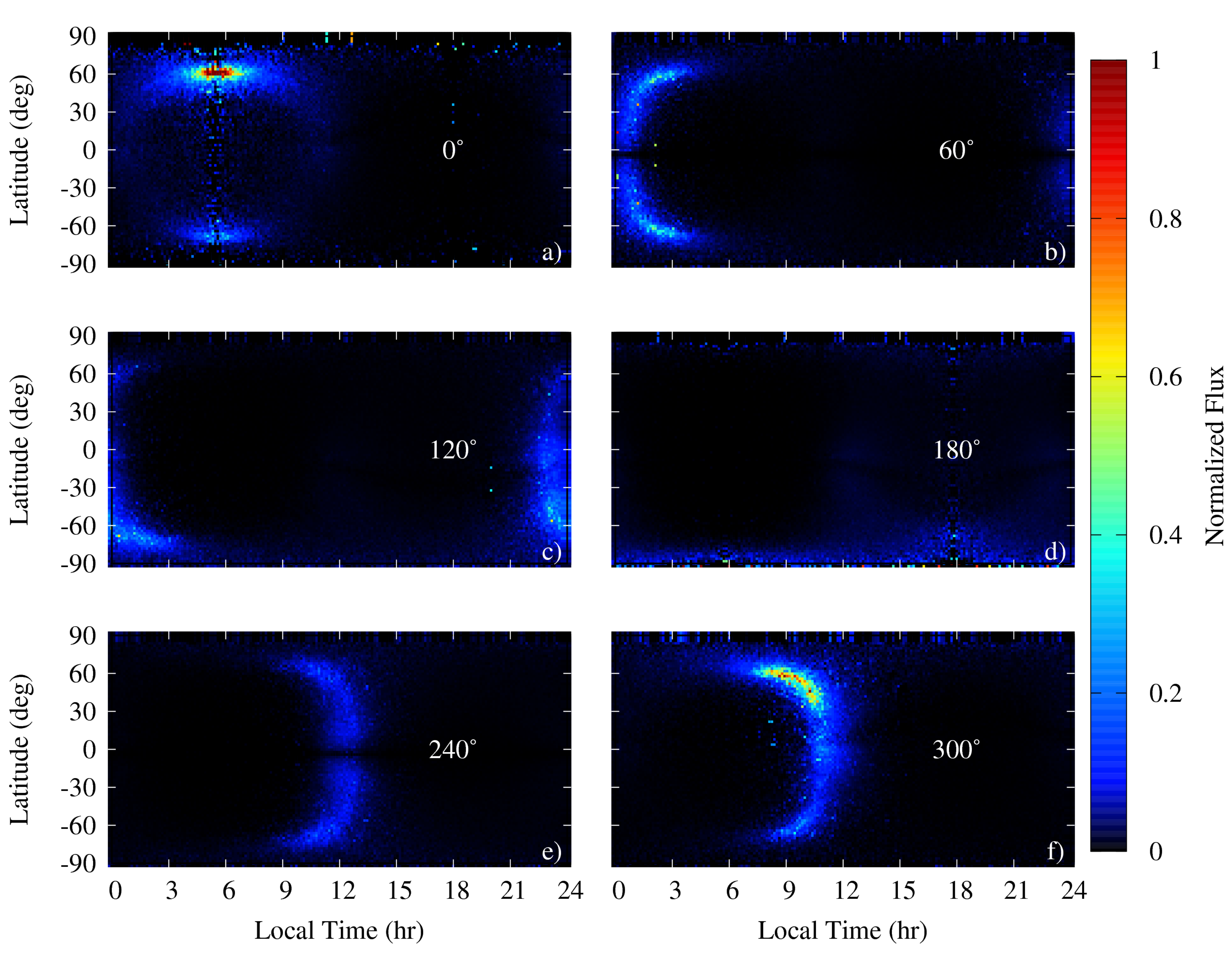}
\caption{Normalized radiant distribution of $D=10~\mu$m JFC meteoroids impacting Mercury's surface for six different TAA (white number at 18 hr, $0^\circ$ in all panels). The mutual meteoroid collisions are not considered in this case. The x-axis represents the local time on Mercury, and it is fixed with regards to the sub-solar point (12 hr). Due to Mercury's eccentricity, the location of the apex (approximately at 6 hr) changes along Mercury's orbit. The latitude is measured from Mercury's orbital plane (not the ecliptic).} 
\label{FIG_JFC_RADIANTS_10}
 \end{figure}

The radiant distributions for JFC meteoroids with $D=10~\mu$m (see Fig. \ref{FIG_JFC_RADIANTS_10}) are similar to those of larger AST meteoroids (see Fig. \ref{FIG_AST_RADIANTS_2000}), a finding which can be attributed to having similar distributions of orbital elements. However, some differences remain which are caused mainly by the resulting wider range of eccentricities and inclinations of JFC meteoroids. At perihelion the meteoroid flux is concentrated around ($6$ hr, $60^\circ$) as a result of the non-zero inclination of Mercury's orbit and the orientation of the Hermean velocity vector (Fig. \ref{FIG_JFC_RADIANTS_10}). 
Similarly to AST meteoroids, there is a shift in the radiant distribution of JFC meteoroids as Mercury moves toward or away from the Sun, caused by the non-zero eccentricity and inclination of Mercury's orbit and a consequent drift of the planet's velocity vector from the ecliptic plane and its perpendicular orientation with regards to the radial vector.

\begin{figure}[h]
\centering
\includegraphics[width=0.9\textwidth]{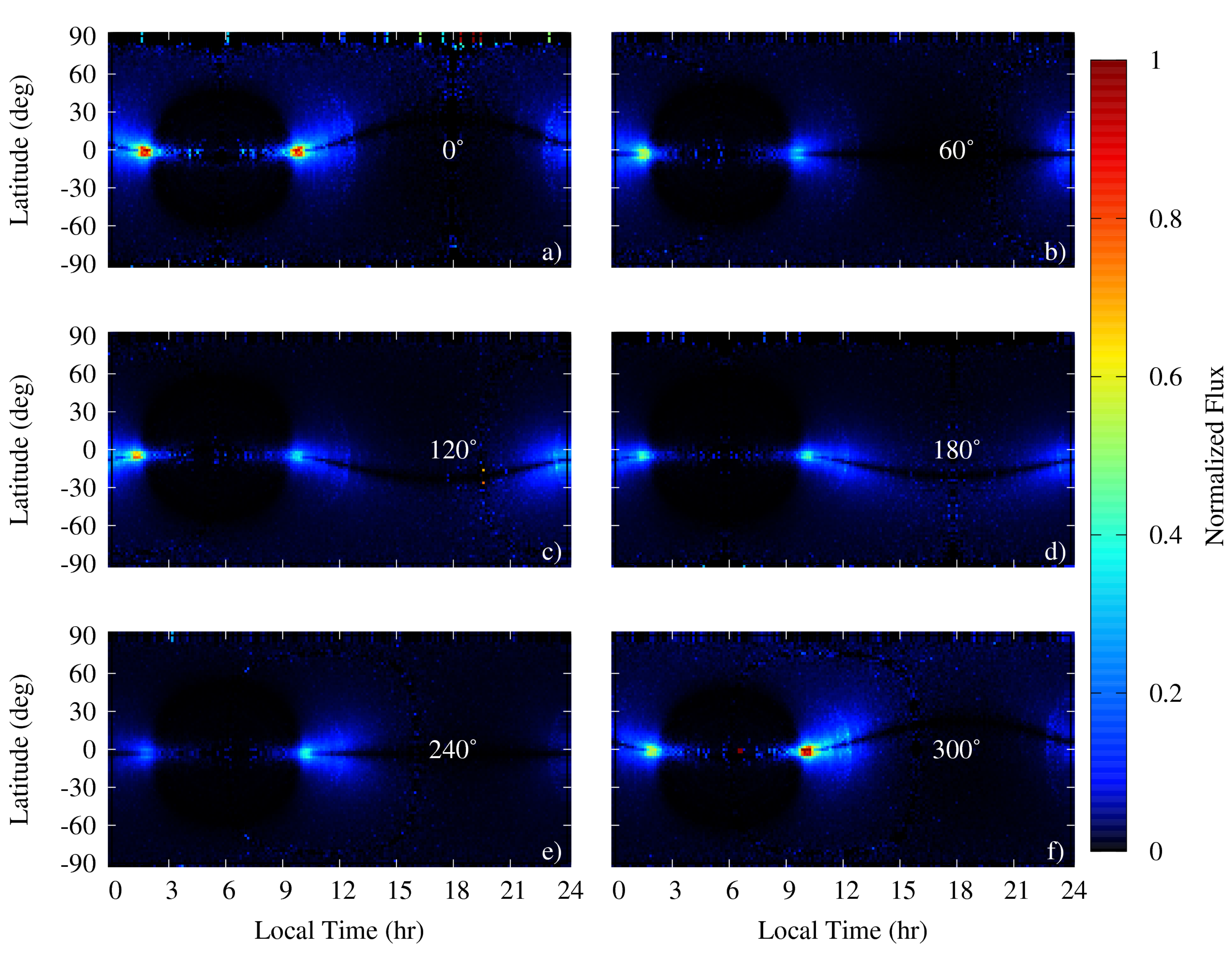}
\caption{The same as Fig. \ref{FIG_JFC_RADIANTS_10} but now for JFC meteoroids with $D=2000~\mu$m and for the nominal collisional lifetime, $F_\mathrm{coll} = 1$}
\label{FIG_JFC_RADIANTS_2000}
 \end{figure}

When no collisions are applied, the changes in the radiant distribution of JFC particles with Mercury's TAA is similar for all particle sizes.
The radiant distribution becomes broader and less concentrated compared to smaller meteoroids. However, when we apply collisions the directionality of JFC impacts changes. Figure \ref{FIG_JFC_RADIANTS_2000} shows an extreme example of a collisionally evolved ($F_\mathrm{coll}=1$) JFC meteoroid population at Mercury with $D=2000~\mu$m. The most striking feature in the collisionally groomed JFC populations is the lack of high latitude particle impacts that was characteristic of the collisionless case. The low eccentricity JFC meteoroids were collisionally destroyed in this case and only the high eccentricity/high velocity (the original) portion of the JFC population remains. Another consequence of collisions is a much smaller helion/antihelion asymmetry with very subtle variations along Mercury's orbit.

\begin{figure}[h]
\centering
\includegraphics[width=0.9\textwidth]{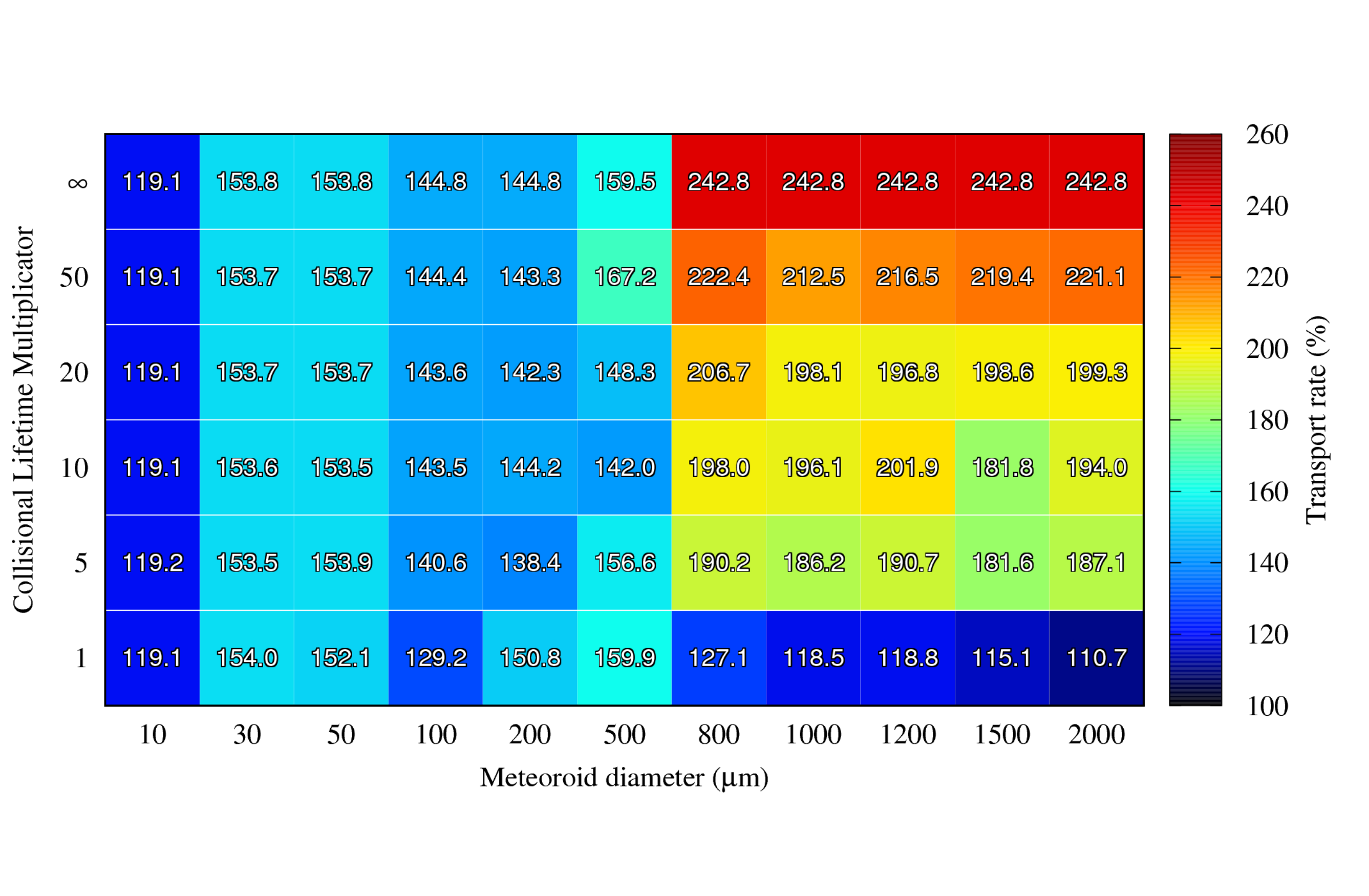}
\caption{The transport rate (i.e., the ratio between the JFC meteoroid flux averaged over Mercury's orbit and the JFC meteoroid flux at Earth) as a function of the meteoroid diameter $D$ and the collisional lifetime multiplier $F_\mathrm{coll}$. A transport rate of 100\% means that the meteoroid flux averaged over Mercury's orbit and the flux at Earth are the same.}
\label{FIG_JFC_EARTH_MERCURY_TRANS}
 \end{figure}

 Due to higher initial eccentricities of JFC meteoroids, the transport process is generally more efficient compared to AST meteoroids (Fig. \ref{FIG_JFC_EARTH_MERCURY_TRANS}). Consequently, even for the shortest collisional lifetimes ($F_\mathrm{coll}=1$) and larger sizes considered in the model, the flux of JFC meteoroids at Mercury is larger than at Earth. This result is a consequence of the combination of two phenomena: a) the high eccentricity JFC meteoroids are able to impact both Mercury and Earth even though their low eccentric counterparts are destroyed, and b) Mercury and Earth are swiping through a different volume with a different velocity, which results in an increase of the collisional probability of the meteoroid cloud with Mercury. For AST meteoroids the transport rate rapidly decreases by a factor of 10 with increasing meteoroid diameter $D$ for shorter collisional lifetimes ($F_\mathrm{coll}=1,5,10,20$). Such an effect is not as pronounced for JFC meteoroids, yielding at most a factor of 3 decrease.

\begin{figure}[h]
\centering
\includegraphics[width=0.9\textwidth]{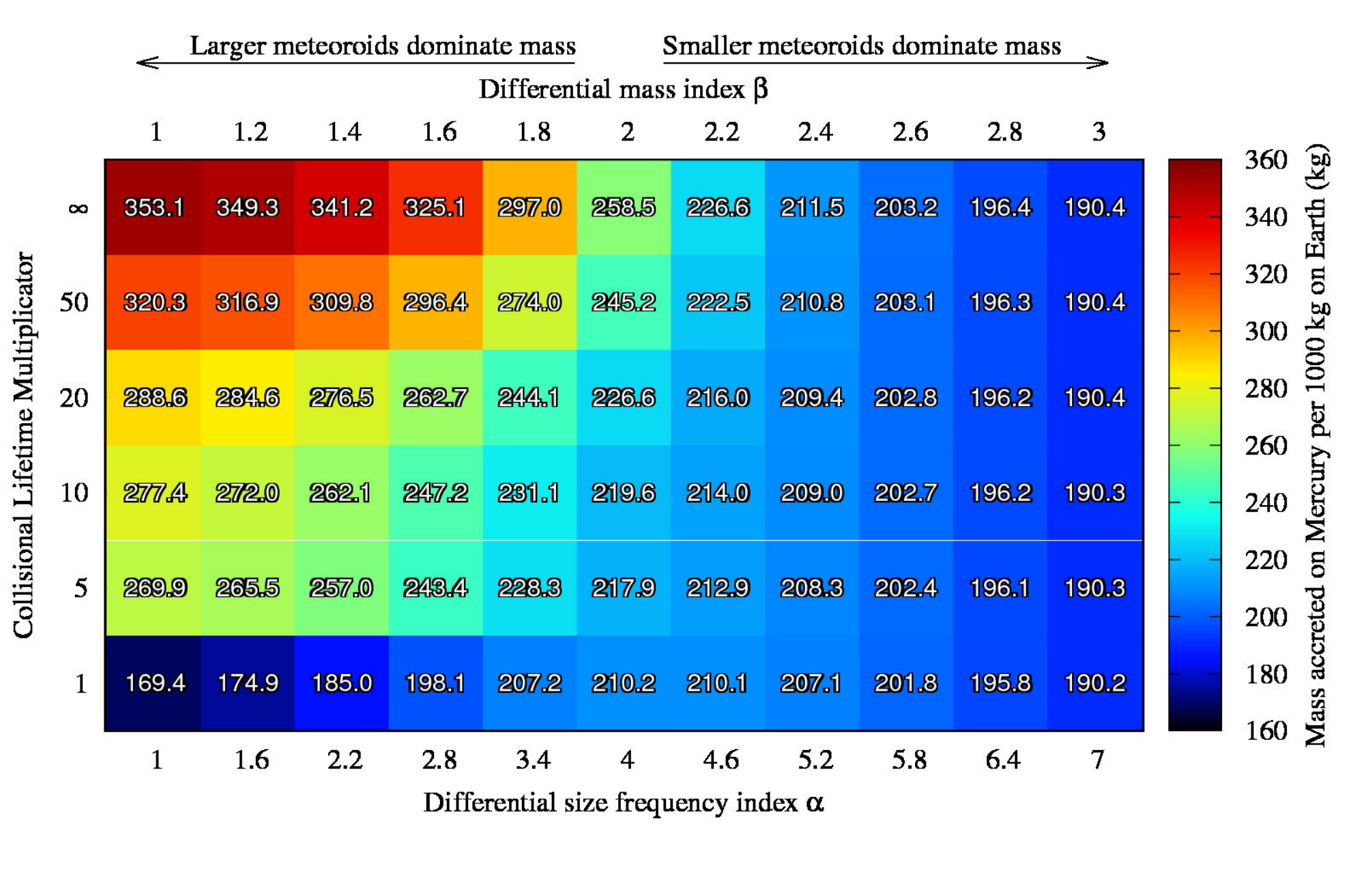}
\caption{The same as Fig. \ref{FIG_AST_MASSINDEX} but now for JFC meteoroids.}
\label{FIG_JFC_MASSINDEX}
 \end{figure}

 Figure \ref{FIG_JFC_MASSINDEX} shows that for each 1000 kg of JFC mass accreted at Earth, approximately 200--300 kg of JFC meteoroids are accreted at Mercury for a wide variety of collisional lifetimes and size-frequency distributions. Similarly to AST meteoroids, the JFC meteoroids provide more mass accreted on Mercury for shallower SFDs/MFDs. This is because of the gravitational focusing effect at Earth that is more pronounced for smaller meteoroids in our model, that are not that susceptible to eccentricity pumping from MMRs.  Even for the most extreme values considered in Fig. \ref{FIG_JFC_MASSINDEX}, the range of accreted mass is quite narrow where the difference between the maximum and minimum is only a factor of 2. This suggests an overall insensitivity of JFC meteoroids to SFD/MFD and collisional lifetimes considered in our model
 
 In summary, the dynamics of JFC meteoroids are more complex compared to their AST counterparts due to the influence of close encounters with Jupiter, MMRs, and secular resonances. Collisions with the zodiacal cloud become significant for larger particles, $D\ge 200~\mu$m, but even for the shortest collisional lifetimes considered here, the mass influx at Mercury does not become negligible due to the low perihelion initial distances of a portion of JFC meteoroids. When comparing the mass influx at Earth and Mercury, JFC meteoroids are less affected than AST meteoroids by uncertainties in the SFD/MFD and collisional lifetime. 
 
 \subsection{Results - Halley type meteoroids}
 \label{sec:HTC}
To simulate Halley type comet meteoroids we adopt the  \citet{Pokorny_etal_2014} model, even though the most recent observations of HTCs suggest that they are more uniformly distributed in inclination space \citep{Wang_Brasser_2014,Nesvorny_etal_2017}, while in the \citet{Pokorny_etal_2014} model  HTCs had inclinations with a median value of $I=55^\circ$. We tested weighting meteoroids in our model based on their initial inclination to reflect the more recent HTC observations and the results did not significantly change. 
 
\begin{figure}[h]
\centering
\includegraphics[width=0.9\textwidth]{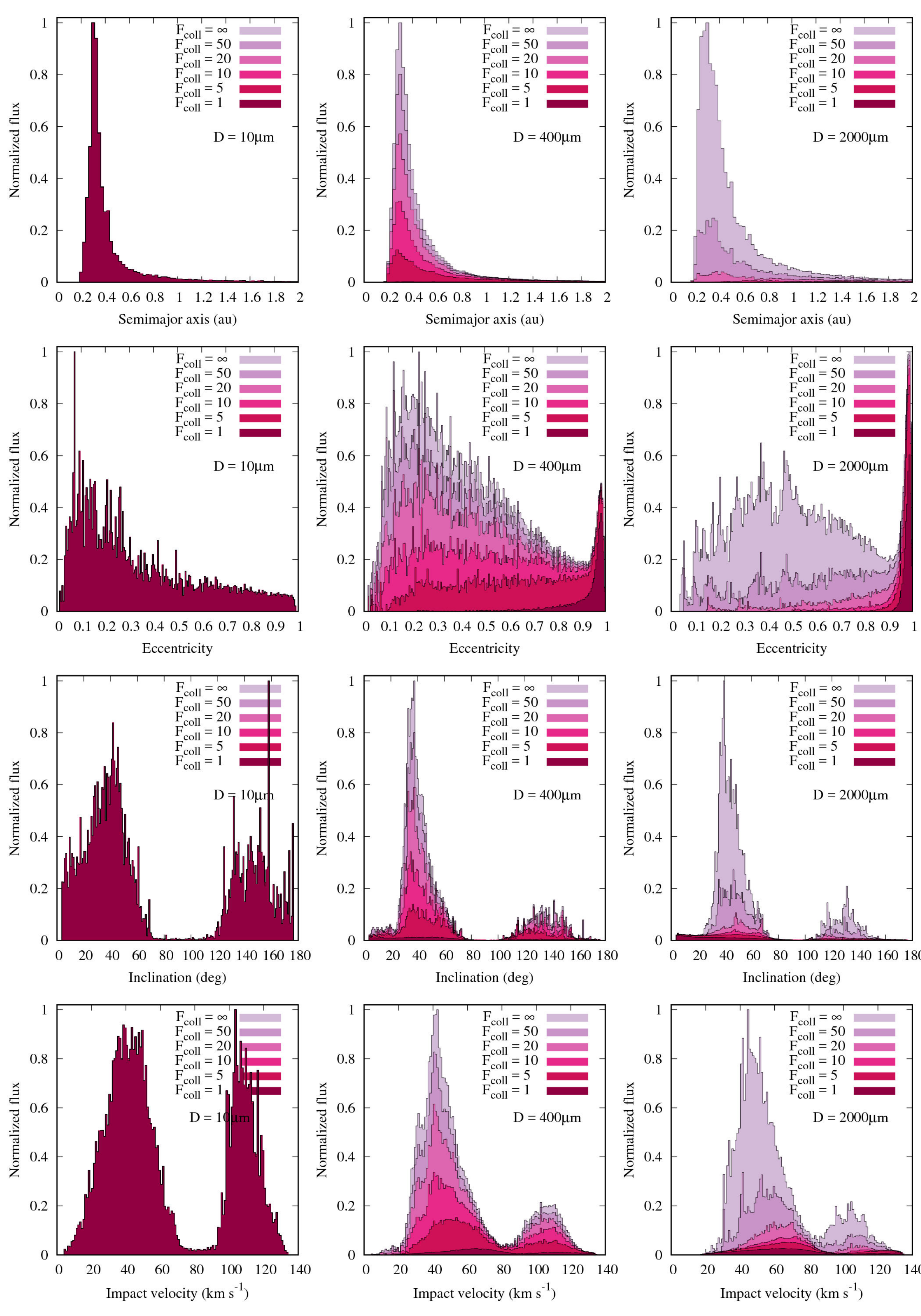}
\caption{Distribution of orbital elements and impact velocities for HTC meteoroids with $D=10,400, \mathrm{~and~2000}~\mu$m for TAA$=0^\circ$. Different collisional lifetime multipliers $F_\mathrm{coll}$ are represented by the range of colors. All distributions are normalized to unity.
}
\label{FIG_HTC_ELEMS_10_400_2000}
 \end{figure}

 HTC meteoroids are dynamically quite different from AST and JFC meteoroid populations because they originate beyond Jupiter/Saturn. All meteoroids in our model start with very high eccentricities ($e>0.9$) and must undergo significant dynamical evolution before their orbits are circularized and thus accreted more effectively by Mercury. 
 The orbits of the smallest HTC meteoroids considered in our model, $D=10~\mu$m, are circularized regardless of the collisional lifetime (Fig. \ref{FIG_HTC_ELEMS_10_400_2000}, left columns). As explained before, this occurs because PR drag is highly efficient for these sizes. 
 For larger HTC meteoroids, $D=100- 400~\mu$m, only the shortest collisional lifetime $F_\mathrm{coll}=1$ is effectively preventing the particles from circularization. With further increases in meteoroid size, longer collisional lifetimes are required for particles to undergo circularization under the influence of PR drag before being collisionally disrupted. For the largest meteoroids considered in our model, $D=2000~\mu$m, even the longest collisional lifetime assumed here, $F_\mathrm{coll} = 50 $, is not sufficiently long enough to ensure full circularization for most of the modelled meteoroids. The inclination distribution of HTC meteoroids (Fig. \ref{FIG_HTC_ELEMS_10_400_2000}, third row) results in two different regimes: a more dominant prograde population ($I<90^\circ$) and  a smaller retrograde population ($I>90^\circ$). For $D=400~\mu$m meteoroids the prograde portion of the HTC population always dominates over the retrograde one regardless of the collisional lifetime factor used, whereas for $D=10~\mu$m the dominance of the prograde trajectories is less pronounced. This dichotomy translates into a bimodal distribution of impact velocities of HTC meteoroids, where the prograde/retrograde transition happens at $V_\mathrm{imp}=80$ km s$^{-1}$ at perihelion (Fig. \ref{FIG_HTC_ELEMS_10_400_2000}, bottom row). This transition shifts to lower $V_\mathrm{imp}$ as Mercury is moving toward its aphelion where the transition occurs around $V_\mathrm{imp}=60$ km s$^{-1}$ due to Mercury's smaller orbital velocity.
 
 

The largest HTC meteoroids in our model, $D=2000~\mu$m, are significantly influenced by mutual collisions with the zodiacal cloud. The dynamical evolution via PR drag is 200 times slower than for the smallest meteoroids in our model. This implies that larger HTC meteoroids have effectively more time to be collisionally destroyed while evolving from high eccentricity orbits. Hence, even for the collision-less case, $F_\mathrm{coll} = \infty$, the largest HTC meteoroids always have a higher excess of high eccentricity orbits ($e>0.95$) compared to their smaller counterparts. 
\replaced{
This results from an effective barrier posed by giant planets (mostly by Jupiter), that prevents particles from entering the inner solar system and scattering them into hyperbolic orbits or accreting them during frequent close encounters.}{This results from an effective barrier posed by giant planets (mostly by
Jupiter), scattering particles into hyperbolic orbits or accreting them
during frequent close encounters, thus preventing them from entering the
inner solar system.}
\added{For smaller meteoroids $D\sim 20~\mu$m the PR drag is able to decrease the meteoroid semimajor axis quickly and thus avoid frequent close encounters with Jupiter and Saturn. Larger meteoroids spend more time in Jupiter and Saturn crossing orbits and thus are more susceptible to planetary scattering.} This effect influences not only the eccentricity of HTC meteoroids but also their inclination, and subsequently the impact velocity distribution.  The dichotomy is still apparent, however the inclination distribution of prograde particles is more dominant suggesting that the giant planet barrier is far more effective in shielding the inner solar system from retrograde HTC particles. \added{Once meteoroids are decoupled from Jupiter (i.e. their aphelion distance is well below perihelion distance of Jupiter; $\sim 4.5$ au) the effect of $F_\mathrm{coll}$ becomes more important and is similar to what we report for JFC meteoroids, where larger meteoroids are more affected by collisions than smaller meteoroids.}
 
\begin{figure}[h]
\centering
\includegraphics[width=0.9\textwidth]{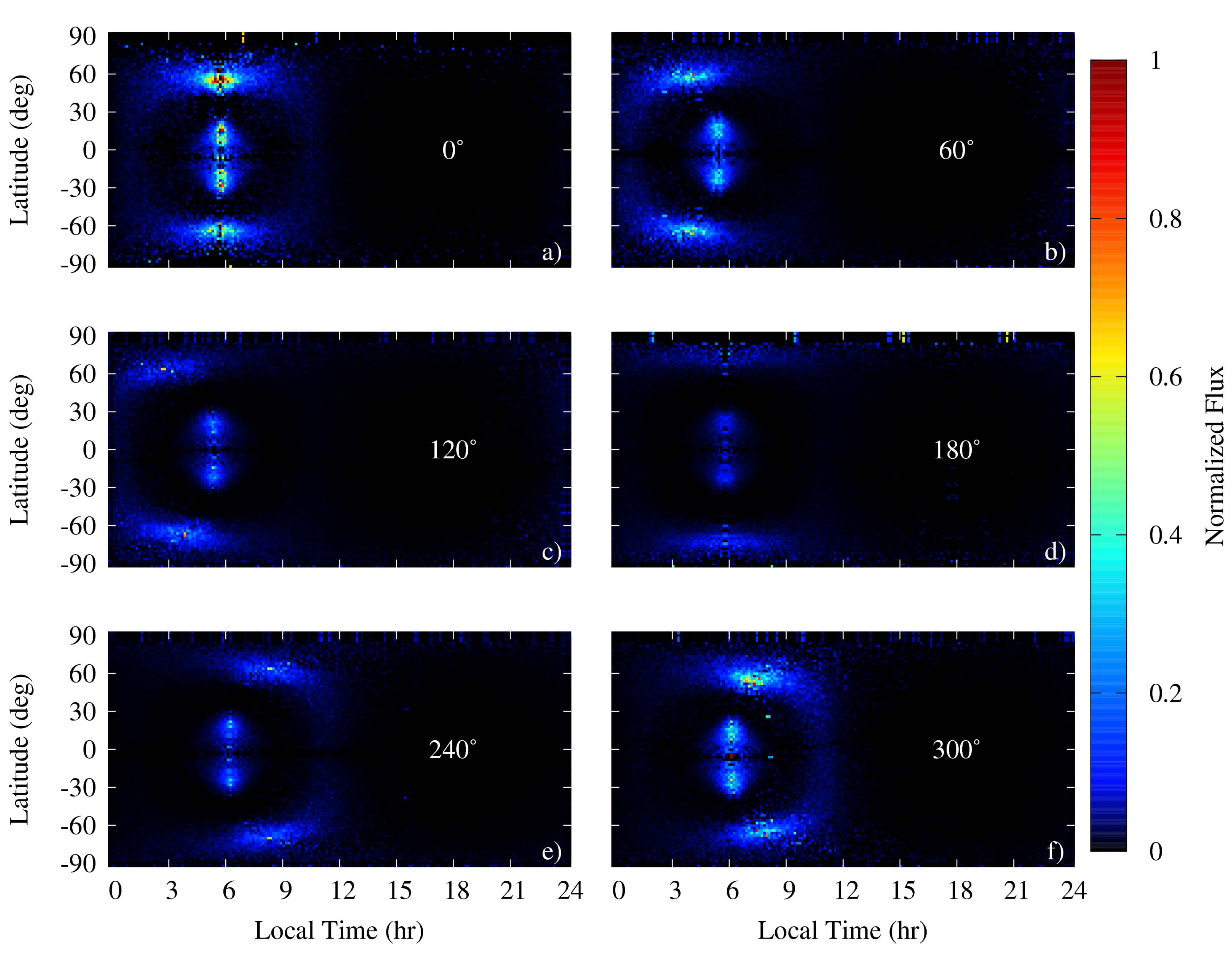}
\caption{Normalized radiant distribution of HTC meteoroids with $D=10~\mu$m impacting Mercury's surface for six different TAA (white number at 18 hr, $0^\circ$ in all panels). The mutual meteoroid collisions are not considered in this case, however applying collisional lifetimes assumed in this manuscript makes no difference for HTC meteoroids of these sizes.} 
\label{FIG_HTC_RADIANTS_10}
 \end{figure}

The inclination dichotomy, the presence of the retrograde particles and the higher median inclination value of the  prograde portion of HTC meteoroids result in unique patterns of radiant distributions unlike the radiant distributions of AST and JFC meteoroids (Fig. \ref{FIG_HTC_RADIANTS_10} for $D=10~\mu$m). The first unique feature is the strong concentration around Mercury's apex direction/dawn terminator ($6\mathrm{ hr}, 0^\circ)$. This concentration is well-known from the Earth meteor radar observation as the apex source \citep[e.g.][]{Campbell-Brown_2008,Janches_etal_2015} and is solely populated by retrograde meteoroids with high impact velocities \citet{Pokorny_etal_2017_APJL}. The apex source is persistent during the entire orbital cycle but the flux decreases with increasing TAA reaching the minimum at aphelion ($\mathrm{TAA}=180^\circ$). The second feature is a north/south concentration centered at the dawn terminator when Mercury is at perihelion. This population of radiants is fed by prograde low eccentricity meteoroids and follows a similar motion to that of AST and JFC meteoroids, moving towards the night side (anti-helion source) as Mercury moves from perihelion toward aphelion, becoming centered at the dawn terminator when Mercury arrives at aphelion, and moving toward the day side when Mercury is moving toward perihelion.

\begin{figure}[h]
\centering
\includegraphics[width=0.9\textwidth]{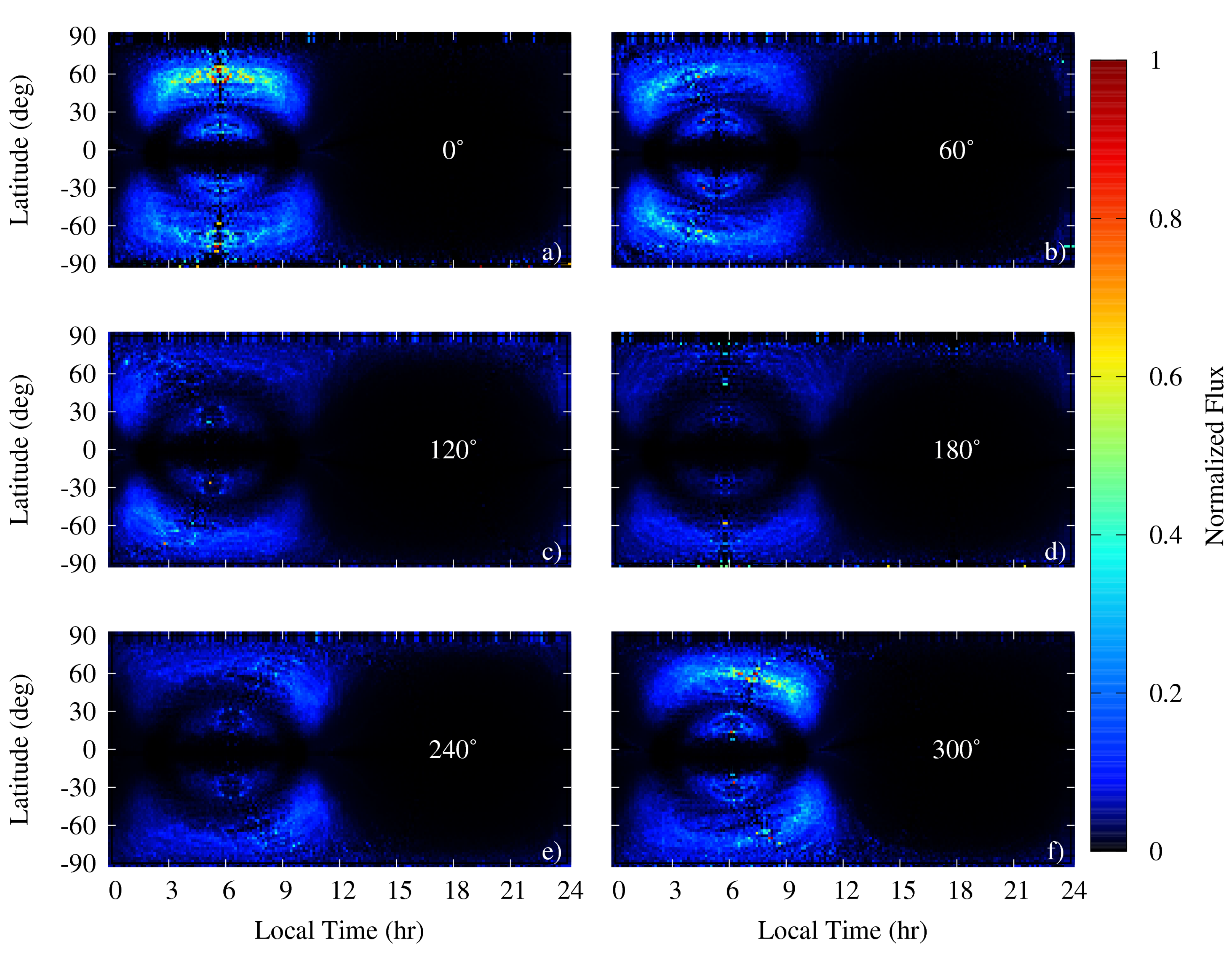}
\caption{The same as Fig. \ref{FIG_HTC_RADIANTS_10} but now for HTC meteoroids with $D=2000~\mu$m.}
\label{FIG_HTC_RADIANTS_2000}
 \end{figure}

For the  largest HTC meteoroids $D=2000~\mu$m, the prograde population dominates the mass flux and the giant planet barrier prevents meteoroids from efficient circularization, resulting in generally more eccentric orbits (Fig. \ref{FIG_HTC_ELEMS_10_400_2000}, right column). 
 This translates into a wider spread of radiants and lower flux in the apex sources (Fig. \ref{FIG_HTC_RADIANTS_2000}). The absence of low inclination HTC meteoroids effectively translates in an absence of particles with equatorial radiants up to latitudes around $\pm 30^\circ$. The variations of the radiant distributions with true anomaly angle show a similar behavior to that previously reported for AST and JFC meteoroids.  

\begin{figure}[h]
\centering
\includegraphics[width=0.9\textwidth]{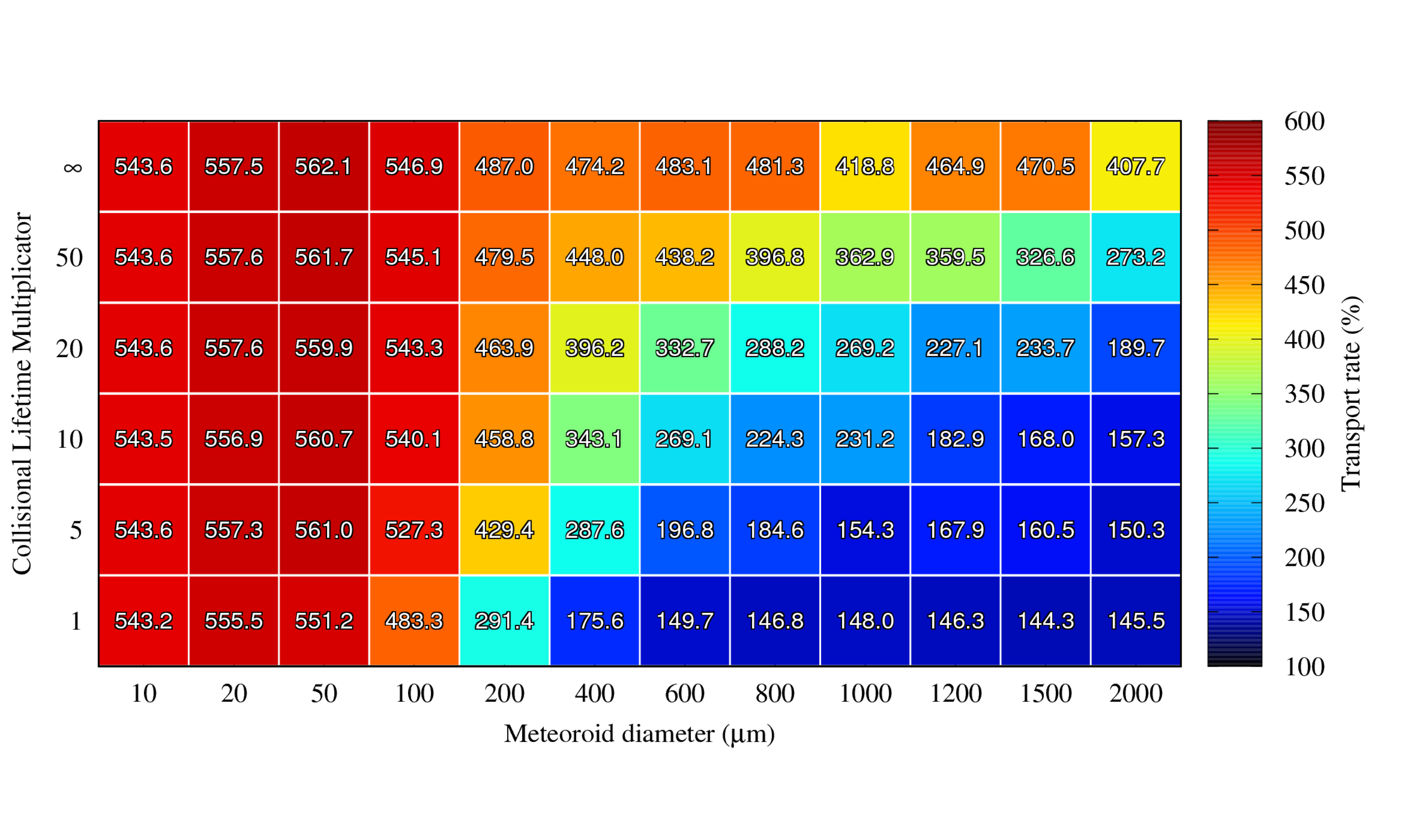}
\caption{The same as Fig. \ref{FIG_AST_EARTH_MERCURY_TRANS} but now for HTC meteoroids.}
\label{FIG_HTC_EARTH_MERCURY_TRANS}
 \end{figure}

 The trend of the transport rate is different from AST and JFC meteoroids (Fig. \ref{FIG_HTC_EARTH_MERCURY_TRANS}). Higher relative velocities of HTC meteoroids at Earth and Mercury lead to smaller gravitational focusing effects for both planets. For the smaller HTC meteoroids ($D<100~\mu$m) the collisional lifetime uncertainty yields no changes in transport rates. The effect of the giant planet barrier is noticeable as a decrease in the transport rate for larger meteoroids $D>200~\mu$m and is accentuated by shorter collisional lifetimes. Though not completely intuitive, a fraction of HTC meteoroids can access the Earth while their perihelion distance is beyond Mercury's aphelion. 
 \replaced{Overall, the transport rates are 50\% for the smaller HTC particles ($D=10-200~\mu$m), somewhat higher than JFC meteoroids, and decreases significantly for larger HTC meteoroids.}{Overall, the transport rates for smaller HTC particles ($D=10-200~\mu$m) are around 500\%, they significantly decrease for larger HTC meteoroids and shorter collisional lifetimes, and are mostly larger than those of JFC meteoroids.}  
 
\begin{figure}[h]
\centering
\includegraphics[width=0.9\textwidth]{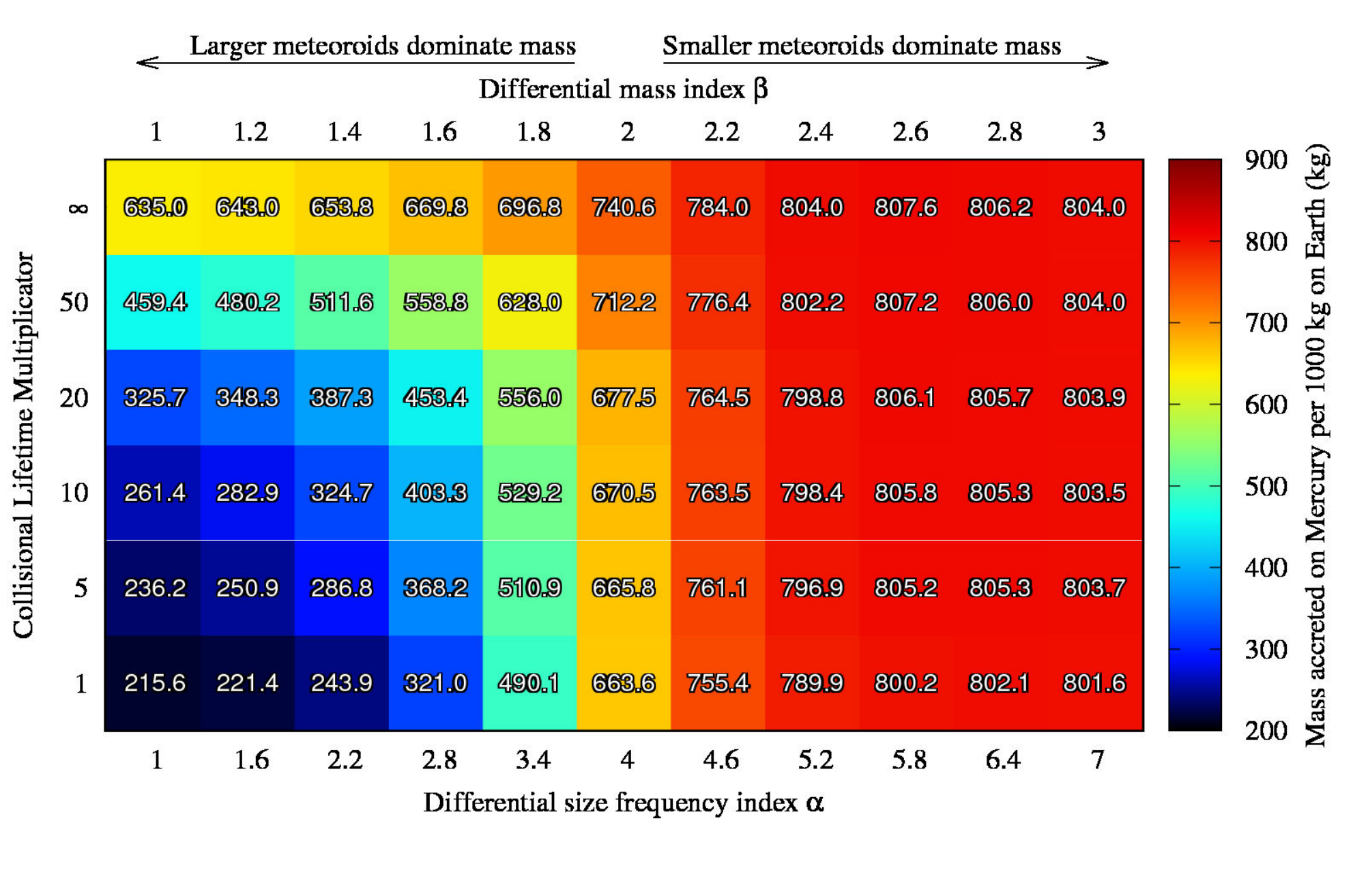}
\caption{The same as Fig. \ref{FIG_AST_MASSINDEX} but now for HTC meteoroids.}
\label{FIG_HTC_MASSINDEX}
 \end{figure}

 Transport rates significantly larger than those of AST and JFC meteoroids result in close to equal Earth vs. Mercury mass accretion rates (700--800 kg at Mercury per 1,000 kg at Earth) for larger mass indices $\alpha>4,~\beta>2$ (Fig. \ref{FIG_HTC_MASSINDEX}). With the decreasing differential mass index of the HTC meteoroid population, the accreted mass ratio decreases, reflecting the lower transport rates of larger HTC meteoroids. The overall decrease is significant even for longer collisional lifetimes (factor of 2--3 for $F_\mathrm{coll}=20,50$), and increases for shorter collisional lifetimes. The area centered around $F_\mathrm{coll}=20, \alpha=4, \beta=2$ is more sensitive than for the case of JFC meteoroids, but it varies by just 20\%. 
 
In conclusion, HTC meteoroids follow a more complex dynamical evolution compared to AST and JFC meteoroids. Due to the presence of the giant planet barrier, the larger HTC meteoroids are easily scattered which prevents them from circularization due to PR drag. HTC meteoroids impact Mercury both from the apex and toroidal directions. The apex direction is populated solely by retrograde HTC meteoroids whose presence, in addition to prograde meteoroids, creates bimodal inclination and impact velocity distributions. HTC meteoroids thus yield much more energy per mass unit than both JFC and AST meteoroids. Moreover, for each kilogram of HTC meteoroids accreted on Earth a similar amount of meteoroids is accreted on Mercury for a wide spectrum of possible configurations of collisional lifetimes and SFDs. 
 
 \subsection{Results - Oort Cloud Comet meteoroids}
 \label{sec:OCC}
 The dynamical evolution of HTC meteoroids shows that the giant planet barrier is difficult to overcome and its effectiveness  is closely related to the meteoroid diameter $D$. In our model OCC meteoroids start with an initial semimajor axis $a_\mathrm{init}>300$ au before they embark on their long journey to the inner solar system (see Section \ref{SEC:Models}). Their characteristic dynamical timescales, i.e. the time period during which the number of meteoroids in the model decreases to $1-1/e=36.8\%$, are approximately 0.58 Myr for $D=10~\mu$m OCC meteoroids, and approximately 1.7 Myr for those with $D=100~\mu$m. The dynamical timescale of OCC meteoroids is on average 30\% shorter for larger initial semimajor axes (in our model $a_\mathrm{init}=1000,3000$ au) due to higher initial eccentricities and more efficient scattering by giant planets. OCC meteoroids with initial $a_\mathrm{init}=300$ au are scattered the least and the dynamical timescale is more strongly driven by PR drag. In our model the same weight is given to all OCC meteoroids regardless of their initial semimajor axis. Due to the easier access of $a_\mathrm{init}=300$ OCC meteoroids, the resulting flux at Mercury is mostly comprised of these meteoroids, whereas meteoroids with $a_\mathrm{init}=1000$ au contribute on average a factor of 5 less in terms of flux than $a_\mathrm{init}=300$ au particles, and $a_\mathrm{init}=3000$ au meteoroids have a factor of 10 lower flux compared to $a_\mathrm{init}=300$ au.

\begin{figure}[h]
\centering
\includegraphics[width=0.9\textwidth]{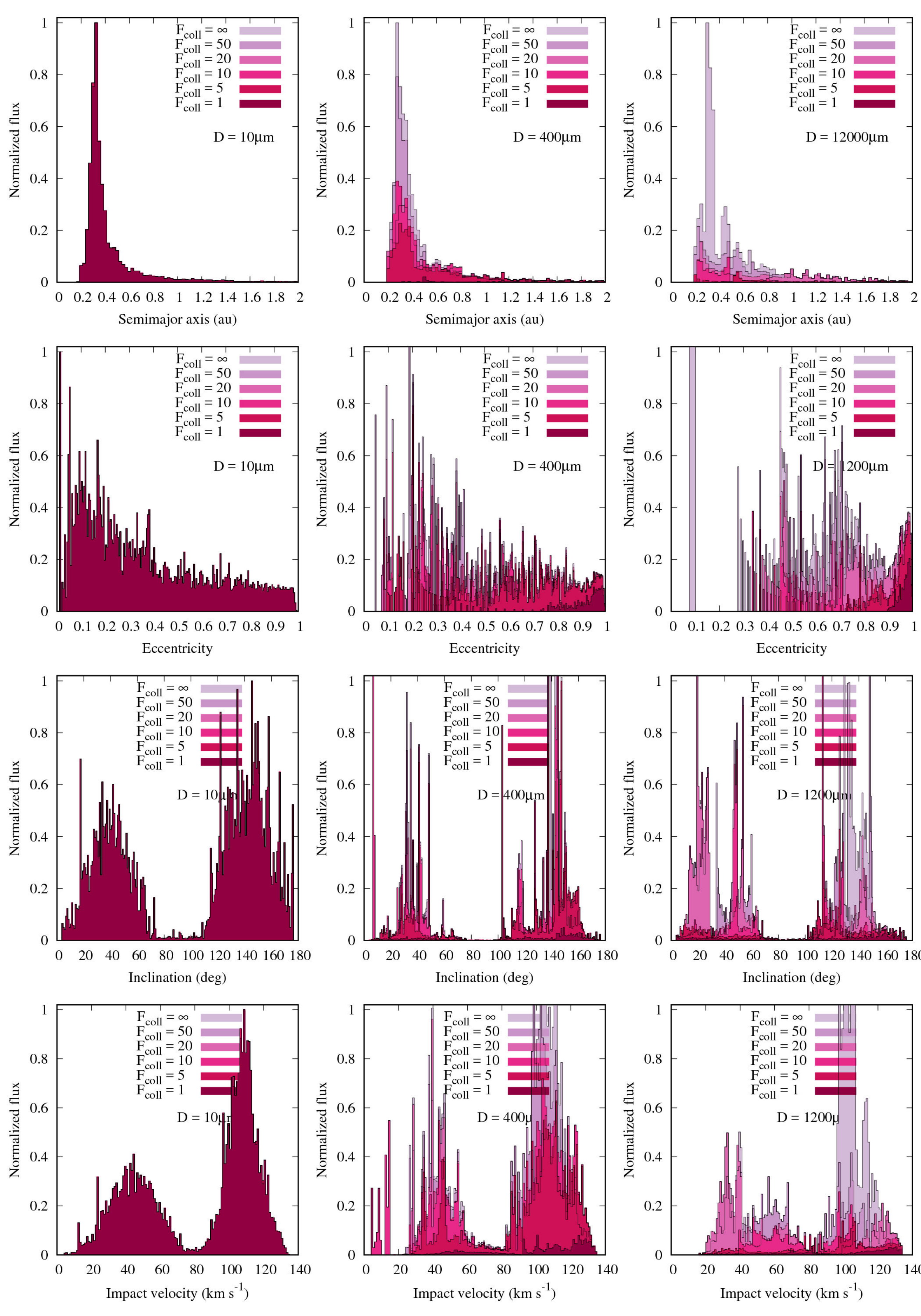}
\caption{Distribution of orbital elements for OCC meteoroids with $D=10,~400,~1200~\mu$m for TAA$=0^\circ$. Different collisional lifetime multipliers $F_\mathrm{coll}$ are represented by the range of colors. All distributions are normalized to unity.
}
\label{FIG_OCC_ELEMS}
 \end{figure}

 Figure \ref{FIG_OCC_ELEMS} shows the semimajor axis, eccentricity, inclination and impact velocity distributions for three different sizes of OCC meteoroids; $D=10,400,1200~\mu$m. The tracing of OCC meteoroids results in noisier distributions than other simulated populations  because only a very small fraction of these meteoroids can penetrate the inner solar system and evolve into orbits with accountable collisional probability with Mercury. This effect is more evident for larger diameters, while for $D=10~\mu$m the statistics are better. Similar results were reported previously by \citet{Nesvorny_etal_2011OCC}. However, those authors simulated an even smaller statistical sample (1000 meteoroids per size bin) and thus motivated us to expand our models to 20 times more meteoroids. The low number of meteoroids able to decouple from Jupiter results in a limited reservoir of possible orbital configurations (records in our model) that undergo further orbital evolution with higher probabilities. In reality, the amount of OCC meteoroids is many orders of magnitude higher than our current modeling capabilities, thus such spikiness is just a model artifact due to low statistics. 
 
 Many features of orbital element distributions of OCC meteoroids are similar to HTC meteoroids. While OCC meteoroids start with much higher eccentricities, their impact probabilities with Mercury are extremely low due to their large values of semimajor axis. Once OCC meteoroids assume orbits with lower semimajor axes, their eccentricity is actually less extreme than for HTC meteoroids resulting in less meteoroids with $e>0.9$. Results for the smallest OCC meteoroids in our model, $D=10~\mu$m, are very similar to those for HTC meteoroids of the same size, with the only difference being that the ratio of prograde to retrograde orbits is in favor of the latter and consequently the distribution of impact velocities for OCCs is dominated by $V_\mathrm{imp}>80$ km s$^{-1}$. The difference stems from HTC meteoroids having initially preferentially prograde orbits compared to an isotropic initial distribution of OCC meteoroids. With increasing size, the overall trends seem to follow similar distribution as the smallest grains. For $D=400~\mu$m meteoroids the effect of collisions starts becoming important, where for the shortest collisional lifetimes $F_\mathrm{coll}=1,5$ OCC meteoroids do not survive long enough to be driven to low eccentricities by PR drag. The ratio between prograde and retrograde meteoroids does not change regardless of the collisional lifetime even for the largest grains, $D=1200~\mu$m, which is different from HTC meteoroid behavior where the retrograde population was decreasing in importance with increasing particle diameter. This might be an important factor in distinguishing HTC and OCC meteoroid populations since their orbital element distributions are very similar.

\begin{figure}[h]
\centering
\includegraphics[width=0.9\textwidth]{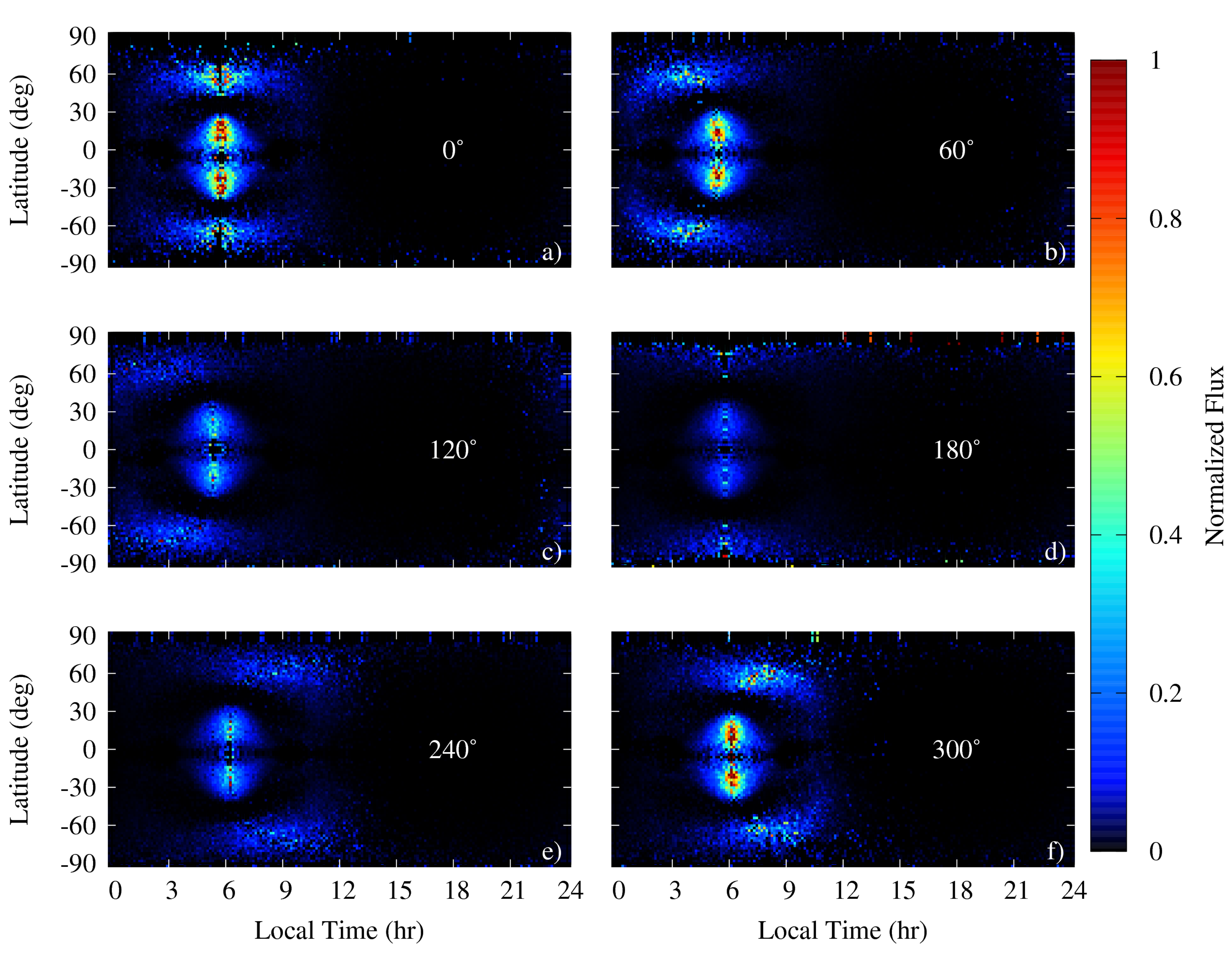}
\caption{Normalized radiant distribution of OCC meteoroids with $D=10~\mu$m impacting Mercury's surface for six different TAA (white number at 18 hr, $0^\circ$ in all panels). The mutual meteoroid collisions are not considered in this case. The x-axis represents the local time on Mercury, and it is fixed with regards to the sub-solar point (12 hr). Due to Mercury's eccentricity, the location of the apex (approximately at 6 hr) changes along Mercury's orbit. The latitude is measured from Mercury's orbital plane (not the ecliptic).} 
\label{FIG_OCC_RADIANTS_10}
 \end{figure}

The orbital element distribution of OCC meteoroids impacting Mercury suggests that the majority of meteoroids will impact from the apex region (6 hr, $0^\circ$). This is true  for all OCC meteoroid sizes and collisional lifetimes considered in this manuscript. Figure \ref{FIG_OCC_RADIANTS_10} shows the normalized distribution of impacting meteoroid radiants for $D=10~\mu$m. The apex source dominates the total flux regardless of Mercury's TAA. Furthermore, the north and south toroidal sources (panel a in Fig. \ref{FIG_OCC_RADIANTS_10})  experience the same motion along Mercury's orbit as previously shown for HTC meteoroids. The apex source, while seemingly always centered at LT = 6 hr, experiences a subtle $\pm30$ minute motion in local time during Mercury's orbital cycle following the direction of the more pronounced motion of lower velocity meteoroids.

\begin{figure}[h]
\centering
\includegraphics[width=0.9\textwidth]{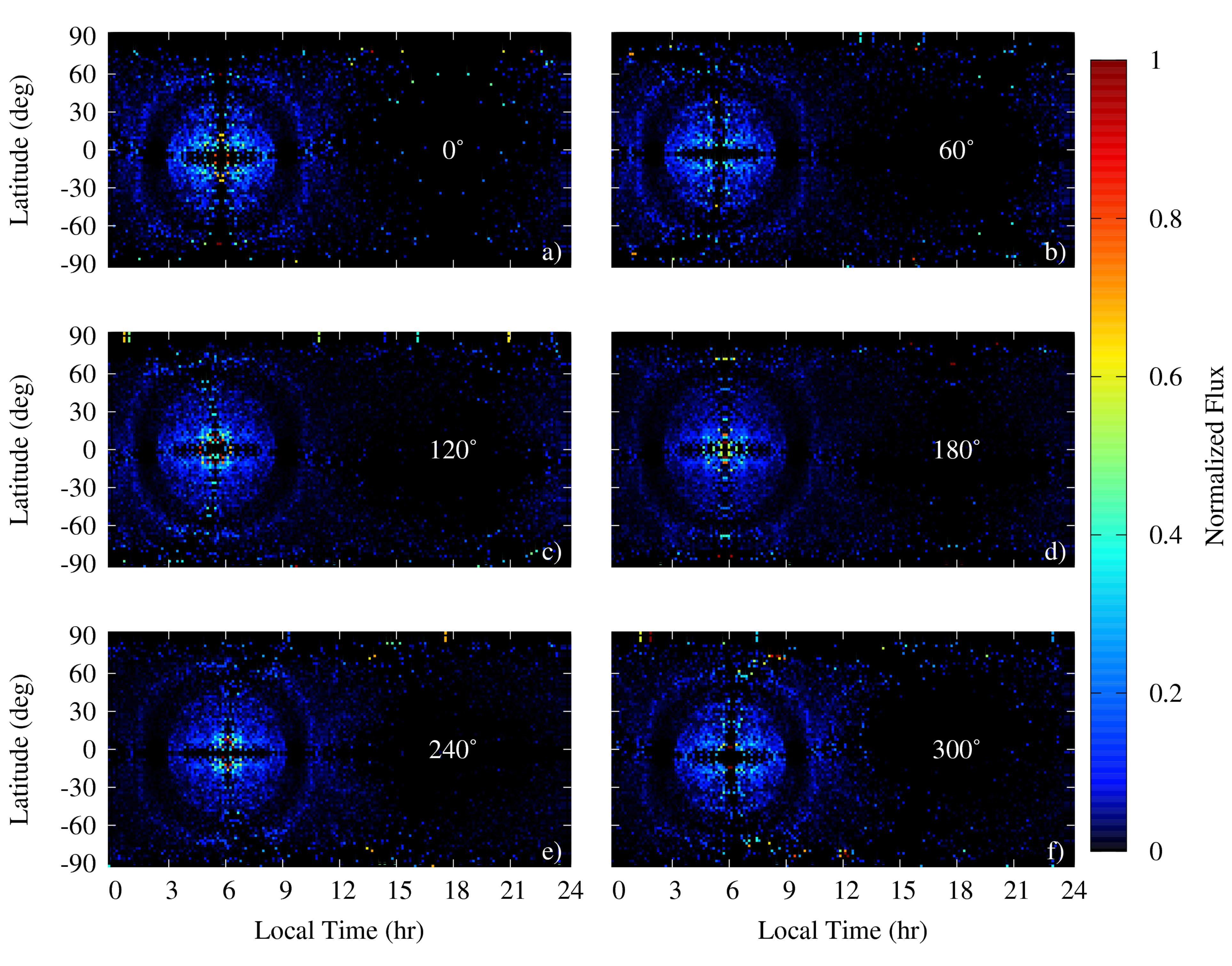}
\caption{The same as Fig. \ref{FIG_OCC_RADIANTS_10} but now for OCC meteoroids with $D=1200~\mu$m and $F_\mathrm{coll}=1$.}
\label{FIG_OCC_RADIANTS_2000}
 \end{figure}

 Due to very similar orbital distributions of OCC meteoroids impacting Mercury, the radiant distributions are very similar for all sizes and collisional lifetimes. The only exception is the most extreme case for the shortest collisional lifetime and the largest meteoroids, $D=1200~\mu$m and $F_\mathrm{coll}=1$, shown in Fig. \ref{FIG_OCC_RADIANTS_2000}. Collisions of OCC meteoroids with the zodiacal cloud effectively prevent all meteoroids from being circularized below $e=0.9$. This results in a very broad distribution of OCC meteoroid radiants with a still prominent apex source but also with a significant contribution from other sources except for the anti-apex source (18 hr, $0^\circ$). Note, that the collisions in this case decrease the flux of OCC meteoroids at Mercury by more than 95\%.

\begin{figure}[h]
\centering
\includegraphics[width=0.9\textwidth]{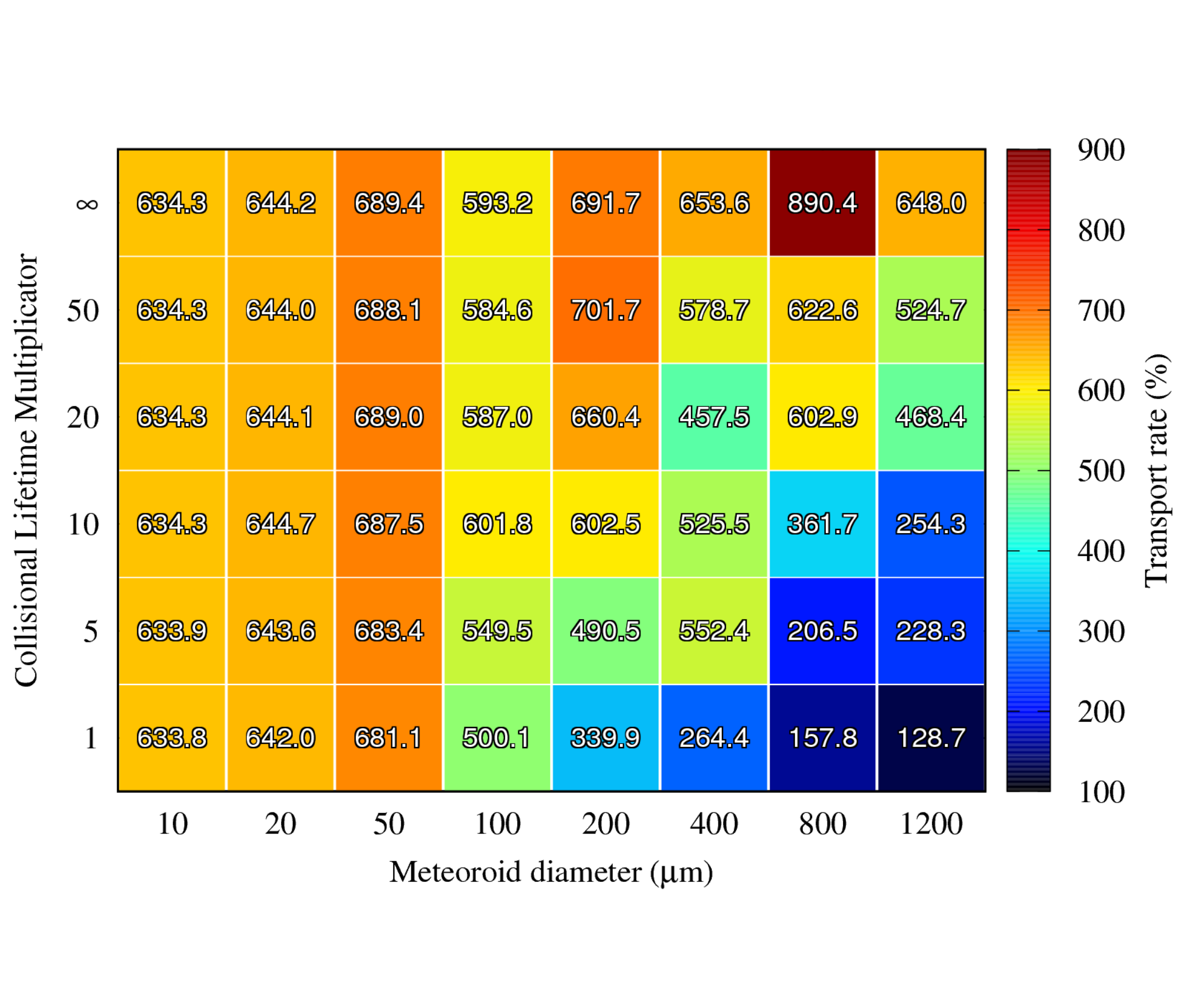}
\caption{The same as Fig. \ref{FIG_AST_EARTH_MERCURY_TRANS} but now for OCC meteoroids.}
\label{FIG_OCC_EARTH_MERCURY_TRANS}
 \end{figure}

 Due to the very high eccentricities and low perihelion distances of OCC meteoroids in the initial distributions, both Earth and Mercury are being impacted by OCC meteoroids since the beginning of their dynamical evolution. The fact that Mercury is orbiting faster and through a more dense medium than the Earth means that more OCC meteoroids impact on Mercury (Fig. \ref{FIG_OCC_EARTH_MERCURY_TRANS}), similar to other meteoroid populations. For grains smaller than $D=100~\mu$m, the transport rate between Earth and Mercury is fairly constant, resulting in seven times higher flux of OCC meteoroids impacting Mercury as compared to Earth, regardless of the collisional lifetime multiplier $F_\mathrm{coll}$. With increasing size the collisions decrease the transport rates by a factor of 2--3. The low number statistics cause noticeable fluctuations of the transport rate values for larger sizes, $D=800,1200~\mu$m, particulary compared to the smooth progression seen for HTC meteoroids (Fig. \ref{FIG_HTC_EARTH_MERCURY_TRANS}).

 By comparing the accreted fluxes at Earth and Mercury, it becomes evident that OCC meteoroids are very effective in transporting meteoroid mass to the inner solar system. For a large spectrum of size frequency and mass indices, $\alpha \ge 4, \beta \ge 2$, the amount of mass accreted at Mercury is actually comparable to that at Earth despite of the planet's seven times smaller surface area. The mass accretion rate is also insensitive to changes of the collisional lifetime, where the mass accretion values are fairly constant for $F_\mathrm{coll} \ge 20$. Only for shorter collisional lifetimes $F_\mathrm{coll}=1,~5$ and large particle dominated SFDs/MFDs ($\alpha<2.2,~\beta<1.4$) the mass accreted at Mercury is significantly smaller when compared to Earth.
 
  \begin{figure}[h]
\centering
\includegraphics[width=0.9\textwidth]{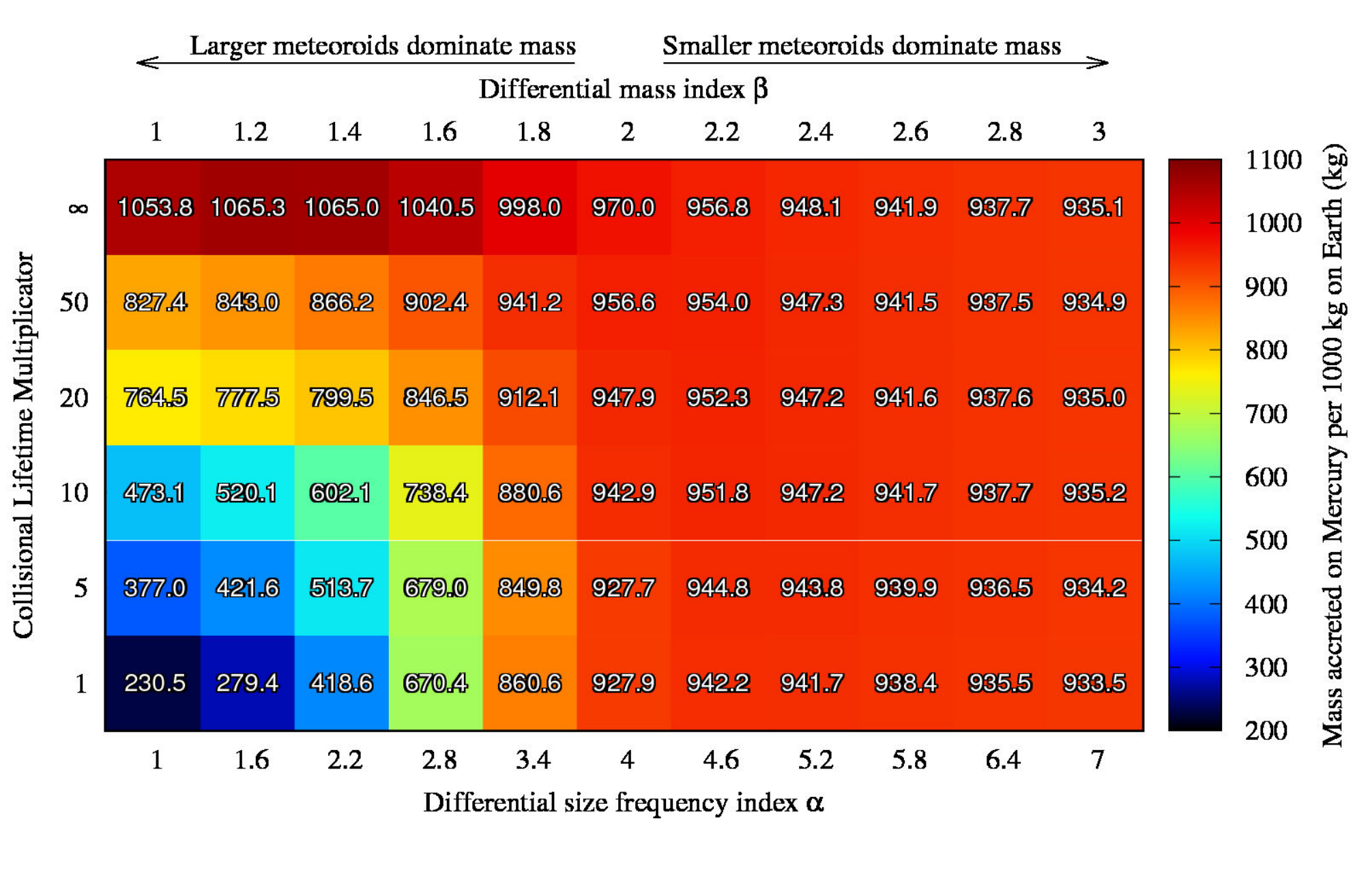}
\caption{The same as Fig. \ref{FIG_AST_MASSINDEX} but now for OCC meteoroids.}
\label{FIG_OCC_MASSINDEX}
 \end{figure}
 
 In summary, our results show that OCC meteoroids are very similar to HTC meteoroids where the only significant difference is their preferably retrograde orbits in comparison to preferentially prograde HTC meteoroids. Regardless of their size or the collisional lifetime, OCC meteoroids preferentially impact Mercury from the apex region showing only a subtle motion during Mercury's orbital path. Consequently, OCC meteoroids have the highest average impact velocities for all meteoroid sources considered here. A very complex dynamical evolution, mainly overcoming the Jupiter barrier, prevents a large fraction of OCC meteoroids from decoupling from Jupiter and obtaining higher impact probability with inner solar system objects \citep[see e.g.,][]{Nesvorny_etal_2011OCC}. Nevertheless, when OCC meteoroids effectively penetrate the inner solar system, the amount of material accreted onto Earth and Mercury is comparable for a large variety of possible modeled parameters based on variations of SFD/MFD and the collisional lifetime.

  \subsection{Results - Preferred solution}
 \label{sec:preferred}
 We now combine all modeled meteoroid populations to describe the characteristics of the Hermean meteoroid environment as a whole. In the previous sections we demonstrated that choices regarding the collisional lifetime and SFD/MFD can somewhat influence the access of different populations to Mercury while maintaining the same amount of meteoroid mass impacting the Earth. The total mass flux at Earth and its partitioning to different populations is itself uncertain. In this section we use the latest meteoroid influx calibration at Earth reported by \citet{Carrillo-Sanchez_etal_2016}, who presented estimates for each meteoroid population assumed here: $\mathrm{AST}=3.7\pm2.1 \mathrm{~t~day}^{-1},\mathrm{JFC}=34.6\pm13.8\mathrm{~t~day}^{-1}, \mathrm{LPC}=5.0\pm2.7 \mathrm{~t~day}^{-1},$ where LPC means long period comets. \citet{Carrillo-Sanchez_etal_2016}, who focused on the atmospheric effects produced by the meteoroid flux, assumed that LPC = HTC because differences between considering HTC or OCC meteoroids are
 negligible since both meteoroid populations at Earth have atmospheric entry velocities large enough ($V_\mathrm{imp}>25$ km s$^{-1}$) to efficiently ablate all metals in the atmosphere \citep{Vondrak_etal_2008,Janches_etal_2009}.
 Here, we will assume 1:1 ratio between HTC and OCC meteoroids since there are undeniably sources of OCC meteoroids observed in the solar system \citep{Marsden_2005} and recently this ratio was used to explain the dust ejecta cloud observed around the Moon \citep{Janches_etal_2018}. Since this choice is rather arbitrary we also tested the sensitivity of our solution for different mixing ratios of HTC and OCC meteoroids. 
 
Besides the influx at Earth, two additional adjustable parameters of our model are: the SFD/MFD of meteoroids at their sources characterized by the indices $\alpha, \beta$, and the collisional lifetime characterized by $F_\mathrm{coll}$. Recent dynamical modeling investigations suggest that the $F_\mathrm{coll}$ must be much higher than unity in order to plausibly explain many features observed at Earth \citep{Nesvorny_etal_2011JFC,Pokorny_etal_2014}. In this section we will use $F_\mathrm{coll}=20$ as our preferred solution based on \citet{Pokorny_etal_2014} and similar to \citet{Nesvorny_etal_2011JFC} and let $F_\mathrm{coll}$ vary between 10 and 50 for all meteoroid populations. Selecting the correct SFD of each meteoroid population is a challenging task since in-situ measurements of SFDs are scarce and limited. In recent years, progress has been made in estimating the SFD/MFD of impact cratering observations on the Moon \citep{Suggs_etal_2014}, sporadic meteor background measurements at Earth \citep{Blaauw_etal_2011,Pokorny_Brown_2016}, and the infra-red observations of the Zodiacal cloud \citep{Planck_2014}, suggesting various values for the sporadic meteoroid environment. In this work we choose $\alpha=4,\beta=2$ as our preferred solution for all meteoroid populations, and we will investigate the sensitivity of our solution for the following ranges of SFD/MFD indices, $\alpha=\left[3.4,4.6\right],\beta=\left[1.8,2.2\right]$, which cover a wide suite of possible solutions and bracket the observed meteoroid environment at Earth \citep{Pokorny_Brown_2016}. Our model has $4\times3 = 12$ free parameters (the mass influx at Earth, SFD/MFD, $F_\mathrm{coll}$ for each of four meteoroid populations), which even for 3 different values for each parameter leaves us with a rather large number of model outcomes (531,441). Thus, we will show the preferred solution and then the most extreme deviations from the preferred solution in order to keep the manuscript in publishable length. 
 
\begin{figure}[h]
\centering
\includegraphics[width=0.9\textwidth]{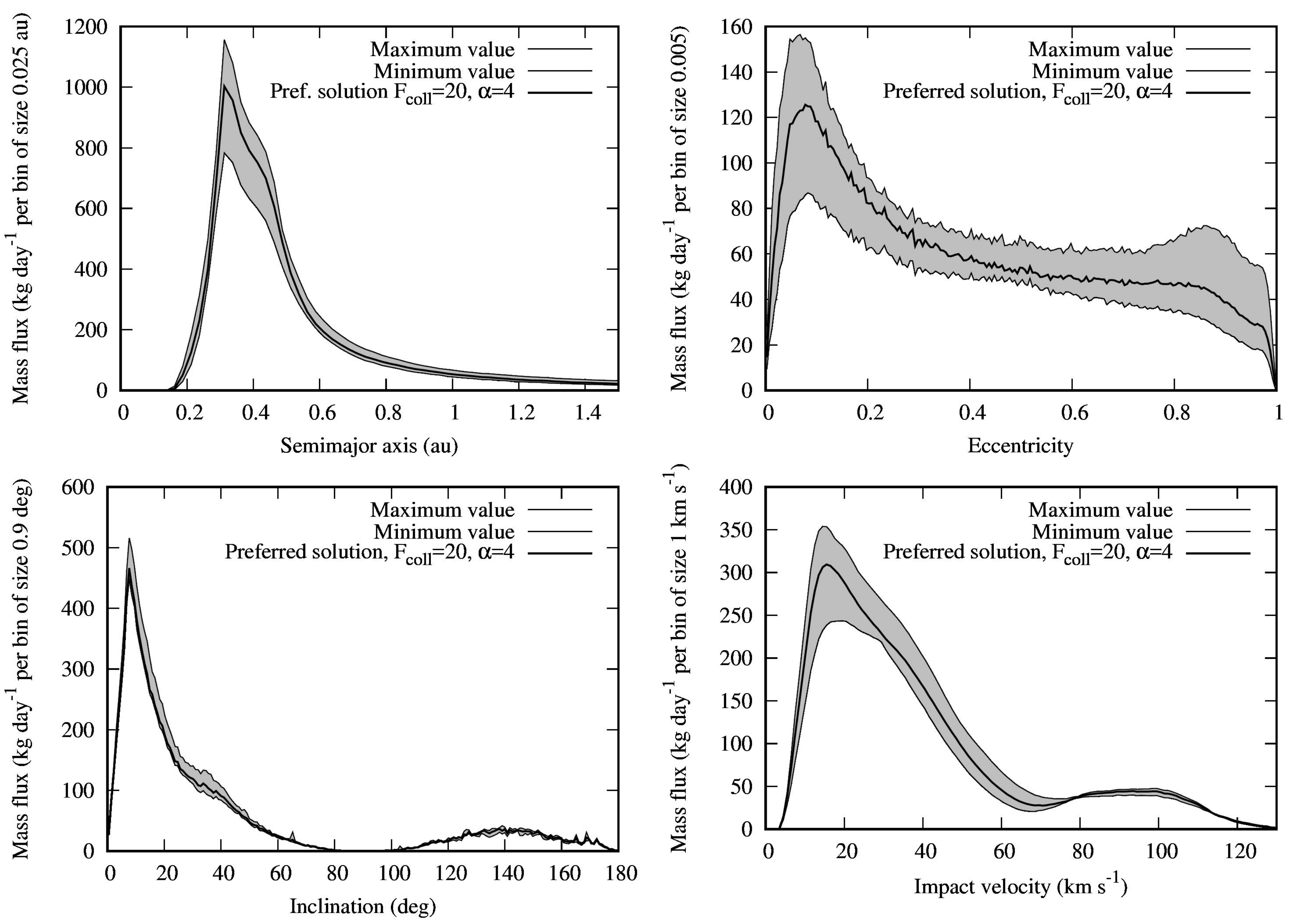}
\caption{Distribution of orbital elements for our preferred solution (thick black lines), $F_\mathrm{coll}=20,\alpha=4.0$, and the confidence interval (gray area and thin black lines representing interval boundaries). This plot represents the total accumulated mass over one Hermean year divided by number of days, i.e. the mass flux for each orbital element averaged over the whole orbit.}
\label{FIG_PREF_ELEMS}
 \end{figure}

 With the mixing ratio proposed by \citet{Carrillo-Sanchez_etal_2016} as well as the estimates of mass transfer rates between the Earth and Mercury reported in previous sections, JFC meteoroids represent the majority of the meteoroid mass accreted on Earth, however as we showed in previous sections, the accretion efficiency of HTC and OCC meteoroids is much higher than those of AST and JFC meteoroids. The actual mass accreted on Mercury averaged over its whole orbit in our preferred solution is: $\mathrm{AST}_M=0.26\pm0.15 \mathrm{~t~day}^{-1},\mathrm{JFC}_M=7.84\pm3.13\mathrm{~t~day}^{-1}, \mathrm{HTC}_M=1.69\pm0.91\mathrm{~t~day}^{-1},\mathrm{~and~ OCC}_M=2.37\pm1.38\mathrm{~t~day}^{-1}$. This results in a mass influx ratio of JFC/(OCC+HTC) $\sim 2$, which is much lower than that at Earth \citep[$\sim 7$,][]{Carrillo-Sanchez_etal_2016}. Figure \ref{FIG_PREF_ELEMS} shows the orbital elements and velocities of impactors accumulated over the entire Mercury year and provides an averaged mass flux in kg/day. For this last calculation we divided the total accumulated mass by a Mercury year, i.e. 88 days. The daily mass influx varies with TAA as will become evident in later figures. The preferred solution is shown by the thick solid black line, and the minimum and maximum values recorded for all possible permutations of our free parameters are shown as thin solid black lines. The gray solid area thus represents our confidence interval. We see several trends in the orbital element distributions.  The $a$ distribution shows that the mass flux is dominated by perihelion particles and decreases toward aphelion, with the wider distribution of semimajor axes reflecting Mercury's range of heliocentric distances. The eccentricity distribution is dominated by low eccentricity meteoroids, with a plateau from $e=0.3$ to 0.9. Interestingly, the confidence interval allows for an almost flat distribution in $e$, however none of the individual solution curves are in fact this flat. Cases with lower size/mass index, $\alpha=3.4,\beta=1.8$, have an increased contribution to the high eccentricity part of the meteoroid populations, while cases with higher size/mass indices emphasize low eccentricity orbits.  The inclination distribution is dominated by the JFC meteoroid portion resulting in a predominantly prograde distribution (10.3 $\mathrm{~t~day}^{-1}$ of prograde meteoroids vs. 1.8 $\mathrm{~t~day}^{-1}$ of retrograde meteoroids). Only the SFD/MFD has a measurable influence on this ratio: for $\alpha=3.4$ the mass of prograde meteoroids increases to 10.6 $\mathrm{~t~day}^{-1}$, whereas for $\alpha=4.6$ the prograde mass decreases to 10.1 $\mathrm{~t~day}^{-1}$. A similar dichotomy that is directly tied to the meteoroid inclination is evident in the impact velocity distribution. Notice that the impact velocity distribution is broader than the meteoroid inclination. This is caused by the changes in Mercury's orbital velocity during its orbit which has no effect on meteoroid inclinations.
 
 The orbital elements of the preferred solution are within the confidence region in Fig. \ref{FIG_PREF_ELEMS} for all possible HTC-to-OCC ratios within LPC = 5 $\mathrm{~t~day}^{-1}$ range, with the inclination and impact velocity distributions being the two exceptions. For the OCC dominating solution, LPC = OCC = 5 $\mathrm{~t~day}^{-1}$ at Earth, there is a 60\% increase in impacting mass from meteoroids with $I>90^\circ$ and $90<V_\mathrm{imp}<120$ km s$^{-1}$, whereas for HTC dominating solution, LPC = HTC = 5 $\mathrm{~t~day}^{-1}$, the model results in a 40\% decrease for the same inclination and impact velocity range.
 
\begin{figure}[h]
\centering
\includegraphics[width=0.9\textwidth]{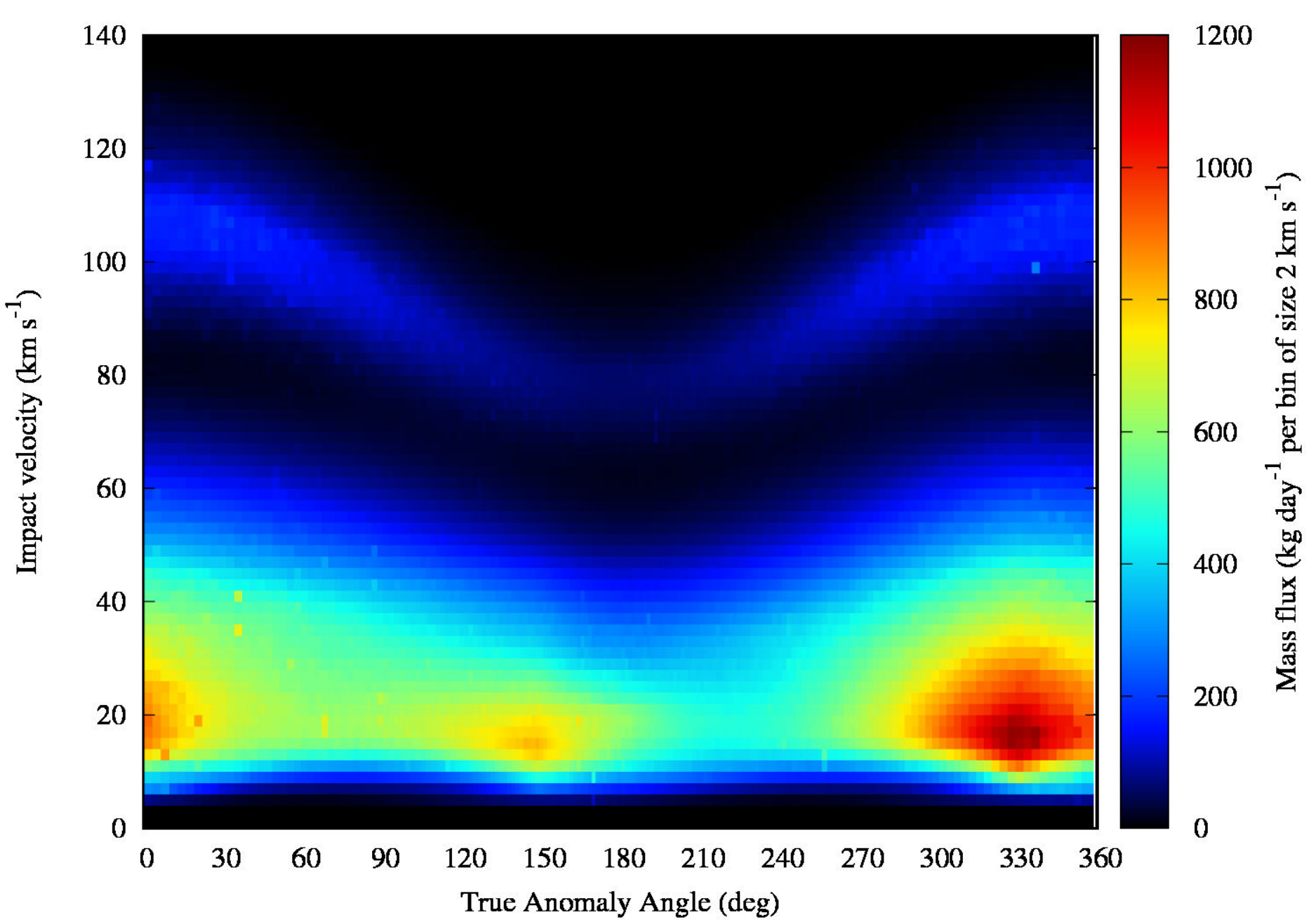}
\caption{The mass influx distribution as a function of Mercury's true anomaly angle (x-axis) and the impact velocity (y-axis). The units are kg/day per 2 km s$^{-1}$ bin, where the non-uniform distribution of TAA in time was taken into account.}
\label{FIG_PREF_VELOCITY}
 \end{figure}

Variations of the impactor velocity with Mercury's true anomaly angle, which are introduced by the planet's eccentricity, are one of the most interesting outputs of the Hermean meteoroid environment model. Figure \ref{FIG_PREF_VELOCITY} shows the distribution of meteoroid impact velocities for our preferred solution for all TAAs color coded in terms of their impacting mass. That is, each impact velocity bin (2 km s$^{-1}$) represents the mass flux from a particular bin in kg day$^{-1}$. Thus the integral over the whole velocity spectrum gives the total mass influx rate in kg day$^{-1}$. The bimodal distribution of impact velocities is present for all TAAs and follows the expected trend where the separation impact velocity (prograde vs. retrograde orbits) decreases with TAA from $V_\mathrm{imp}=80$ km s$^{-1}$ at perihelion (TAA = $0^\circ$) to $V_\mathrm{imp}=65$ km s$^{-1}$ at aphelion (TAA = $180^\circ$). The mass flux provided by both fast and slow populations decreases when Mercury is moving toward its aphelion with the local maximum around TAA = $150^\circ$ for the slower impact velocities. Such a local maximum is not present for faster meteoroids. The global minimum for the faster meteoroids occurs at aphelion (TAA = $180^\circ$), while for slower meteoroids it occurs after the aphelion passage at TAA = $210^\circ$. The global maximum for slower meteoroids is at TAA = $326^\circ$, while  faster meteoroids have the maximum mass influx rate at perihelion.

One of the goals of this work is to provide the absolute impact vaporization flux. Figure \ref{FIG_PREF_VELOCITY} provides the highest resolution data set for estimating the absolute impact vaporization flux reported to date. By simply applying a function that transforms the impacting meteoroid mass/volume and the impact velocity $V_\mathrm{imp}$ into the impact vaporization mass/volume, we can obtain the absolutely calibrated impact vaporization flux for any segment of Mercury's orbit. Moreover, our fine binning in the impact velocity allows us to estimate the effect that velocity cutoffs/thresholds have on vaporization efficiency rates to simulate the behavior of different chemical species. Here, we consider a rather simple but still applicable relation between the impacting volume,  impact velocity and amount of vaporized volume, given by \citet{Cintala_1992}:
\begin{equation}
    \mathcal{V}_\mathrm{vapor} = \mathcal{V}_\mathrm{Met} (c+d V_\mathrm{imp} + eV_\mathrm{imp}^2),
    \label{EQ_VAPOR}
\end{equation}
where $\mathcal{V}_\mathrm{vapor}$ is the volume vaporized by a meteoroid with volume $\mathcal{V}_\mathrm{Met}$, and $c=-3.33$, $d=-0.0102$ km$^{-1}$ s, $e=0.0604$ km$^{-2}$ s$^2$ are empirically determined constants from \citet{Cintala_1992}. These constants are valid for an iron projectile on a target substrate with temperature of 400 K. The variation with different target temperatures is insignificant (a few percent) according to \citet{Cintala_1992}. Changing the projectile material to diabase yields on average 10\% differences depending on the meteoroid impact velocity, where diabase provides more impact vaporization for $V_\mathrm{imp}<50$ km s$^{-1}$. By assuming a bulk density of the vaporized surface $\rho_\mathrm{vapor} = 2000$ kg m$^{-3}$, we can determine the impact vaporization rate as $\mathcal{R}_\mathrm{vapor} = \sum \mathcal{V}_\mathrm{Met} \rho_\mathrm{vapor} (c+d V_\mathrm{imp} + eV_\mathrm{imp}^2)$, where the sum is over all recorded meteoroids for a particular time period. Consequently, the impact vaporization flux is the impact vaporization rate divided by a surface area $A$, $\mathcal{F}_\mathrm{vapor}= \mathcal{R}_\mathrm{vapor}/A$. If not specified differently, we consider $A=7.477\times10^{17}$ cm$^2$, the surface area of Mercury. To simulate a threshold for vaporization, which is comparable to the sound speed in the medium, we consider that impactors with $V_\mathrm{imp}<7.5$ km s$^{-1}$ have zero contribution to the total vapor \citep{Cintala_1992}.
 
\begin{figure}[h]
\centering
\includegraphics[width=0.9\textwidth]{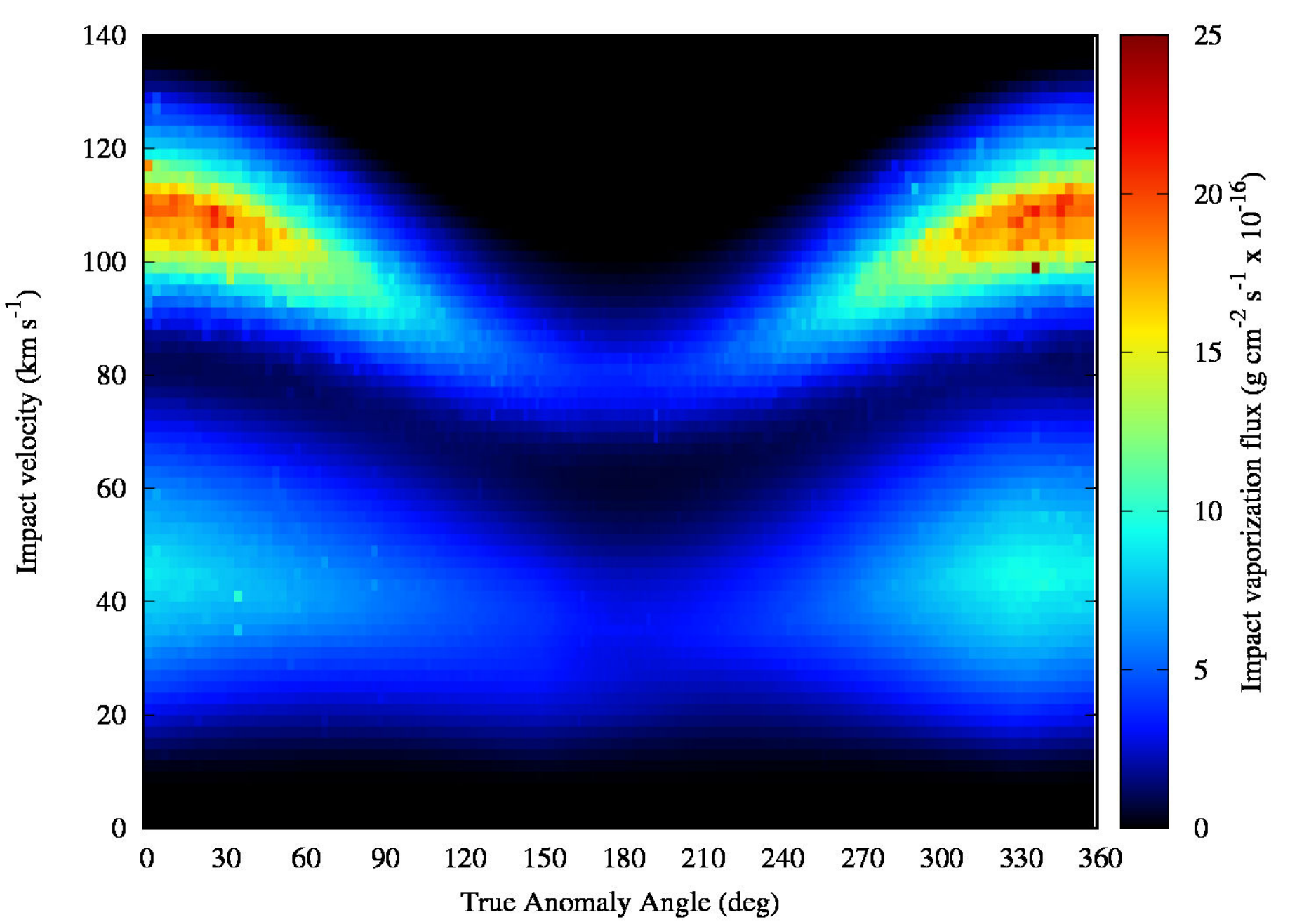}
\caption{The total vaporization flux as a function of true anomaly angle (x-axis), and the impact velocity (y-axis). The units are g cm$^{-2}$ s$^{-1}$ per 2 km s$^{-1}$ bin.}
\label{FIG_PREF_VAPOR_VELOCITY}
 \end{figure}

Figure \ref{FIG_PREF_VAPOR_VELOCITY} shows the impact vaporization flux $\mathcal{F}_\mathrm{vapor}$ in g cm$^{-2}$ s$^{-1}$ per velocity bin (2 km s$^{-1}$) for the whole range of TAAs. It is evident that the dependence of vaporization efficiency of the soil on the projectile velocity enhances the relative importance of the high velocity component of the Hermean meteoroid environment. Even though the mass influx ratio between the slower and faster component is approximately 6:1, the higher impact velocities allow the faster meteoroids to dominate the total impact vaporization flux. Note that our calibration allows for less than 5 $\mathrm{~t~day}^{-1}$ of retrograde meteoroids at Earth and yet, compared to more than 35 $\mathrm{~t~day}^{-1}$ of slower meteoroids at Earth, the long period comet meteoroids dominate the impact vaporization fluxes. The importance of this fast (retrograde) meteoroid component on the impact-driven portion of the Hermean exosphere was first emphasized by \citet{Pokorny_etal_2017_APJL}. 

\begin{figure}[h]
\centering
\includegraphics[width=0.9\textwidth]{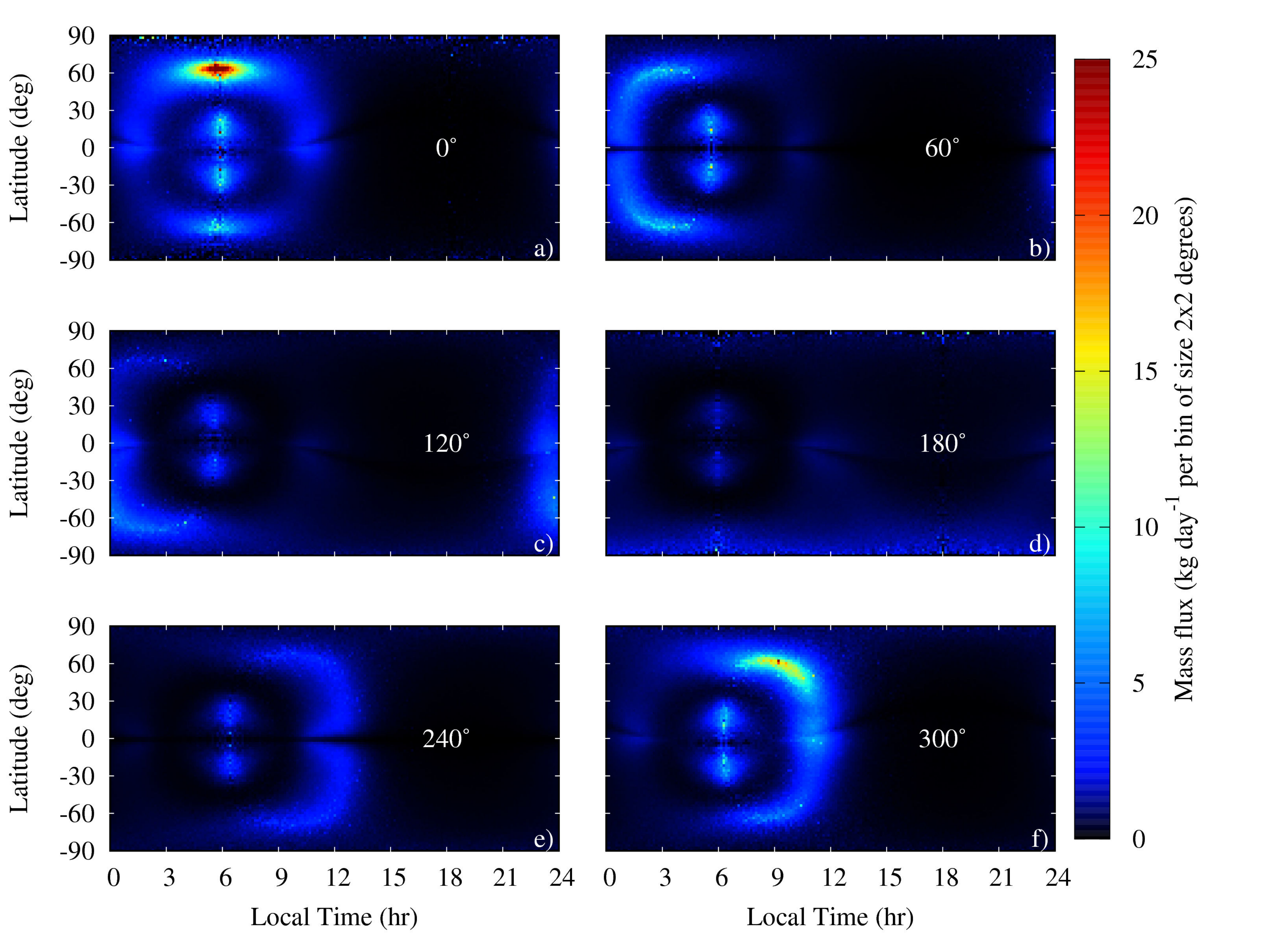}
\caption{Radiant distribution of meteoroids impacting Mercury for six (6) different true anomaly angles for our preferred solution. The units in each of $2 \times 2 $ degree bins are kg/day. 
}
\label{FIG_PREF_RADIANTS}
 \end{figure}

 In our preferred solution the arrival direction of meteoroids is mainly influenced by JFC meteoroids with a smaller contribution from the apex source (Fig. \ref{FIG_PREF_RADIANTS}). At perihelion (TAA = $0^\circ$) the meteoroids are mostly arriving from the north toroidal source (associated with JFC meteoroids) and from the ring structure, with the mass influx in the south being less than the mass flux arriving from northern latitudes. At TAA = 60$^\circ$ the radiant distribution is almost perfectly symmetric over the ecliptic, with the night side/anti-helion source dominating the mass flux. The maximum mass flux is at latitudes around $\pm60^\circ$, whereas the apex source is weaker compared to its activity during perihelion passage. As Mercury moves further away from perihelion, TAA = 120$^\circ$, the original ring structure is pushed more toward the night side and becomes weaker. At this point, the apex source is centered at LT = 5:40 hr and most of day side impacts disappear. At aphelion, TAA = 180$^\circ$, the overall flux reaches its minimum value and is enhanced at southern latitudes. The apex flux is symmetric around LT = 6hr. At TAA = 240$^\circ$, the overall flux increases and the ring structure is almost entirely on the day side and near the sub-solar terminator (12 hr). As Mercury moves closer to perihelion, TAA = 300$^\circ$, the north toroidal concentration in the day side (10 hr, 60$^\circ$) dominates once again the mass flux of meteoroids. 
 
\begin{figure}[h]
\centering
\includegraphics[width=0.9\textwidth]{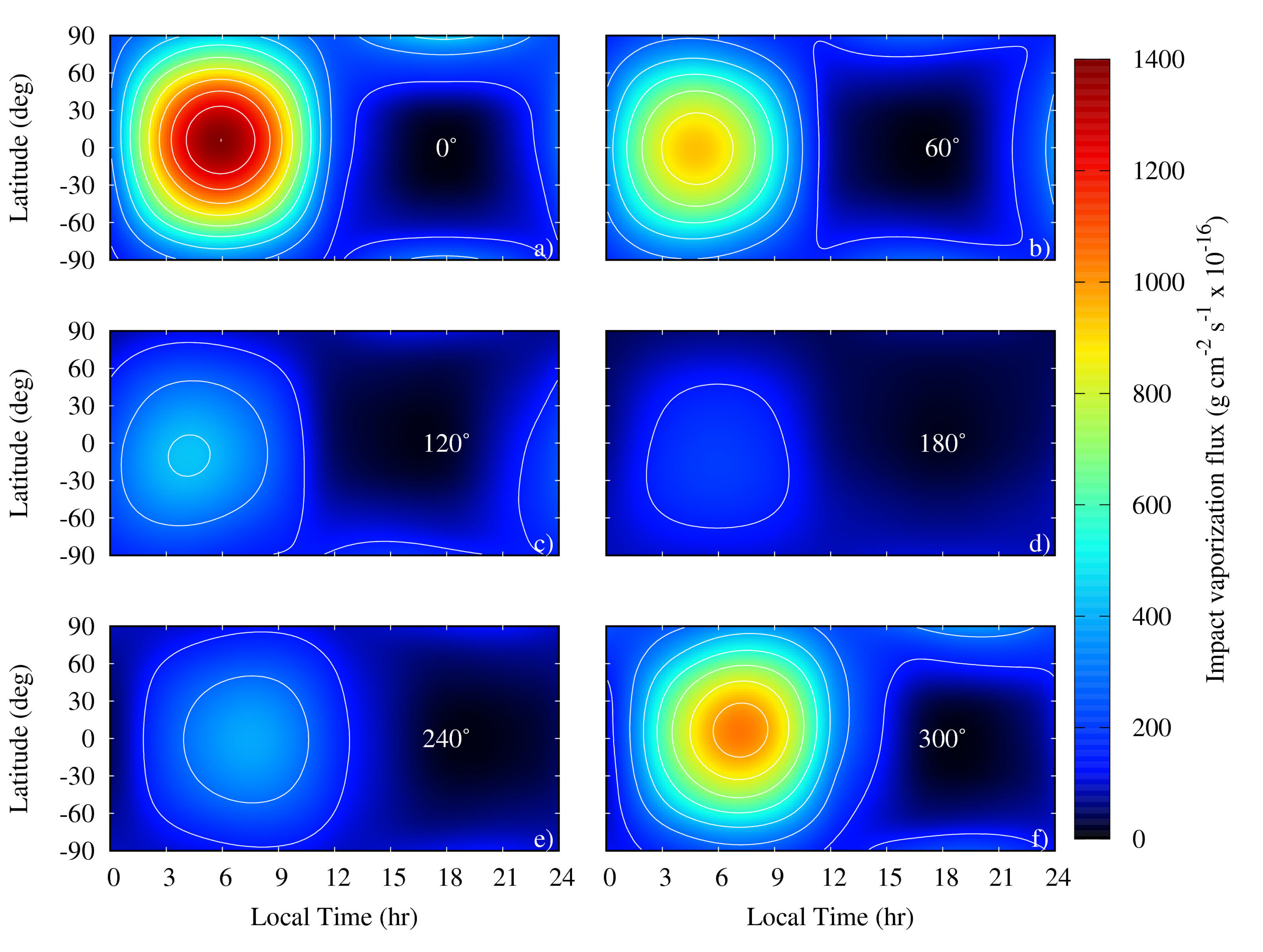}
\caption{The impact vaporization flux on Mercury for six (6) different true anomaly angles for our preferred solution. The units are g cm$^{-2}$ s$^{-1}$ $\times 10^{-16}$ and are not subject to any binning or mesh size. The contour levels represent 10\% increments. The TAA value for each particular map is labeled with white text at 18 hr, 0$^\circ$.}
\label{FIG_PREF_SURFACE}
 \end{figure}

Now, we combine the distribution of radiants with the impact velocities and calculate the dependence of the impact vaporization flux $\mathcal{F}$ from Mercury's surface as a function of local time, latitude, and the planet's true anomaly angle. These maps were constructed by first converting the meteoroid flux from steradians (Fig. \ref{FIG_PREF_RADIANTS}) to the mass flux from apparent zenith per unit area by applying a factor $1/\cos(\mathrm{latitude})$, which accounts for the smaller surface area of patches at higher latitudes. The resulting zenith flux allows us to calculate the flux to each surface patch. We assume that the flux decreases as $\cos(\mathcal{A})$, where $\mathcal{A}$ is the angular distance from the zenith for a particular surface patch, and we only consider positive values of the flux (i.e., only points within the hemisphere focused on the zenith receive any meteoroids from this direction). The zenith flux does not diverge at the poles in our model since we recorded no meteoroids impacting exactly at either of the poles. With the meteoroid flux per surface area we incorporate the impactor velocity and using Eq. \ref{EQ_VAPOR} we calculate the absolute impact vaporization flux for each Mercury's surface patch. This calculation assumes that the incidence angle of a meteoroid with respect to the local vertical does not influence the total amount of vapor produced (i.e., only the flux is reduced from the zenith, not the vapor yield). The resulting impact vaporization flux on Mercury's surface is shown in Figure \ref{FIG_PREF_SURFACE}, where the color bar shows the absolute impact vaporization flux in g cm$^{-2}$ s$^{-1}$ $\times 10^{-16}$. We added 10\% contour levels of the maximum value to emphasize the movement and shape of underlying features. A comparison with the radiant distribution of the meteoroid flux (Fig. \ref{FIG_PREF_RADIANTS}) shows that the impact vaporization flux $\mathcal{F}_\mathrm{vapor}$ is closer to the apex region than the radiant distribution would suggest. This is a consequence of the impact velocity dichotomy, where the apex source is populated by meteoroids several times faster on average compared to other directions. The difference in vaporization flux between the dawn (6 hr) and dusk (18 hr) terminators is significant for all TAA of Mercury's orbit, consistent with the clear dawn/dusk asymmetry in Mercury's exosphere observed by MESSENGER \citep{Burger_etal_2014,Killen_Hahn_2015,Merkel_etal_2017}. 

\begin{figure}[h]
\centering
\includegraphics[width=0.9\textwidth]{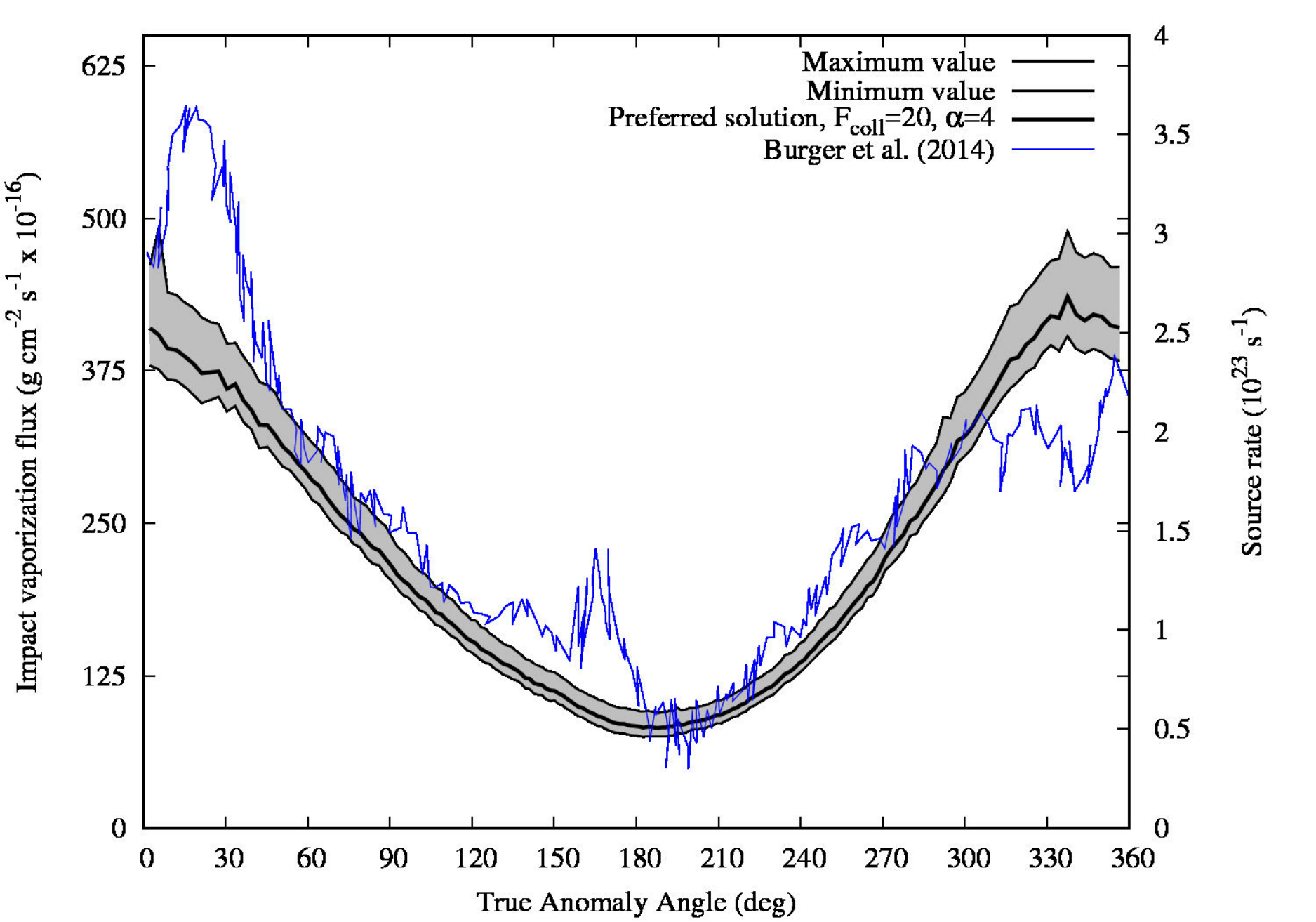}
\caption{The average impact vaporization flux in absolute units with respect to the true anomaly angle of our preferred solution (solid thick black line), the confidence interval (gray area and thin black lines representing interval boundaries), and the exospheric Ca source rate from \citet{Burger_etal_2014} (blue solid line using the secondary y-axis).} 
\label{FIG_PREF_BURGER}
 \end{figure}

By integrating the impact vaporization flux over the whole Hermean surface and dividing its own surface area we obtain the average impact vaporization flux $\mathcal{F}$ (Figure \ref{FIG_PREF_BURGER}). Our preferred solution (thick solid line in Fig. \ref{FIG_PREF_BURGER}) predicts a global maximum value at TAA = 337$^\circ$ and a global minimum at TAA = 188$^\circ$. The global maximum and minimum are assumed for the same values of TAA for all solutions in our confidence interval (gray area with thin solid black lines as borders).
The ratio between the maximum and minimum $\mathcal{F}_{\mathrm{TAA}=337^\circ}/\mathcal{F}_{\mathrm{TAA}=188^\circ}=436/82=5.29\pm0.07$, where the impact vaporization flux averaged over one Mercury orbit is $\mathcal{F}_{\mathrm{orbit}}=200 \pm 16 \times 10^{-16}$ g cm$^{-2}$ s$^{-1}$. The predicted impact vaporization flux has a similar dependence with TAA to source rates for Mercury's exospheric Calcium presented by \citet{Burger_etal_2014} with a few exceptions: enhancements at TAA = 20$^\circ$ and TAA = 170$^\circ$ that are thought to  originate from 2P/Encke \citep{Christou_etal_2015,Killen_Hahn_2015} and TAA from 315$^\circ$ to 350$^\circ$ where our model predicts 30-40\% higher impact vaporization fluxes than \citet{Burger_etal_2014}.
 
 \subsection{Discussion - free parameters}
 In this section we explore the effect that the most extreme variations from the preferred solution have on the overall results.
 The mixing ratios for individual meteoroid populations at Earth are still being refined and tested when new measurements become available. The calibration by \citet{Carrillo-Sanchez_etal_2016} shows that individual populations have uncertainly $>50\%$, which would significantly alter the absolute calibration of our model. However, using results we provided for individual meteoroid populations it is possible to reconstruct a majority of our results with a new calibration without performing a full-fledged dynamical model. 
 A major conclusion of this manuscript is the overall insensitivity of individual meteoroid populations to changes in both collisional lifetime setting $F_\mathrm{coll}$ and the size-frequency distribution for various combinations of free parameters. The radiant maps, orbital element distributions and mass transportation ratios hold their characteristics very well for any combination of $F_\mathrm{coll} \ge 10$ and $\alpha>3.4, \beta>1.8$. Moreover, no matter how we set the free parameters in our model, there will always be a strong dawn/dusk asymmetry, since the impacts around the dusk terminator are restricted only to meteoroids on circular orbits close to Mercury's aphelion (TAA = 180$^\circ$), because Mercury's orbital velocity compared to the meteoroids' orbital velocity is smaller while their velocity vectors are aligned. 

\begin{figure}[h]
\centering
\includegraphics[width=0.9\textwidth]{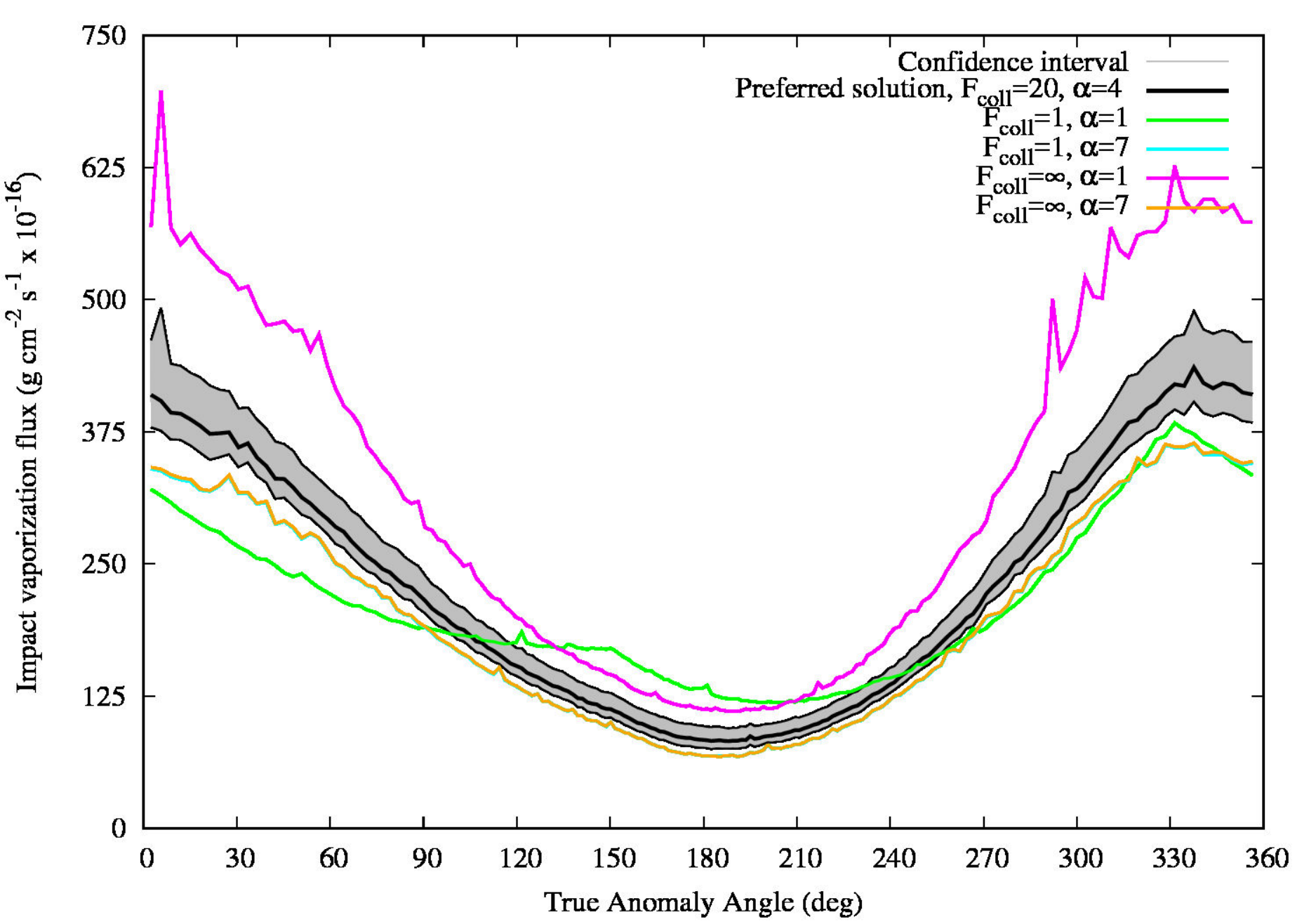}
\caption{The same as Fig. \ref{FIG_PREF_BURGER}, but now for more extreme values of the collisional lifetime multiplier, $F_\mathrm{coll}$ and the SFD/MFD slope indices $\alpha,\beta$. For the shortest collisional lifetimes, $F_\mathrm{coll}=1$ and mass dominated by largest meteoroids, $\alpha=1,\beta=1$ (solid green line) the impact vaporization curve is more flat compared to the preferred solution. For the SFD/MFD dominated by smallest meteoroids,$\alpha=7,\beta=3$ the collisional lifetime has no influence on results (cyan and orange solid lines) and the impact vaporization curve is on average 10\% below the confidence interval. No collisions and SFD/MFD dominated by largest grains $F_\mathrm{coll}=\infty$,$\alpha=1,\beta=1$ provides 10\% increase in the impact vaporization on average. The differences at perihelion are significantly higher compared to aphelion. }
\label{FIG_DISC_EXTREME}
 \end{figure}

To shed light on the sensitivity of the model, we calculated several most extreme solutions and compared them with our preferred solution in terms of the impact vaporization flux with respect to TAA (Fig. \ref{FIG_DISC_EXTREME}). In the previous sections we showed that the mass accreted at Mercury and at Earth is fairly constant for individual populations for the differential SFD/MFD indices $\alpha \ge 4,\beta \ge 2$. Unsurprisingly, when the meteoroid complex is dominated by the smallest meteoroids  $\alpha=7,\beta=3$, the impact vaporization fluxes are similar regardless of the collisional lifetime (orange and cyan solid lines in Fig. \ref{FIG_DISC_EXTREME}).  When, on the other hand, the SFD/MDF is dominated by the largest grains $\alpha=1,\beta=1$, the collisional lifetime multiplier $F_\mathrm{coll}$ significantly shapes the results. When no collisions are assumed, $F_\mathrm{coll}=\infty$, the average impact vaporization flux increases above the confidence interval significantly. Around Mercury's perihelion the impact vaporization flux increases to 25\% higher values as compared to the preferred solution. This is a consequence of higher mass transfer rates of JFC and OCC meteoroids for these SFD/MFD (solid magenta line in Fig. \ref{FIG_DISC_EXTREME}). At aphelion the impact vaporization flux is a few percent above the confidence interval. The ratio between the maximum and minimum value along the whole orbit is $\mathcal{R}_\mathrm{max}/\mathcal{R}_\mathrm{min}=5.45$ (we omitted one outlying point in this analysis at TAA = 6.3$^\circ$). If, on the other hand, the shortest collisional lifetimes in our model are assumed, $F_\mathrm{coll}=1$, we obtain a much flatter solution than our preferred model (solid green line), where the ratio between the maximum and minimum impact vaporization flux is $\mathcal{F}_\mathrm{max}/\mathcal{R}_\mathrm{min}=3.22$. 

Note, that for all settings of $F_\mathrm{coll}$ and $\alpha,\beta$ we keep the amount of mass accreted on Earth constant. Increasing the collisional lifetime and changing the SFD/MFD does not thus change the mass accreted on Earth, however, the production rates required for the source populations to generate the amount of accreted meteoroids change dramatically. For instance, for very short collisional lifetimes and large grain dominated SFD/MFD, the asteroid belt might require to constantly produce an extreme amount of meteoroids, which might not be possible based on our current knowledge. Since this issue greatly exceeds the focus of this paper we leave it for future work. 
 
\begin{figure}[h]
\centering
\includegraphics[width=0.9\textwidth]{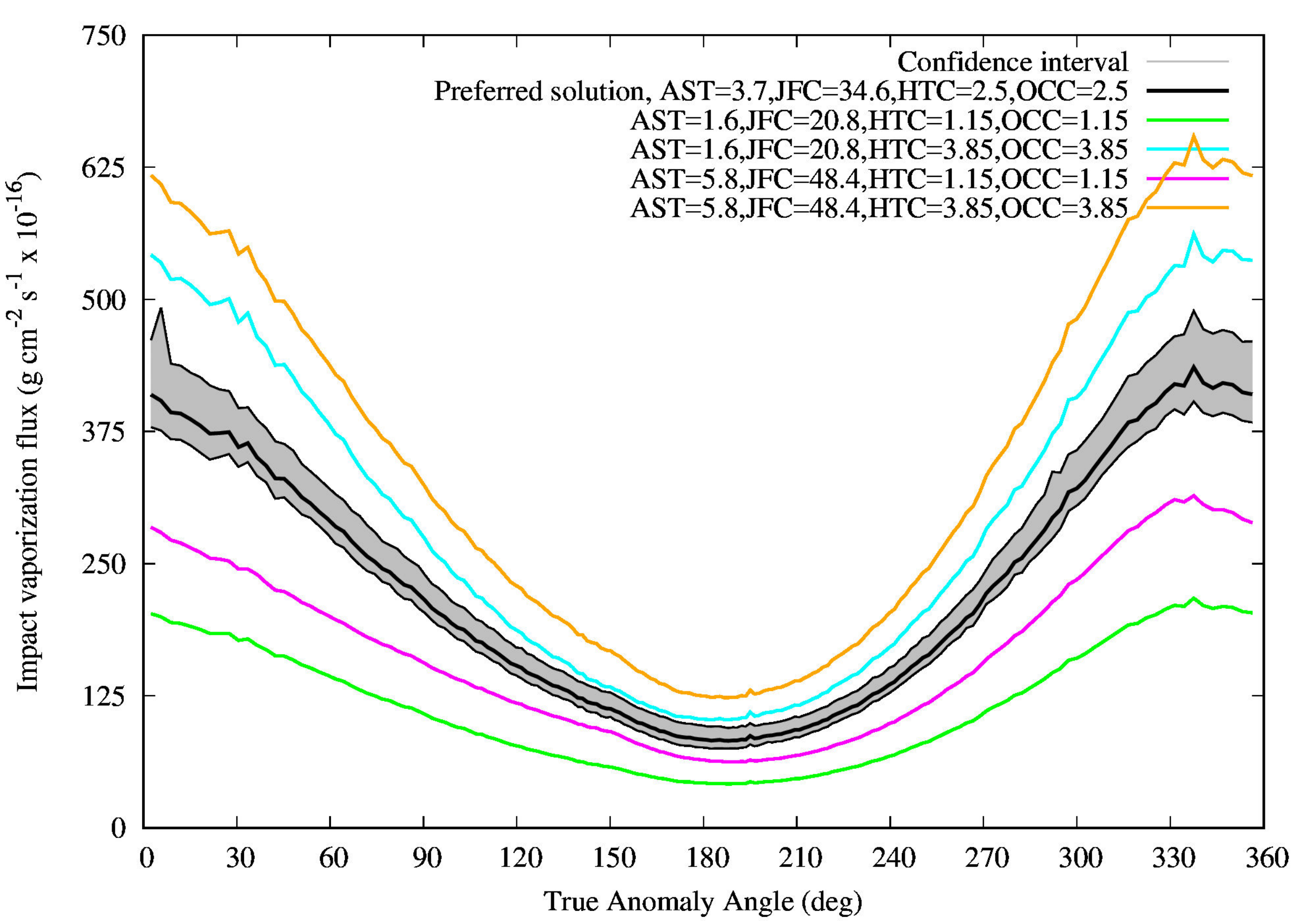}
\caption{The same as Fig. \ref{FIG_PREF_BURGER}, but now for different mixing ratios permitted by \citet{Carrillo-Sanchez_etal_2016}. The lowest mass influx on Earth, $\mathrm{AST}~=1.6\mathrm{~t~day}^{-1},\mathrm{JFC}~=20.8\mathrm{~t~day}^{-1},\mathrm{HTC}~=1.15\mathrm{~t~day}^{-1},\mathrm{OCC}~=1.15\mathrm{~t~day}^{-1}$ (solid green line) gives on average 48\% smaller impact vaporization flux compared to the preferred solution, while mixed inner and outer solar \added{system} sources, $\mathrm{AST}~=1.6\mathrm{~t~day}^{-1},\mathrm{JFC}~=20.8\mathrm{~t~day}^{-1},\mathrm{HTC}~=3.85\mathrm{~t~day}^{-1},\mathrm{OCC}~=3.85\mathrm{~t~day}^{-1}$ (solid cyan line) and $\mathrm{AST}~=5.8\mathrm{~t~day}^{-1},\mathrm{JFC}~=48.4\mathrm{~t~day}^{-1},\mathrm{HTC}~=1.15\mathrm{~t~day}^{-1},\mathrm{OCC}~=1.15\mathrm{~t~day}^{-1}$ \replaced{(solid orange line)}{magenta} give values within the confidence interval of our preferred solution, and the highest mass influx $\mathrm{AST}~=5.8\mathrm{~t~day}^{-1},\mathrm{JFC}~=48.4\mathrm{~t~day}^{-1},\mathrm{HTC}~=3.85\mathrm{~t~day}^{-1},\mathrm{OCC}~=3.85\mathrm{~t~day}^{-1}$ (solid orange line) provides on average 33\% higher impact vaporization flux. }
\label{FIG_DISC_CARRILLO}
 \end{figure}

 To test the sensitivity of our preferred model for Mercury to uncertainties of the mass influx of different meteoroid populations at Earth given by \citet{Carrillo-Sanchez_etal_2016}, we fix the free parameters $F_\mathrm{coll}=20, \alpha=4, \beta=2$ and alternate the mixing ratios of different populations (Fig. \ref{FIG_DISC_CARRILLO}). The lowest mass influx on Earth within the \citet{Carrillo-Sanchez_etal_2016} range, $\mathrm{AST}~=1.6\mathrm{~t~day}^{-1},\mathrm{JFC}~=20.8\mathrm{~t~day}^{-1},\mathrm{HTC}~=1.15\mathrm{~t~day}^{-1},\mathrm{OCC}~=1.15\mathrm{~t~day}^{-1}$, is represented by the solid green line and gives on average 48\% smaller impact vaporization flux compared to the preferred solution. The upper boundary on the other hand gives on average 33\% higher impact vaporization flux than our preferred solution, $\mathrm{AST}~=5.8\mathrm{~t~day}^{-1},\mathrm{JFC}~=48.4\mathrm{~t~day}^{-1},\mathrm{HTC}~=3.85\mathrm{~t~day}^{-1},\mathrm{OCC}~=3.85\mathrm{~t~day}^{-1}$ (solid orange line). We also picked two intermediate states where the inner solar system sources, the AST and JFC meteoroids, and the outer solar system sources, the HTC and OCC meteoroids, were given a combination of the lowest and highest permitted values  $\mathrm{AST}~=1.6\mathrm{~t~day}^{-1},\mathrm{JFC}~=20.8\mathrm{~t~day}^{-1},\mathrm{HTC}~=3.85\mathrm{~t~day}^{-1},\mathrm{OCC}~=3.85\mathrm{~t~day}^{-1}$ (solid cyan line) and $\mathrm{AST}~=5.8\mathrm{~t~day}^{-1},\mathrm{JFC}~=48.4\mathrm{~t~day}^{-1},\mathrm{HTC}~=1.15\mathrm{~t~day}^{-1},\mathrm{OCC}~=1.15\mathrm{~t~day}^{-1}$ (solid orange line).
We test only the sensitivity with respect to different abundances of individual meteoroid populations and their uncertainty intervals estimated by \citet{Carrillo-Sanchez_etal_2016} and omit further sensitivity to other free parameters. We conclude that even for the extreme values permitted by \citet{Carrillo-Sanchez_etal_2016} our model does not vary by more than a factor of 2 with respect to our preferred solution

  One of the outstanding questions about our choice of free parameters is whether they are degenerate or oversimplified. The value of the collisional lifetime multiplier $F_\mathrm{coll}$ is kept the same for all meteoroid populations in our model, however this might not be entirely correct due to the different nature of meteoroids from different sources, e.g. the main-belt asteroid meteoroids might be stronger, harder to break compared to cometary meteoroids especially from the Oort cloud. Introducing these and testing these factors is beyond the scope of this study mainly because $F_\mathrm{coll}$ is unconstrained for most meteoroid populations.
 Our selection of a single-power law SFD/MFD might seem to be an oversimplification given that recent dynamical models \citep{Nesvorny_etal_2010,Nesvorny_etal_2011JFC,Pokorny_etal_2014} were using three parametric broken-power law SFD/MFD. 
 By investigating a wide range of the SFD/MFD single power-law indices we show how the results change if the meteoroid population is dominated by smaller meteoroids (high values of $\alpha,\beta$) and then gradually the importance lies in larger meteoroids (lower values of $\alpha,\beta$). Due to mostly smooth transitions between larger/smaller modes we assume that models with more complicated power-laws would lie somewhere in the range of models presented here. Another simplification lies in keeping the SFD/MFD the same for all meteoroid populations. While the possibility of alternating SFD/MFD of individual populations in our model is viable, due to added complexity we keep this element for future investigations. 

 
Our calibrated model appears to grossly overestimate observed vaporization fluxes. To quantify by how much, we assumed a Mg fraction of $f_\mathrm{Mg}=0.17$ in the high-Mg regions of Mercury \citep{Lawrence_etal_2017}, and converted the peak vaporization flux, $\mathcal{F}=436 \times 10^{-16}$ g cm$^{-2}$ s$^{-1}$, to the number of Mg atoms expected to be vaporized from the surface, $\mathcal{N}_\mathrm{Mg}=1.8 \times 10^{8}$ Mg atoms cm$^{-2}$ s$^{-1}$, using $\mathcal{N}_\mathrm{Mg}=N_\mathrm{A}\mathcal{F}f_\mathrm{Mg}/m_\mathrm{Mg}$, where $N_\mathrm{A}=6.022\times10^{23}\mathrm{~mol}^{-1}$ is the Avogadro constant, and $m_\mathrm{Mg}=24.305$ is the atomic weight of magnesium. This estimate exceeds the inferred Mg source rate from MESSENGER measurements, $8 \times 10^{5}$ Mg atoms cm$^{-2}$ s$^{-1}$ \citep{Merkel_etal_2017}, by more than a factor of 100, whereas the variation of Mg on the surface is only about a factor of two \citep{Lawrence_etal_2017}. 
Using results in Fig \ref{FIG_PREF_VAPOR_VELOCITY} and applying a simple impact velocity cut-off where the impact vaporization does no longer contribute for $V_\mathrm{imp}> 17$ km s$^{-1}$ provides  $\mathcal{N}_\mathrm{Mg}$ comparable to rates observed by MESSENGER for a wide range of TAAs.
It seems clear that high-velocity impacts, especially due to impactors in retrograde orbits, lead to ionization, but the ionized fraction is uncertain. The impact charge generated by the incidence of a dust particle on a target increases as  $mV_\mathrm{imp}^{3.5}$ \citep{Auer_2001}, which is faster than the increase with impactor velocity of the total impact vapor from \citet{Cintala_1992}. Therefore, the ionized fraction will increase with impactor velocity $V_\mathrm{imp}$. Choosing different parameters for impactor and target from \citet{Cintala_1992} does not yield significantly different results for the total vapor, on average less than 10\%. Thus, since our transport rate calculations show that it is impossible to decrease the flux on Mercury by more than a factor of two for any reasonable combination of collisional lifetime and SFD/MFD and at the same time explain the arrival rates of meteoroids at Earth, our findings suggest that the vaporization function is questionable at v$>20$ km/s, possibly due to losses to ionization and other phases like condensates \citep{Berezhnoy_Klumov_2008}.
 

 \subsection{Conclusions}
 
 In this manuscript, we presented modeled results of the orbital element distributions, impacting meteoroid radiant maps, transportation rates and ratio between mass accreted on Mercury and Earth for meteoroids originating in the main-belt, Jupiter family comets, Halley-type comets, and Oort cloud comets. Our model includes the effects of meteoroid collisions with the Zodiacal cloud through \citet{Steel_Elford_1986} collisional model refined for \citet{Cremonese_etal_2012} meteoroid flux estimates.
 
 The main-belt meteoroids and Jupiter family comet meteoroids impact Mercury preferentially with lower eccentricities $e<0.2$ and small inclinations $I<30^\circ$ which results in impact velocities $V_\mathrm{imp}<70$ km s$^{-1}$ at perihelion and $V_\mathrm{imp}<50$ km s$^{-1}$ at aphelion. Approximately 7\% of mass accreted at Earth originating from main-belt asteroids is recorded at Mercury. This is due to low eccentricity orbits of main-belt meteoroids resulting in low relative impact velocities with both Earth and Mercury. Such low velocities are efficiently attracted by Earth's gravity while this effect is much smaller for Mercury. Jupiter family comet meteoroids have a broader distribution of eccentricities, which weakens the gravitational focusing and leads to higher mass accretion at Mercury as compared to that at Earth, $\sim 23\%$.
 
 The meteoroids produced by Halley-type and Oort cloud comets impact Mercury with a flat eccentricity distribution and a bimodal distribution of orbital inclinations of prograde and retrograde orbits. Retrograde meteoroids are predominantly impacting from the apex direction, have impact velocities $95<V_\mathrm{imp}<120$ km s$^{-1}$ at perihelion and $75<V_\mathrm{imp}<90$ km s$^{-1}$ at aphelion and are less influenced by Mercury's orbital motion. Both long-period comet meteoroid populations have 
 significantly higher Mercury to Earth mass accretion ratios as compared to Jupiter family comet meteoroids, $\sim 70\%$ for Halley-type comet meteoroids and $\sim 90\%$ for Oort cloud comet meteoroids,

 Our preferred solution for a dynamical model of meteoroids at Mercury based on calibration from \citet{Carrillo-Sanchez_etal_2016} provides the following values of accreted mass averaged over the entire Hermean orbit; main-belt asteroid meteoroids $\mathrm{AST}_M=0.26\pm0.15 \mathrm{~t~day}^{-1},$ Jupiter family comet meteoroids $\mathrm{JFC}_M=7.84\pm3.13\mathrm{~t~day}^{-1},$ Halley-type comets $\mathrm{HTC}_M=1.69\pm0.91\mathrm{~t~day}^{-1},$ and Oort cloud comets $\mathrm{ OCC}_M=2.37\pm1.38\mathrm{~t~day}^{-1}$. This results in a mass influx ratio of short/long-period comet meteoroids $\sim 2$, which is much lower than that at Earth \citep[$\sim 7$,][]{Carrillo-Sanchez_etal_2016}.
 The vaporization flux averaged over one Mercury orbit then yields $\mathcal{F}_\mathrm{orbit}=200\pm16 \times 10^{-16}$ g cm$^{-2}$ s$^{-1}$, with maximum at TAA = 337$^\circ$, $\mathcal{F}_\mathrm{max}=436\pm57\times 10^{-16}$ g cm$^{-2}$ s$^{-1}$ and minimum at TAA = 188$^\circ$, $\mathcal{F}_\mathrm{min}=82\pm12\times 10^{-16}$ g cm$^{-2}$ s$^{-1}$. Our model provides impact vaporization fluxes of the same order of magnitude as \citet{Borin_etal_2009,Borin_etal_2017}. However, this similarity is misleading because we predict a lower mass flux of meteoroids at Mercury (Mercury/Earth ratio less than 1 vs 35 in their model) accompanied by higher speed distributions.  The variations of the impact vaporization flux and the impact directionality also contradict the result of \citet{Borin_etal_2009,Borin_etal_2017} because our model predicts larger perihelion-to-aphelion ratio in the impact vaporization rate. Furthermore, we predict that meteoroid impacts onto Mercury provide a significant source of surface ions. Only $\sim 1\%$ of the estimated total vapor appears to contribute neutrals to Mercury's exosphere.
 
 Our preferred solution shows a strong dawn/dusk asymmetry in both meteoroid impact direction distribution and the impact vaporization pattern on the surface. The impact vaporization pattern undergoes significant movements during Mercury's orbit, being centered at the dawn terminator (6 hr) at perihelion and aphelion and moving towards the night-side during the first leg (maximum displacement of the center is $~3$ hr) and being shifted toward the day-side in the second leg when Mercury is moving back to its pericenter. 
 The impact vaporization flux integrated over the whole surface has a similar pattern to source rates for Calcium presented by \citet{Burger_etal_2014} with a few exceptions: enhancements at TAA = 20$^\circ$ and TAA = 170$^\circ$ that are thought to originate from 2P/Encke \citep{Christou_etal_2015,Killen_Hahn_2015} and TAA from 315$^\circ$ to 350$^\circ$ where our model predicts 30-40\% higher relative impact vaporization flux than \citet{Burger_etal_2014}.


%

\acknowledgments
 The work was supported with NASA's SSO and LDAP awards. \added{We would like to thank David Asher for an excellent review that helped to improve the manuscript}. PP would like to thank Marc Kuchner and Laura Lenki\'{c} for enlightening discussions, useful comments, and thorough reads of the manuscript.

\bibliography{Papers}

\begin{thebibliography}{}
\expandafter\ifx\csname natexlab\endcsname\relax\def\natexlab#1{#1}\fi
\providecommand{\url}[1]{\href{#1}{#1}}

\bibitem[{{Auer}(2001)}]{Auer_2001}
{Auer}, S. 2001, Interplanetary Dust, Edited by E.~Gr{\"u}n, B.A.S.~Gustafson,
  S.~Dermott, and H.~Fechtig.~Astronomy and Astrophysics Library.~2001, 804 p.,
  ISBN: 3-540-42067-3.~ Berlin: Springer, 2001, p.~385, 385

\bibitem[{{Berezhnoy} \& {Klumov}(2008)}]{Berezhnoy_Klumov_2008}
{Berezhnoy}, A.~A., \& {Klumov}, B.~A. 2008, \icarus, 195, 511

\bibitem[{{Blaauw} {et~al.}(2011){Blaauw}, {Campbell-Brown}, \&
  {Weryk}}]{Blaauw_etal_2011}
{Blaauw}, R.~C., {Campbell-Brown}, M.~D., \& {Weryk}, R.~J. 2011, \mnras, 412,
  2033

\bibitem[{{Borin} {et~al.}(2016{\natexlab{a}}){Borin}, {Cremonese}, {Bruno}, \&
  {Marzari}}]{Borin_etal_2016b}
{Borin}, P., {Cremonese}, G., {Bruno}, M., \& {Marzari}, F. 2016{\natexlab{a}},
  \icarus, 264, 220

\bibitem[{{Borin} {et~al.}(2016{\natexlab{b}}){Borin}, {Cremonese}, \&
  {Marzari}}]{Borin_etal_2016a}
{Borin}, P., {Cremonese}, G., \& {Marzari}, F. 2016{\natexlab{b}}, \aap, 585,
  A106

\bibitem[{{Borin} {et~al.}(2009){Borin}, {Cremonese}, {Marzari}, {Bruno}, \&
  {Marchi}}]{Borin_etal_2009}
{Borin}, P., {Cremonese}, G., {Marzari}, F., {Bruno}, M., \& {Marchi}, S. 2009,
  \aap, 503, 259

\bibitem[{{Borin} {et~al.}(2017){Borin}, {Cremonese}, {Marzari}, \&
  {Lucchetti}}]{Borin_etal_2017}
{Borin}, P., {Cremonese}, G., {Marzari}, F., \& {Lucchetti}, A. 2017, \aap,
  605, A94

\bibitem[{{Burger} {et~al.}(2014){Burger}, {Killen}, {McClintock}, {Merkel},
  {Vervack}, {Cassidy}, \& {Sarantos}}]{Burger_etal_2014}
{Burger}, M.~H., {Killen}, R.~M., {McClintock}, W.~E., {et~al.} 2014, \icarus,
  238, 51

\bibitem[{{Burns} {et~al.}(1979){Burns}, {Lamy}, \& {Soter}}]{Burns_etal_1979}
{Burns}, J.~A., {Lamy}, P.~L., \& {Soter}, S. 1979, \icarus, 40, 1

\bibitem[{{Campbell-Brown}(2008)}]{Campbell-Brown_2008}
{Campbell-Brown}, M.~D. 2008, \icarus, 196, 144

\bibitem[{{Carrillo-S{\'a}nchez} {et~al.}(2016){Carrillo-S{\'a}nchez},
  {Nesvorn{\'y}}, {Pokorn{\'y}}, {Janches}, \&
  {Plane}}]{Carrillo-Sanchez_etal_2016}
{Carrillo-S{\'a}nchez}, J.~D., {Nesvorn{\'y}}, D., {Pokorn{\'y}}, P.,
  {Janches}, D., \& {Plane}, J.~M.~C. 2016, \grl, 43, 11

\bibitem[{{Christou} {et~al.}(2015){Christou}, {Killen}, \&
  {Burger}}]{Christou_etal_2015}
{Christou}, A.~A., {Killen}, R.~M., \& {Burger}, M.~H. 2015, \grl, 42, 7311

\bibitem[{{Cintala}(1992)}]{Cintala_1992}
{Cintala}, M.~J. 1992, \jgr, 97, 947

\bibitem[{{Cremonese} {et~al.}(2012){Cremonese}, {Borin}, {Martellato},
  {Marzari}, \& {Bruno}}]{Cremonese_etal_2012}
{Cremonese}, G., {Borin}, P., {Martellato}, E., {Marzari}, F., \& {Bruno}, M.
  2012, \apjl, 749, L40

\bibitem[{{Dohnanyi}(1969)}]{Dohnanyi_1969}
{Dohnanyi}, J.~S. 1969, \jgr, 74, 2531

\bibitem[{{Fulle} {et~al.}(1995){Fulle}, {Colangeli}, {Mennella}, {Rotundi}, \&
  {Bussoletti}}]{Fulle_etal_1995}
{Fulle}, M., {Colangeli}, L., {Mennella}, V., {Rotundi}, A., \& {Bussoletti},
  E. 1995, \aap, 304, 622

\bibitem[{{Fulle} {et~al.}(2016){Fulle}, {Marzari}, {Della Corte}, {Fornasier},
  {Sierks}, {Rotundi}, {Barbieri}, {Lamy}, {Rodrigo}, {Koschny}, {Rickman},
  {Keller}, {L{\'o}pez-Moreno}, {Accolla}, {Agarwal}, {A'Hearn}, {Altobelli},
  {Barucci}, {Bertaux}, {Bertini}, {Bodewits}, {Bussoletti}, {Colangeli},
  {Cosi}, {Cremonese}, {Crifo}, {Da Deppo}, {Davidsson}, {Debei}, {De Cecco},
  {Esposito}, {Ferrari}, {Giovane}, {Gustafson}, {Green}, {Groussin},
  {Gr{\"u}n}, {Gutierrez}, {G{\"u}ttler}, {Herranz}, {Hviid}, {Ip},
  {Ivanovski}, {Jer{\'o}nimo}, {Jorda}, {Knollenberg}, {Kramm}, {K{\"u}hrt},
  {K{\"u}ppers}, {Lara}, {Lazzarin}, {Leese}, {L{\'o}pez-Jim{\'e}nez},
  {Lucarelli}, {Mazzotta Epifani}, {McDonnell}, {Mennella}, {Molina},
  {Morales}, {Moreno}, {Mottola}, {Naletto}, {Oklay}, {Ortiz}, {Palomba},
  {Palumbo}, {Perrin}, {Rietmeijer}, {Rodr{\'{\i}}guez}, {Sordini}, {Thomas},
  {Tubiana}, {Vincent}, {Weissman}, {Wenzel}, {Zakharov}, \&
  {Zarnecki}}]{Fulle_etal_2016}
{Fulle}, M., {Marzari}, F., {Della Corte}, V., {et~al.} 2016, \apj, 821, 19

\bibitem[{{Green} {et~al.}(2007){Green}, {McBride}, {Colwell}, {McDonnell},
  {Tuzzolino}, {Economou}, {Clark}, {Sekanina}, {Tsou}, \&
  {Brownlee}}]{Green_etal_2007}
{Green}, S.~F., {McBride}, N., {Colwell}, M.~T.~S.~H., {et~al.} 2007, Dust in
  Planetary Systems, 643, 35

\bibitem[{{Greenberg}(1982)}]{Greenberg_1982}
{Greenberg}, R. 1982, \aj, 87, 184

\bibitem[{{Grun} {et~al.}(1985){Grun}, {Zook}, {Fechtig}, \&
  {Giese}}]{Grun_etal_1985}
{Grun}, E., {Zook}, H.~A., {Fechtig}, H., \& {Giese}, R.~H. 1985, \icarus, 62,
  244

\bibitem[{{Janches} {et~al.}(2009){Janches}, {Dyrud}, {Broadley}, \&
  {Plane}}]{Janches_etal_2009}
{Janches}, D., {Dyrud}, L.~P., {Broadley}, S.~L., \& {Plane}, J.~M.~C. 2009,
  \grl, 36, L06101

\bibitem[{Janches {et~al.}(2018)Janches, Pokorný, Sarantos, Szalay, Horányi,
  \& Nesvorný}]{Janches_etal_2018}
Janches, D., Pokorný, P., Sarantos, M., {et~al.} 2018, Geophysical Research
  Letters, n/a, 2017GL076065.
\newblock \url{http://dx.doi.org/10.1002/2017GL076065}

\bibitem[{{Janches} {et~al.}(2015){Janches}, {Close}, {Hormaechea},
  {Swarnalingam}, {Murphy}, {O'Connor}, {Vandepeer}, {Fuller}, {Fritts}, \&
  {Brunini}}]{Janches_etal_2015}
{Janches}, D., {Close}, S., {Hormaechea}, J.~L., {et~al.} 2015, \apj, 809, 36

\bibitem[{{Kessler}(1981)}]{Kessler_1981}
{Kessler}, D.~J. 1981, \icarus, 48, 39

\bibitem[{{Killen} \& {Hahn}(2015)}]{Killen_Hahn_2015}
{Killen}, R.~M., \& {Hahn}, J.~M. 2015, \icarus, 250, 230

\bibitem[{{Lawrence} {et~al.}(2017){Lawrence}, {Peplowski}, {Beck}, {Feldman},
  {Frank}, {McCoy}, {Nittler}, \& {Solomon}}]{Lawrence_etal_2017}
{Lawrence}, D.~J., {Peplowski}, P.~N., {Beck}, A.~W., {et~al.} 2017, \icarus,
  281, 32

\bibitem[{{Levison} \& {Duncan}(2013)}]{Levison_Duncan_2013}
{Levison}, H.~F., \& {Duncan}, M.~J. 2013, {SWIFT: A solar system integration
  software package}, Astrophysics Source Code Library, , , ascl:1303.001

\bibitem[{{Levison} {et~al.}(2006){Levison}, {Duncan}, {Dones}, \&
  {Gladman}}]{Levison_etal_2006}
{Levison}, H.~F., {Duncan}, M.~J., {Dones}, L., \& {Gladman}, B.~J. 2006,
  \icarus, 184, 619

\bibitem[{{Love} \& {Brownlee}(1993)}]{Love_Brownlee_1993}
{Love}, S.~G., \& {Brownlee}, D.~E. 1993, Science, 262, 550

\bibitem[{{Marchi} {et~al.}(2005){Marchi}, {Morbidelli}, \&
  {Cremonese}}]{Marchi_etal_2005}
{Marchi}, S., {Morbidelli}, A., \& {Cremonese}, G. 2005, \aap, 431, 1123

\bibitem[{{Marsden}(2005)}]{Marsden_2005}
{Marsden}, B.~G. 2005, \araa, 43, 75

\bibitem[{{Merkel} {et~al.}(2017){Merkel}, {Cassidy}, {Vervack}, {McClintock},
  {Sarantos}, {Burger}, \& {Killen}}]{Merkel_etal_2017}
{Merkel}, A.~W., {Cassidy}, T.~A., {Vervack}, R.~J., {et~al.} 2017, \icarus,
  281, 46

\bibitem[{{Nesvorn{\'y}} {et~al.}(2011{\natexlab{a}}){Nesvorn{\'y}}, {Janches},
  {Vokrouhlick{\'y}}, {Pokorn{\'y}}, {Bottke}, \&
  {Jenniskens}}]{Nesvorny_etal_2011JFC}
{Nesvorn{\'y}}, D., {Janches}, D., {Vokrouhlick{\'y}}, D., {et~al.}
  2011{\natexlab{a}}, \apj, 743, 129

\bibitem[{{Nesvorn{\'y}} {et~al.}(2010){Nesvorn{\'y}}, {Jenniskens}, {Levison},
  {Bottke}, {Vokrouhlick{\'y}}, \& {Gounelle}}]{Nesvorny_etal_2010}
{Nesvorn{\'y}}, D., {Jenniskens}, P., {Levison}, H.~F., {et~al.} 2010, \apj,
  713, 816

\bibitem[{{Nesvorn{\'y}} {et~al.}(2017){Nesvorn{\'y}}, {Vokrouhlick{\'y}},
  {Dones}, {Levison}, {Kaib}, \& {Morbidelli}}]{Nesvorny_etal_2017}
{Nesvorn{\'y}}, D., {Vokrouhlick{\'y}}, D., {Dones}, L., {et~al.} 2017, \apj,
  845, 27

\bibitem[{{Nesvorn{\'y}} {et~al.}(2011{\natexlab{b}}){Nesvorn{\'y}},
  {Vokrouhlick{\'y}}, {Pokorn{\'y}}, \& {Janches}}]{Nesvorny_etal_2011OCC}
{Nesvorn{\'y}}, D., {Vokrouhlick{\'y}}, D., {Pokorn{\'y}}, P., \& {Janches}, D.
  2011{\natexlab{b}}, \apj, 743, 37

\bibitem[{{\"{O}pik}(1951)}]{Opik_1951}
{\"{O}pik}, E.~J. 1951, Proc.~R.~Irish Acad.~Sect.~A, vol.~54, p.~165-199
  (1951)., 54, 165

\bibitem[{{Planck Collaboration} {et~al.}(2014){Planck Collaboration}, {Ade},
  {Aghanim}, {Armitage-Caplan}, {Arnaud}, {Ashdown}, {Atrio-Barandela},
  {Aumont}, {Baccigalupi}, {Banday}, \& et~al.}]{Planck_2014}
{Planck Collaboration}, {Ade}, P.~A.~R., {Aghanim}, N., {et~al.} 2014, \aap,
  571, A14

\bibitem[{{Pokorn{\'y}} \& {Brown}(2016)}]{Pokorny_Brown_2016}
{Pokorn{\'y}}, P., \& {Brown}, P.~G. 2016, \aap, 592, A150

\bibitem[{{Pokorn{\'y}} {et~al.}(2017){Pokorn{\'y}}, {Sarantos}, \&
  {Janches}}]{Pokorny_etal_2017_APJL}
{Pokorn{\'y}}, P., {Sarantos}, M., \& {Janches}, D. 2017, \apjl, 842, L17

\bibitem[{{Pokorn{\'y}} \&
  {Vokrouhlick{\'y}}(2013)}]{Pokorny_Vokrouhlicky_2013}
{Pokorn{\'y}}, P., \& {Vokrouhlick{\'y}}, D. 2013, \icarus, 226, 682

\bibitem[{{Pokorn{\'y}} {et~al.}(2014){Pokorn{\'y}}, {Vokrouhlick{\'y}},
  {Nesvorn{\'y}}, {Campbell-Brown}, \& {Brown}}]{Pokorny_etal_2014}
{Pokorn{\'y}}, P., {Vokrouhlick{\'y}}, D., {Nesvorn{\'y}}, D.,
  {Campbell-Brown}, M., \& {Brown}, P. 2014, \apj, 789, 25

\bibitem[{{Poppe}(2016)}]{Poppe_2016}
{Poppe}, A.~R. 2016, \icarus, 264, 369

\bibitem[{{Rickman} {et~al.}(2014){Rickman}, {Wi{\'s}niowski}, {Wajer},
  {Gabryszewski}, \& {Valsecchi}}]{Rickman_etal_2014}
{Rickman}, H., {Wi{\'s}niowski}, T., {Wajer}, P., {Gabryszewski}, R., \&
  {Valsecchi}, G.~B. 2014, \aap, 569, A47

\bibitem[{{Rotundi} {et~al.}(2015){Rotundi}, {Sierks}, {Della Corte}, {Fulle},
  {Gutierrez}, {Lara}, {Barbieri}, {Lamy}, {Rodrigo}, {Koschny}, {Rickman},
  {Keller}, {L{\'o}pez-Moreno}, {Accolla}, {Agarwal}, {A'Hearn}, {Altobelli},
  {Angrilli}, {Barucci}, {Bertaux}, {Bertini}, {Bodewits}, {Bussoletti},
  {Colangeli}, {Cosi}, {Cremonese}, {Crifo}, {Da Deppo}, {Davidsson}, {Debei},
  {De Cecco}, {Esposito}, {Ferrari}, {Fornasier}, {Giovane}, {Gustafson},
  {Green}, {Groussin}, {Gr{\"u}n}, {G{\"u}ttler}, {Herranz}, {Hviid}, {Ip},
  {Ivanovski}, {Jer{\'o}nimo}, {Jorda}, {Knollenberg}, {Kramm}, {K{\"u}hrt},
  {K{\"u}ppers}, {Lazzarin}, {Leese}, {L{\'o}pez-Jim{\'e}nez}, {Lucarelli},
  {Lowry}, {Marzari}, {Epifani}, {McDonnell}, {Mennella}, {Michalik}, {Molina},
  {Morales}, {Moreno}, {Mottola}, {Naletto}, {Oklay}, {Ortiz}, {Palomba},
  {Palumbo}, {Perrin}, {Rodr{\'{\i}}guez}, {Sabau}, {Snodgrass}, {Sordini},
  {Thomas}, {Tubiana}, {Vincent}, {Weissman}, {Wenzel}, {Zakharov}, \&
  {Zarnecki}}]{Rotundi_etal_2015}
{Rotundi}, A., {Sierks}, H., {Della Corte}, V., {et~al.} 2015, Science, 347,
  aaa3905

\bibitem[{{Steel} \& {Elford}(1986)}]{Steel_Elford_1986}
{Steel}, D.~I., \& {Elford}, W.~G. 1986, \mnras, 218, 185

\bibitem[{{Suggs} {et~al.}(2014){Suggs}, {Moser}, {Cooke}, \&
  {Suggs}}]{Suggs_etal_2014}
{Suggs}, R.~M., {Moser}, D.~E., {Cooke}, W.~J., \& {Suggs}, R.~J. 2014,
  \icarus, 238, 23

\bibitem[{{Vokrouhlick{\'y}} {et~al.}(2012){Vokrouhlick{\'y}}, {Pokorn{\'y}},
  \& {Nesvorn{\'y}}}]{Vokrouhlicky_etal_2012}
{Vokrouhlick{\'y}}, D., {Pokorn{\'y}}, P., \& {Nesvorn{\'y}}, D. 2012, \icarus,
  219, 150

\bibitem[{{Vondrak} {et~al.}(2008){Vondrak}, {Plane}, {Broadley}, \&
  {Janches}}]{Vondrak_etal_2008}
{Vondrak}, T., {Plane}, J.~M.~C., {Broadley}, S., \& {Janches}, D. 2008,
  Atmospheric Chemistry \& Physics, 8, 7015

\bibitem[{{Wang} \& {Brasser}(2014)}]{Wang_Brasser_2014}
{Wang}, J.-H., \& {Brasser}, R. 2014, \aap, 563, A122

\bibitem[{{Wetherill}(1967)}]{Wetherill_1967}
{Wetherill}, G.~W. 1967, \jgr, 72, 2429

\end{thebibliography}



\end{document}